\documentclass[aps,prc,twocolumn,longbibliography,reprint,superscriptaddress,amsmath,nofootinbib]{revtex4-1}

\usepackage[colorlinks, urlcolor=blue,linkcolor=blue, anchorcolor=blue, citecolor=blue]{hyperref}
\usepackage[utf8]{inputenc}
\usepackage{lineno}
\hypersetup{bookmarksopen=true}

\usepackage{amssymb}
\usepackage{amsmath}
\usepackage{amsfonts}
\usepackage{bm}
\usepackage{mathcomp}
\usepackage{graphicx}
\usepackage{subfigure}
\usepackage{multirow}
\usepackage[dvipsnames]{xcolor}
\usepackage{soul}

\graphicspath{{fig/}}

\newcommand{\D}{\Delta}
\newcommand{\plus}{{(+)}}
\newcommand{\hatbarphi}{\tilde{\bar{\phi}}_0}
\newcommand{\hatxi}{\tilde{\xi}}
\newcommand{\bes}{{\sc BEShydro}}
\newcommand{\fm}{fm$^{-1}$}

\newcommand{\V}{n}
\newcommand{\n}{n}
\newcommand{\ed}{e}

\newcommand{\peq}{p}

\newcommand{\cp}{c_p}
\newcommand{\cpz}{c_{p,0}}

\begin{document}
\title{Fluctuation dynamics near the QCD critical point}
\author{Lipei Du}
\affiliation{Department of Physics, The Ohio State University, Columbus, OH 43210}
\author{Ulrich Heinz}
\affiliation{Department of Physics, The Ohio State University, Columbus, OH 43210}
\affiliation{Institut f\"ur Theoretische Physik, J.~W.~Goethe-Universit\"at, Max-von-Laue-Str.~1, D-60438 Frankfurt am Main, Germany}
\author{Krishna Rajagopal}
\affiliation{Center for Theoretical Physics, Massachusetts Institute of Technology, Cambridge, MA 02139}
\author{Yi Yin}
\affiliation{Center for Theoretical Physics, Massachusetts Institute of Technology, Cambridge, MA 02139}
\affiliation{Quark Matter Research Center, Institute of Modern Physics,
Chinese Academy of Sciences, Lanzhou, Gansu, 073000, China}
\affiliation{University of Chinese Academy of Sciences, Beijing, 100049, China}

\date{\today}

\begin{abstract}
    The evolution of non-hydrodynamic slow processes near the QCD critical point is explored with the novel {\sc hydro+} framework, which extends the conventional hydrodynamic description by coupling it to additional explicitly evolving slow modes
    describing long wavelength fluctuations. Their slow relaxation is controlled by critical behavior of the correlation length and is independent from gradients of matter density and pressure that control the evolution of the hydrodynamic quantities. In this exploratory study we follow the evolution of the slow modes on top of a simplified QCD matter background, allowing us to clearly distinguish, and study both separately and in combination, the main effects controlling the dynamics of critical slow modes. In particular, we show how the evolution of the slow modes depend on their wave number, the expansion of and advection by the fluid background, and the behavior of the correlation length. Non-equilibrium contributions from the slow modes to bulk matter properties that affect the bulk dynamics (entropy, pressure, temperature and chemical potential) are  discussed and found to be small.
\end{abstract}

\maketitle


\newpage
%
\section{Introduction}
\label{sec:intro}
\vspace*{-2mm}
%

Is there a critical point (CP) in the QCD phase diagram that connects the rapid but smooth crossover between color confined (hadron gas) and color deconfined matter (quark-gluon plasma) at small baryon chemical potential $\mu$ \cite{Aoki:2006we, Bazavov:2009zn, Borsanyi:2010cj, Bazavov:2011nk} with a first order phase transition at large $\mu$ \cite{Stephanov:2007fk,Bzdak:2019pkr}? Much recent effort has gone into answering this still unresolved question \cite{Berges:1998rc,Halasz:1998qr,Stephanov:1998dy,Stephanov:1999zu,Rajagopal:2000wf,Hatta:2003wn,Stephanov:2008qz,Fukushima:2010bq,Stephanov:2011pb,Luo:2017faz,Busza:2018rrf,Bzdak:2019pkr,Rajagopal:2019xwg}. One of the most promising suggestions for searching for signatures of the CP is through critical fluctuations in relativistic heavy-ion collisions \cite{Stephanov:1998dy,Stephanov:1999zu,Hatta:2003wn,Stephanov:2008qz,Stephanov:2011pb,Luo:2017faz}, based on the idea that near a critical point the correlation length for fluctutions of the order parameter of the phase transition should diverge (see e.g. \cite{RevModPhys.49.435}). Complications with this suggestion arise from the fact that the hot and dense QCD matter created in such heavy-ion collisions evolves very rapidly, expanding within a time period of several 10 ioctoseconds from energy densities of up to hundreds of GeV/fm$^3$ to much less than 1\,GeV/fm$^3$ (see e.g. the reviews \cite{Heinz:2013th,Busza:2018rrf}), giving those critical fluctuations not enough time to reach thermodynamic equilibrium \cite{Berdnikov:1999ph}. This implies that reliable predictions of observable fluctuation signatures require complex dynamical simulations of the non-equilibrium dynamics of the critical fluctuation modes, coupled with a comprehensive dynamical evolution package for the medium in which these fluctuations arise \cite{Nahrgang:2011mv, Nahrgang:2011mg,Mukherjee:2015swa,Akamatsu:2016llw,Murase:2016rhl,Herold:2016uvv,Stephanov:2017ghc,Hirano:2018diu,Singh:2018dpk,Herold:2018ptm,Nahrgang:2018afz,Yin:2018ejt}.

We here use the newly developed {\sc hydro+} framework \cite{Stephanov:2017ghc} to study the coupled dynamics of out-of-equilibrium fluctuations and the bulk hydrodynamic evolution. This is the second such paper, with the first \cite{Rajagopal:2019xwg} having studied a slightly simpler limit in which the CP is close to the temperature axis at $\mu{\,=\,}0$, avoiding the need for dynamically evolving also the net baryon density of the underlying dynamic medium. Here we consider the more generic and likely situation that the CP, if it exists, is located a relatively large baryon chemical potential $\gtrsim 400$\,MeV \cite{Bazavov:2017dus,Mukherjee:2019eou,Giordano:2019gev} which requires simultaneous evolution of the net baryon density and baryon diffusion current of the expanding fluid. 

Although we recently developed the (3+1)-dimensional \bes\ code \cite{Du:2019obx} for this task (see also related work in \cite{Denicol:2018wdp}), we will not use it in the present work but rather replace it by an analytically solvable model, ideal Gubser flow \cite{Gubser:2010ze, Gubser:2010ui}, as our background for the non-equilibrium fluctuation dynamics. As already noted in Ref.~\cite{Rajagopal:2019xwg}, the non-equilibrium evolution of critical fluctuations is affected by several different physical mechanisms: (i) the space-time evolution of the thermodynamic properties of the background fluid; (ii) the advection of critical fluctuations from the inside to the outside of the fireball by radial hydrodynamic flow; and (iii) the critical dynamics of the correlation length for these fluctuations in fluid cells that pass close to the CP. Using the analytically known ideal Gubser flow for the hydrodynamic background allows us to surgically isolate, and study these effects both separately and in combination, without giving up on the simultaneous existence of both longitudinal {\it and} transverse flow in the expanding medium, by applying appropriate analytic manipulations to the background.\footnote{%
    In Ref.~\cite{Rajagopal:2019xwg} the transverse expansion was modelled numerically instead of analytically, but this approach can not easily be generalized to large non-zero net baryon density short of developing a comprehensive (3+1)-dimensional hydrodynamic approach such as \bes \cite{Du:2019obx}.}
The disadvantage of our approach here is that, unlike Ref.~\cite{Rajagopal:2019xwg}, it does not allow to account for the back-reaction of the non-equilibrium fluctuation dynamics on the underlying medium flow because this would destroy the Gubser symmetry that enables the analytic treatment in the first place. Inclusion of back-reaction effects will therefore be left for a future comprehensive study with \bes+, a generalization of \bes\ \cite{Du:2019obx} that couples \bes~to the non-equilibrium evolution of the critical slow modes discussed in the present paper. However, based on the findings of \cite{Rajagopal:2019xwg} and the results presented here we expect these back-reaction effects to be very small and not of primary phenomenological interest: Even after accounting for the real QCD CP being located at large non-zero baryon chemical potential, which alters the critical behavior of the correlation length and changes the space-time evolution of the equilibrium values and relaxation rates of the critical fluctuations, we still find back-reaction effects that are never larger than $\sim10^{-4}$ times the background hydrodynamic fields.   

The paper is organized as follows: We start by briefly reviewing in Sec.~\ref{sec:fluct} the dynamics of hydrodynamic fluctuations, specifically that of the slowest mode near the QCD critical point, following the {\sc hydro+} \cite{Stephanov:2017ghc} and ``hydro-kinetic" \cite{Akamatsu:2018vjr} approaches. In Sec.~\ref{sec:quasieos} we summarize the corrections to the bulk properties of the system that are induced by the non-equilibrium evolution of the slow fluctuating degrees of freedom \cite{Stephanov:2017ghc} and describe their computation at non-zero net baryon density. Details of the dynamical setup used in this work are specified in Sec.~\ref{sec:frame}. Sec.~\ref{sec:results} is the heart of this paper: By studying various simplified limits of the expansion of the underlying thermalized medium in which the slow modes are allowed to evolve, we are able to isolate and discuss separately the effects of expansion, advection and critical correlations on the fluctuation dynamics, before studying their combined effects in a more realistic model where the medium expands both longitudinally and transversally according to ideal Gubser flow \cite{Gubser:2010ze, Gubser:2010ui}. We discuss the radial and temporal profiles of the slow modes as a function of their wave number $Q$ and study their combined effect on the total entropy content of the expanding fireball. Our key findings are summarized in Sec.~\ref{sec:conclusion}. --- The present work focuses on identifying the different mechanisms affecting the dynamics of slow fluctuation modes near the critical point and discussing their interplay in an analytical model for the fireball expansion (ideal Gubser flow) that incorporates crucial features of realistic heavy-ion collision dynamics. A full (3+1)-dimensional study without simplifying geometrical and dynamical assumptions that, in addition, fully includes the back-reaction of the slow-mode dynamics on the bulk evolution is left for the future. The validation of the (3+1)-dimensional \bes+\ code developed in this work that will be used in this future analysis is documented in the Appendix.  

In this work we use a metric $g^{\mu\nu}$ with negative signature as well as natural units where $\hbar = c = 1$.

%
\vspace*{-2mm}
\section{Fluctuation dynamics}
\label{sec:fluct}
\vspace*{-2mm}
%
After briefly discussing hydrodynamic fluctuations and their correlations in Sec. \ref{sec:hydrofluct}, we focus in Sec. \ref{sec:equifluct} on the equilibrium value of the slowest mode near the QCD critical point. Out-of-equilibrium fluctuations and their equations of motion are studied in Sec. \ref{sec:nonequi} where we also compare the different models used in this work with each other and with those introduced in Refs.~\cite{Stephanov:2017ghc, Akamatsu:2018vjr, Rajagopal:2019xwg}.

\subsection{Hydrodynamic fluctuations}\label{sec:hydrofluct}
\vspace*{-2mm}
%

We begin by introducing one- and two-point functions of hydrodynamic quantities, following the treatment in Refs.~\cite{Stephanov:2017ghc,An:2019osr,An:2019csj}. One-point functions of hydrodynamic quantities $\Psi_m(t, \bm x)$ are denoted as $\langle\Psi_m(t, \bm x)\rangle$ where $m$ labels energy density, net baryon density, components of the flow velocity, etc., and $\langle\dots\rangle$ indicates the ensemble average over the classically statistically fluctuating thermal ensemble describing the fluid cell at point $(t,\bm x)$. The ensemble-averaged quantities $\langle\Psi_m(t, \bm x)\rangle$ vary only slowly  with $\bm x$ on a scale of inhomogeneity $\ell$ and evolve deterministically according to hydrodynamic evolution equations \cite{Stephanov:2017ghc, An:2019osr,An:2019csj}. 

Next we define equal-time two-point functions (also referred to as (equal-time) correlation functions or correlators) 
in the local fluid rest frame (LRF),\footnote{%
    \label{fn1}  
    In expanding fluids the LRF is a function of space and time, and formulating the equal-time condition in the LRF covariantly requires some care \cite{An:2019osr, An:2019csj}. We will write the equal-time (in the LRF) correlators non-covariantly, using LRF coordinates. For a covariant treatment see
    \cite{An:2019osr, An:2019csj}.
    }
\begin{eqnarray}
\label{eq:correlator}
    &&G_{mn}(t,\bm x_1, \bm x_2)
    = \langle\delta\Psi_m(t, \bm x_1)\delta\Psi_n(t, \bm x_2)\rangle
\\\nonumber
    &&\quad\equiv \langle\Psi_m(t, \bm x_1) \Psi_n(t, \bm x_2)\rangle - \langle\Psi_m(t, \bm x_1)\rangle\langle\Psi_n(t, \bm x_2)\rangle\,,
\end{eqnarray}
where
\begin{equation}
    \delta\Psi_m(t, \bm x) \equiv \Psi_m(t, \bm x)-\langle\Psi_m(t, \bm x)\rangle\,,
\label{eq:fluctuation}
\end{equation}
are the fluctuations of $\Psi_m(t, \bm x)$ \cite{Stephanov:2017ghc,An:2019osr,An:2019csj}. In this paper our interest will focus on the ``slowest critical mode'' \cite{Stephanov:2017ghc}, i.e. the slowest-evolving correlator associated with critical fluctuations near the QCD critical point, and therefore mostly suppress the matrix subscripts $m,n$ from here onward. 

Expressing the correlator (\ref{eq:correlator}) as a function of the mid-point $\bm x \equiv (\bm x_1{+}\bm x_2)/2$ and separation $\Delta\bm x \equiv (\bm x_1{-}\bm x_2)$ we can write 
\begin{equation}
     G(x, \Delta\bm x)=\Bigl\langle
     \delta\Psi\bigl(t, {\bm x}{+}{\textstyle\frac{1}{2}}\Delta\bm x \bigr)
     \delta\Psi\bigl(t, {\bm x}{-}{\textstyle\frac{1}{2}}
     \Delta\bm x\bigr)\Bigr\rangle
\end{equation}
where $x \equiv (t, \bm x)$ \cite{Stephanov:2017ghc, An:2019osr, An:2019csj}. (Remember that $\Delta\bm{x}$ is the two-point separation in the LRF.)   

The following discussion relies on a separation of scales between the (microscopic) correlation length $\xi$ and the (macroscopic) hydrodynamic (in-)homogeneity length $\ell$: Assuming that approximate thermal equilibrium is reached over regions of size $\ell$ (i.e. a length scale over which the macroscopic hydrodynamic quantities can be considered approximately constant), this length $\ell$ will also define the range of ${\bm x}$ over which $G(x, \Delta\bm x)$, considered as a function of the mid-point coordinate, will vary appreciably. As a function of the separation $\Delta{\bm x}$, on the other hand, the correlator $G(x, \Delta\bm x)$ typically falls off exponentially, with the decay length given by the correlation length $\xi$. We will follow \cite{Stephanov:2017ghc, An:2019osr, An:2019csj} and assume $\xi\ll\ell$, i.e. $G(x, \Delta\bm x)$ varies much more rapidly with $\Delta{\bm x}$ than with ${\bm x}$. In this so-called {\it thermodynamic limit} $V \sim \ell^3 \gg \xi^3$ the fluctuations are small, and their probability distribution is approximately Gaussian, a fact which will be used for calculating the equilibrium value of the correlators \cite{Stephanov:2017ghc,landau2013statistical} ({\it cf.} Eq.~(\ref{eq:phiQeq}) below).

Given the scale separation $\xi\ll\ell$ it is convenient to introduce the mixed Fourier transform (i.e. Wigner transform) with respect to the separation vector $\Delta\bm x$ for the correlator \cite{Stephanov:2017ghc,An:2019osr,An:2019csj}:
\begin{eqnarray}
  &&W_{\bm Q}(x) =  \int d^3(\Delta \bm x)\, G(x, \Delta\bm x)\, e^{i\bm Q \cdot \Delta\bm x} 
\\\nonumber
  &&= \int_{\Delta \bm x}\!\!
  \Bigl\langle
     \delta\Psi\bigl(t, {\bm x}{+}{\textstyle\frac{1}{2}}\Delta\bm x \bigr)
     \delta\Psi\bigl(t, {\bm x}{-}{\textstyle\frac{1}{2}}
     \Delta\bm x\bigr)\Bigr\rangle
    e^{i\bm Q\cdot \Delta\bm x}\,. 
\label{eq:wigner}
\end{eqnarray}
$W_{\bm Q}(x)$ can be thought of as a mode distribution function, similar to the particle phase-space distribution function $f(x,\bm Q)$ in kinetic theory, where the mode index $\bm Q$ specifies the wave vector of the mode in the LRF \cite{Stephanov:2017ghc}.\footnote{%
    As shown in \cite{Stephanov:2017ghc, An:2019osr}, a specific linear combination of the matrix elements $W^{mn}_{\bm Q}(x)$ can be identified with the phase-space distribution function of phonons with momentum $\bm Q$ at point $x$ and shown to satisfy a Boltzmann equation.}
The scale separation implies an infrared momentum cutoff for $\bm Q$: $\ell^{-1} \ll Q \sim \xi^{-1}$, with $Q$ being the magnitude of $\bm Q$ (see Sec.~\ref{sec:nonequi}).

%
\subsection{Equilibrium fluctuations}
\label{sec:equifluct}

In the {\sc hydro+} framework \cite{Stephanov:2017ghc} attention is focused on the most slowly evolving correlator associated with critical fluctuations. Near the QCD critical point, Ref.~\cite{Stephanov:2017ghc} identifies the slowest mode as the diffusion of fluctuations in the entropy per baryon at fixed pressure, $\delta(s/n)$.\footnote{%
    As shown in \cite{Stephanov:2017ghc, Akamatsu:2018vjr} the fluctuation of $s/n$ is a diffusive eigenmode of linearized hydrodynamics whose evolution decouples from that of other hydrodynamic fluctuations.}
In this work we consider a special partial-equilibrium state where, except for the slowest mode, all other fluctuations have achieved a sufficient degree of local equilibrium on length scales $\ell$ \cite{Stephanov:2017ghc} that they can be described by dissipative fluid dynamics. Since the slowest mode needs more time to thermalize it will be described dynamically. This special case captures the situation near a critical point where the correlation length $\xi$ becomes large and the relaxation rates for different fluctuation modes, scaling with different critical exponents, are suppressed by different powers of $\xi$. 

The relaxation rate for $\delta(s/n)$ satisfies $\Gamma_Q \propto (\lambda_T/\cp)\,Q^2$, with $Q$ being the wave number \cite{Stephanov:2017ghc,Akamatsu:2018vjr}. $\delta(s/n)$ is the slowest critical mode near the QCD critical point since the heat capacity $\cp$ is the most rapidly divergent equilibrium susceptibility, diverging  quadratically with the correlation length, $\cp \propto \xi^2$, while the heat conductivity $\lambda_T$ in the numerator diverges only linearly, $\lambda_T\propto\xi$ (see, e.g., Refs. \cite{Stephanov:2017ghc,An:2019csj}). Thus the relaxation rate for $Q\sim \xi^{-1}$ is $\Gamma_\xi\propto \xi^{-3}$, which is the most strongly suppressed relaxation rate of all critical modes. Consequently, the specific entropy fluctuations $\delta(s/n)$ are the first to fall out of equilibrium when the system passes through the critical region \cite{Akamatsu:2018vjr}.

$\delta(s/n)$ can be written in terms of the fluctuations of the hydrodynamic variables $\delta\Psi_m(x)$, and its correlator
\begin{equation}
    \phi_{\bm Q}(x){\,\sim\!}\int_{\Delta \bm x} \!\!
    \Bigl\langle
     \delta\frac{s}{n}\bigl(t, {\bm x}{+}{\textstyle\frac{1}{2}}\Delta\bm x \bigr)
     \delta\frac{s}{n}\bigl(t, {\bm x}{-}{\textstyle\frac{1}{2}}
     \Delta\bm x\bigr)\Bigr\rangle\,
     e^{i\bm Q \cdot\Delta\bm x}\
\label{eq:phiQ}
\end{equation}
is therefore some combination of the matrix elements $W^{mn}_{\bm Q}(x)$. The normalization of $\phi_{\bm Q}$ is irrelevant (see discussions after Eq.~(\ref{eq:deltas})) and therefore left arbitrary. $\phi_{\bm Q}$ are the non-hydrodynamic slow degrees of freedom (d.o.f.) near the QCD critical point, added in {\sc hydro+} as additional dynamical variables \cite{Stephanov:2017ghc}.\footnote{%
    Even though they represent the fluctuations of the single thermodynamic field $s/n$, we refer to them in the plural as {\it critical slow modes} since they are labeled by a continuous spectral index $\bm Q$ representing their wave number in the LRF.}
In Sec.~\ref{sec:quasieos}, we will see how these additional d.o.f. affect the bulk properties of the system, e.g. its entropy and pressure. 

Due to the separation of scales $\ell^{-1} \ll Q \sim \xi^{-1}$, the values of the hydrodynamic fields $e$, $n$, $u^\mu$ are approximately constant over the range $|\Delta \bm x|\lesssim 1/|\bm Q|$ over which the integral (\ref{eq:phiQ}) receives non-vanishing contributions, and the spatial correlator in the integrand can be taken as approximately translation invariant. Assuming also spatial isotropy in the local rest frame this implies \cite{Stephanov:2017ghc} that the dependence of the equilibrium correlator $\bar\phi_{\bm Q}$ on $\bm Q$ and $\xi$ involves only the magnitude $Q=|\bm Q|$ and must enter in the critical regime near the critical point through a universal scaling function $f_2$,
\begin{eqnarray}
\label{phi_eq0}
    \bar\phi_{\bm Q} &=& \int_{\Delta \bm x} \left\langle\delta \frac sn\left(\Delta\bm x\right)\delta \frac sn\left(\bm 0\right)\right\rangle e^{i\bm Q \cdot\Delta\bm x}
\\
    &=& \bar\phi_0 f_2(Q\xi,\Theta) \label{phi_eq}
\end{eqnarray}
where $\Theta=\Theta(e,n)$ is a scaling variable \cite{Stephanov:2017ghc} and both sides of the equation (although not explicitly indicated) depend on the position $x=(t,\bm x)$ through the local values of $e$ and $n$. The equilibrium value $\bar\phi_0$ of the static mode is obtained as the $Q{\,\to\,}0$ limit (the hydrodynamic limit) of the first of these equations (Eq. (\ref{phi_eq0})),
\begin{equation}
    \bar\phi_0(x) = V\left\langle\left(\delta \frac sn(x)\right)^2\right\rangle = \frac{\cp}{n^2}(x)\,, 
\label{eq:phiQeq}
\end{equation}
where the homogeneity volume around the point $x$, $V{\,\sim\,}\ell^3$, arises from integrating out $\Delta \bm x$ in Eq.~(\ref{phi_eq0}) and then using the Gaussian probability distribution of fluctuations in the thermodynamic limit \cite{landau2013statistical, Stephanov:2017ghc}.  
Here the heat capacity at constant pressure $c_p$ is given by \cite{Akamatsu:2018vjr}
\begin{equation}
    \cp = nT\left(\frac{\partial (s/n)}{\partial T}\right)_p.
\end{equation}
The scaling function is normalized to $f_2(0,\Theta)=1$. We follow \cite{Stephanov:2017ghc} and neglect its $\Theta$ dependence, $f_2(Q\xi,\Theta)\to f_2(Q\xi)$, and  assume $f_2$ to have the Ornstein-Zernike form \cite{RevModPhys.49.435}\footnote{%
    In Refs.~\cite{Akamatsu:2018vjr,Rajagopal:2019xwg}, the function is taken as $1/(1+x^{2-\eta})$ where $\eta$ is a critical exponent ($\approx 0.036$ in the 3D Ising model). Here we set it to zero for simplicity as was also done in Refs.~\cite{Stephanov:2017ghc, Akamatsu:2018vjr, Rajagopal:2019xwg}.}
\begin{equation}
    f_2(x) = \frac{1}{1+x^2}\,.
\label{eq:oz}
\end{equation}
With the simplifications made above, the full $Q$ and $\xi$ dependence of the equilibrium value of the slow mode is given by
\begin{equation}
    \bar\phi_Q = \bar\phi_0 f_2(Q\xi) = \left[\frac{\cpz}{n^2}\left(\frac{\xi}{\xi_0}\right)^2\right]\frac{1}{1+(Q\xi)^2}\,;
\label{eq:eqPhiQ_full}
\end{equation}
here $\cpz$ denotes the ``non-critical'' value of the heat capacity in a system with ``non-critical correlation length'' $\xi_0$. Far away from the critical point when $\xi\to\xi_0\ll Q^{-1}$, the mode spectrum (\ref{eq:eqPhiQ_full}) becomes independent of $Q$ and reduces to Eq.~(\ref{eq:phiQeq}). 

Equations (\ref{phi_eq}, \ref{eq:oz}) show that in thermal equilibrium the shape of the spectrum $\phi_Q$ of critical slow modes (i.e. their dependence on the wave number $Q$) is controlled by the evolution of the correlation length $\xi$ (through $f_2(Q\xi)$) as the fluid cell $x$ passes through the critical region of the phase diagram. In addition to the $Q$-dependence, the critical behaviour of $\xi$ also affects the zero mode $\bar\phi_0$ in Eq.~(\ref{eq:phiQeq}) through the critical growth of the heat capacity $\cp\propto\xi^2$.
Note that the critical behavior of $c_{p}$ near the QCD critical point is different from that of the specific heat (at fixed volume) $c_{V}$ in the 3D Ising model. The latter diverges as $\xi^{\alpha}$ where the critical  exponent is $\alpha\approx 0.11$.
In contrast, 
since the order parameter field at the QCD critical point is a linear combination that includes a baryon density component, 
$c_{p}$ diverges more strongly, as $\xi^{2}$
(see for example Sec.~IIB of Ref.~\cite{Akamatsu:2018vjr} for more details).

%
\subsection{Non-equilibrium fluctuations}
\label{sec:nonequi}
%

Through its dependence on the local hydrodynamic variables the equilibrium mode spectrum $\bar\phi_Q$ evolves dynamically. Expansion of the fluid on the macroscopic scale drives both the background fluid and the slow mode spectrum out of thermal equilibrium while microscopic interactions push the system back towards (a modified) local equilibrium state. The evolution of the dissipative non-equilibrium corrections to the hydrodynamic variables is handled by viscous fluid dynamics; here we discuss the out-of-equilibrium evolution of the critical slow modes.

The authors of Ref.~\cite{Stephanov:2017ghc} first introduced a set of relaxation equations for these modes (which they called ``fluctuation kinetic equations'' or briefly ``kinetic equations'', see also \cite{An:2019osr}) using an educated ansatz that was later refined in the form of so-called ``hydro-kinetic equations'',  derived within the ``hydro-kinetic approach'' to fluctuating hydrodynamics \cite{Akamatsu:2018vjr,Martinez:2018wia,An:2019csj}, first in the hydrodynamic regime ($Q\ll \xi^{-1}$) and then generalized to the scaling regime near a critical point ($\xi^{-1}\ll Q \ll \ell_0^{-1}$, where $\ell_0$ is a microscopic length scale, e.g. $\ell_0\sim1/T$ in a conformal system). By interpolating between these two regimes they could also cover the region $Q{\,\sim\,}\xi^{-1}$. For completeness we here briefly summarize the results of \cite{Akamatsu:2018vjr}, calling the slow-mode evolution equations simply their ``equations of motion''.

%
\subsubsection{Dynamics in the hydrodynamic regime}
\label{sec:awaycp}
%

In the hydrodynamic regime ($Q\ll\xi^{-1}$) slow-mode dynamics is governed by the relaxation equations \cite{Stephanov:2017ghc,Akamatsu:2018vjr}
\begin{equation}
\label{eq:phi_hydro}
    D\phi_Q = -\Gamma_Q\,(\phi_Q - \bar\phi_0)\,,
\end{equation}
where $D\equiv u^\mu d_\mu$ (with $d_\mu$ denoting the covariant derivative which reduces to a simple partial derivative for scalar fields such as $\phi_Q$) is the LRF time derivative\footnote{%
    Note that $D$ is the LRF time derivative {\it at fixed wave number $\bm Q$ in the LRF}. In fully covariant notation this constraint requires replacing the covariant derivative by the ``confluent derivative" \cite{An:2019osr,An:2019csj} (see also foonote~\ref{fn1}).}
and $\bar\phi_0$ (i.e. $\bar\phi_Q$ in the hydrodynamic limit) is given by Eq.~(\ref{eq:phiQeq}). The $Q$-dependent relaxation rate $\Gamma_Q$ takes the form \cite{Stephanov:2017ghc, Akamatsu:2018vjr}
\begin{equation}
    \Gamma_Q \equiv 2 D_p Q^2 = 2\left(\frac{\lambda_T}{\cp}\right)Q^2\,,
\label{eq:gammaQ_hydro}
\end{equation}
where the factor 2 arises from the fact that this describes the relaxation of a two-point function 
and $D_p$ is a diffusion coefficient, related to the thermal conductivity $\lambda_T$ though the Wiedemann-Franz type relation $D_p = \lambda_T/\cp$,
consistent with the expectation that the relaxation rate is proportional to the transport coefficient ($\lambda_T$) over the susceptibility ($\cp$). Note that Eq.~(\ref{eq:gammaQ_hydro}) is the equilibrium value of the relaxation rate \cite{Rajagopal:2019xwg} and ignores off-equilibrium corrections to the relaxation process \cite{Akamatsu:2018vjr}.

A few comments are in order here. First, when the system evolves, the relaxation of the slow modes (originally from the fluctuations of entropy per baryon, both conserved quantities in an ideal fluid) requires transport of conserved quantities through diffusion \cite{An:2019osr}. This explains why the equilibration rate for $\phi_Q$ is proportional to the diffusion coefficient $D_p$. Second, since the relaxation is through diffusive processes, the relaxation rate is proportional to $Q^2$. Third, in the hydrodynamic regime $Q\ll\xi^{-1}$, $\phi_Q$ relaxes for all wave numbers $Q$ to the {\it static} ($Q = 0$) equilibrium value $\bar\phi_0$. 

As observed in Ref.~\cite{Akamatsu:2018vjr}, since the equilibrium value $\bar\phi_0$ evolves as Eq.~(\ref{eq:phiQeq}) when the system evolves, its fractional change per unit time is controlled by the scalar expansion rate $\theta$ of the system,
\begin{equation}
\label{eq:expansion}
    \theta \equiv d_\mu u^\mu = \partial_\mu u^\mu + \Gamma^\lambda_{\lambda\mu}u^\mu\,,
\end{equation}
where $\Gamma^\lambda_{\lambda\mu}$ are the Christoffel symbols \cite{carroll2004spacetime}. Deviations from the equilibrium value, on the other hand, decay with the relaxation rate $\Gamma_Q \propto Q^2$, i.e. short wave length modes equilibrate faster than long wave length modes. For any given wave number $Q$, the dynamics of $\phi_Q$ thus results from the competition between the expansion of the system (controlled by the expansion rate $\theta$ of the background fluid) and the relaxation of the slow modes (controlled by the relaxation rate $\Gamma_Q$), and there should exist a (dynamical) wave number scale $Q_\mathrm{neq}$ which separates approximately thermalized fluctuation modes from those that strongly deviate from equilibrium \cite{Akamatsu:2018vjr}. 

This competition between expansion and relaxation towards equilibrium is generic and persists in the critical scaling regime near the critical point, as will be seen formally in the following subsection and studied numerically in Sec.~\ref{sec:results}. To characterize this competition quantitatively it is convenient to introduce the ``critical Knudsen number" for the slow modes as the ratio between the scalar expansion and relaxation rates:
\begin{equation}
    \textrm{Kn}(Q)\equiv\theta/\Gamma_Q\,.
\label{eq:kn}
\end{equation}
Slow modes with large critical Knudsen numbers will lag behind, not being able to follow the hydrodynamic evolution of the equilibrium value $\bar\phi_Q(x)$. Very roughly, the scale $Q_\mathrm{neq}$ defined in the preceding paragraph should correspond to critical Knudsen numbers of order 1.

%
\subsubsection{Dynamics in the scaling regime}
\label{sec:nearcp}
%

To extend the slow-mode evolution equation (\ref{eq:phi_hydro}) from the hydrodynamic regime $Q\ll\xi^{-1}$ to the scaling regime $\xi^{-1}\ll Q \ll \ell_0^{-1}$ (including the transition region $Q \sim \xi^{-1}$) we generalize it  \cite{Akamatsu:2018vjr} by replacing on the right hand side the equilibrium value $\bar\phi_0(x)$ for the zero mode without critical scaling by the $Q$-dependent equilibrium value $\bar\phi_Q(x)$ from Eq.~(\ref{eq:eqPhiQ_full}), 
\begin{equation}
\label{eq:phi_scaling}
    D\phi_Q = -\Gamma_Q\,(\phi_Q - \bar\phi_Q)\,,
\end{equation}
and the relaxation rate $\Gamma_Q$ by the $\xi$-dependent expression
\begin{equation}
    \Gamma_Q = \Gamma_\xi f_\Gamma(Q\xi) 
\end{equation}
where 
\begin{eqnarray}
\label{eq:fgamma}
    &&f_\Gamma(Q\xi) \equiv (Q\xi)^2 \Bigl(1+(Q\xi)^2\Bigr)\,,
\\
\label{eq:Gamma_xi}
    &&\Gamma_\xi \equiv 2\left(\frac{\lambda_T}{\cpz\xi^2}\right)
    \left(\frac{\xi_0}{\xi}\right)^2\,.
\end{eqnarray}
The factor $(\xi_0/\xi)^2$ in (\ref{eq:Gamma_xi}) accounts for the critical scaling of $\cp$; the extra factor $1/\xi^2$ in (\ref{eq:Gamma_xi}) compensates with a factor $\xi^2$ in $f_\Gamma$ such that $f_\Gamma/\xi^2 \to Q^2$ in the hydrodynamic limit $Q\ll\xi^{-1}$. Equations (\ref{eq:phi_scaling})-(\ref{eq:Gamma_xi}) reduce to Eqs.~(\ref{eq:phi_hydro},\ref{eq:gammaQ_hydro}) in the hydrodynamic limit $\xi\to\xi_0\ll Q^{-1}$. Note that the factor $Q^2$ in Eqs.~(\ref{eq:gammaQ_hydro},\ref{eq:fgamma}) strongly reduces the thermalization rate for slow modes with small wave numbers $Q{\,\ll\,}\xi^{-1}$, even without critical enhancement of the correlation length. In the critical region, where $\xi$ becomes large, the $\xi$-dependence of $\Gamma_\xi$ in (\ref{eq:Gamma_xi}) causes additional ``critical slowing down'' of relaxation processes even for modes with more typical wave numbers $Q\sim\xi^{-1}$: $\Gamma_\xi \propto \xi^{-z}$ with critical exponent $z=4$.

This model for the evolution of the critical slow modes agrees with the one derived in Refs.~\cite{Stephanov:2017ghc, Akamatsu:2018vjr} and recently studied in \cite{Rajagopal:2019xwg} up to the following differences: First, the authors of Ref. \cite{Stephanov:2017ghc} formulate the slow-mode evolution equation (\ref{eq:phi_scaling}) in terms of a relaxation rate $\Gamma_Q$ that differs from the one used here and in \cite{Akamatsu:2018vjr, Rajagopal:2019xwg} by a factor $\phi_Q/\bar\phi_Q$ (which approaches unity in equilibrium). Second, the assumed scaling behavior with the correlation length $\xi$ depends on the assumed universality class of the critical point: following the classification of Ref. \cite{RevModPhys.49.435}, Ref.~\cite{Rajagopal:2019xwg} uses ``model A", Ref.~\cite{Akamatsu:2018vjr} and the present work use ``model B", while Ref.~\cite{Stephanov:2017ghc} uses ``model H". Let us briefly summarize the differences in the dynamical critical exponent $z$ introduced above and in the shape of the scaling functions $f_\Gamma(Q\xi)$ resulting from these model choices.

\begin{figure}[!tp]
    \hspace*{-3mm}
    \includegraphics[width=0.35\textwidth]{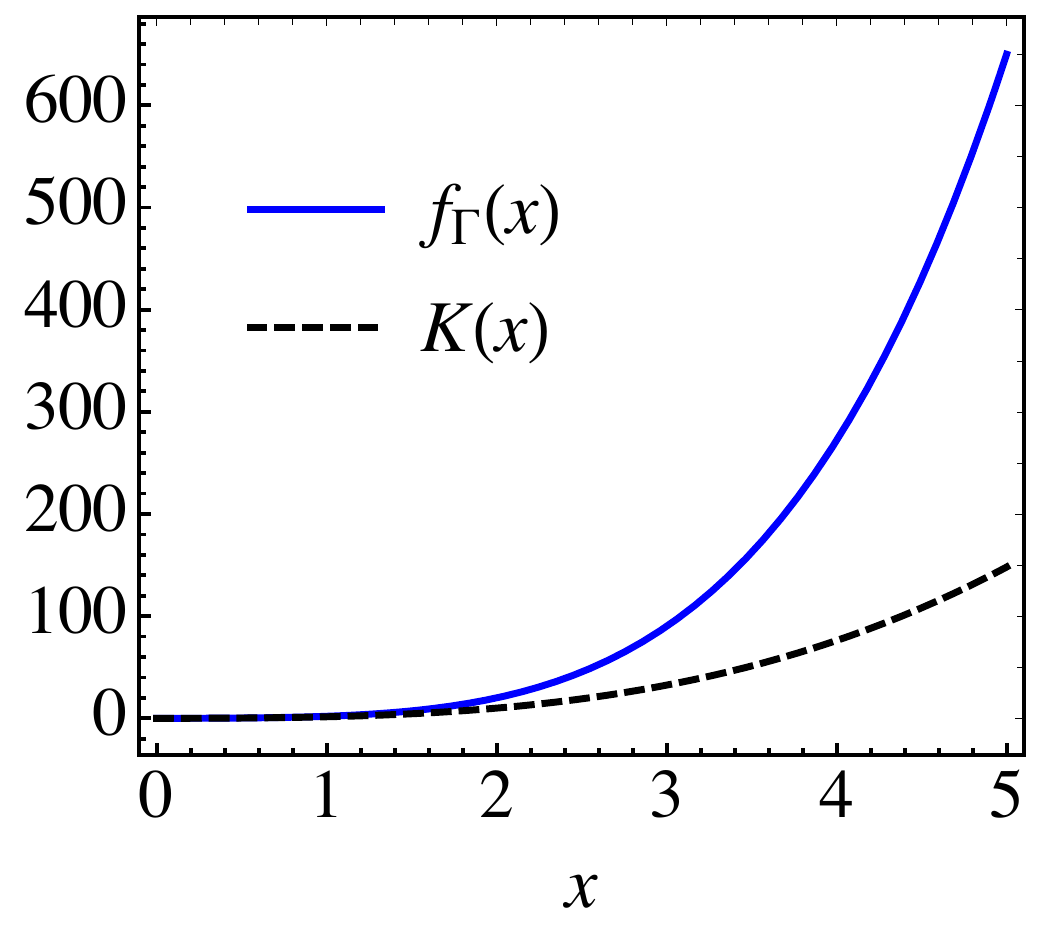}
    \caption{Comparison between the function $f_\Gamma(Q\xi)$ (\ref{eq:fgamma}) and the Kawasaki function defined in footnote \ref{footnote:kawasaki}. Large-$Q$ modes relax faster with $f_\Gamma$ than with $K$.}
    \label{fig:gammaq}
\end{figure}

Ref.~\cite{Rajagopal:2019xwg} studies dynamics at $\mu \approx0$ where the order parameter is the non-conserved chiral condensate ({\it model A}). Ref.~\cite{Akamatsu:2018vjr} and this work place the critical point at large non-zero $\mu$ but ignore the critical behavior of $\lambda_T$ ({\it model B}). {\it Model H}, used in Ref.~\cite{Stephanov:2017ghc}, correctly describes the dynamical universality class of the QCD critical point where the order parameter is a combination of the chiral condensate and baryon density.
\begin{itemize}
\item 
Model A has $z{\,=\,}2$, $\Gamma_Q = \Gamma_\xi f_\Gamma(Q\xi) \propto \xi^{-z}\cdot C \propto \xi^{-z}$ when $Q \lesssim \xi^{-1}$, and $\Gamma_Q \propto \xi^{-z}\cdot(Q\xi)^z \propto Q^z$ when $Q \gg \xi^{-1}$. 
\item
For model B $z{\,=\,}4$, $\Gamma_Q = \Gamma_\xi f_\Gamma(Q\xi) \propto \xi^{-z}\cdot(Q\xi)^2 \propto Q^2\xi^{2-z}$ when $Q \lesssim \xi^{-1}$, and $\Gamma_Q \propto \xi^{-z}\cdot(Q\xi)^4 \propto Q^4\xi^{4-z}$ when $Q \gg \xi^{-1}$. 
\item
Finally, in model H $z{\,=\,}3$, $\Gamma_Q = \Gamma_\xi f_\Gamma(Q\xi) \propto \xi^{-z}\cdot(Q\xi)^2 \propto Q^2\xi^{2-z}$ when $Q \lesssim \xi^{-1}$, and $\Gamma_Q \propto \xi^{-z}\cdot(Q\xi)^3 \propto Q^3\xi^{3-z}$ when $Q \gg \xi^{-1}$.\footnote{%
    Ref.~\cite{Stephanov:2017ghc} uses the Kawasaki function \cite{KAWASAKI19701} for $f_\Gamma(Q\xi)$ according to mode-coupling theory, $K(x) = \frac{3}{4}[1+x^2+(x^3{-}x^{-1})\tanh^{-1}(x)]$, where $K(x) \to x^2$ for $x{\,\ll\,}1$ and $K(x) \to (3\pi/8)x^3$ for $x{\,\gg\,}1$. For $x\ll1$, $K(x)$ has the same limit as $f_\Gamma(x)$ in Eq.~(\ref{eq:fgamma}) but the two disagree in the opposite limit $x\gg1$ limit (see Fig. \ref{fig:gammaq}).\label{footnote:kawasaki}}
\end{itemize}
In this exploratory study we follow Ref.~\cite{Akamatsu:2018vjr} and use model B, but it would be straightforward to implement model H instead. In contrast to model A, these two models both feature a $Q$-dependent slow-mode relaxation rate $\Gamma_Q\propto Q^2$ which delays the relaxation of low-$Q$ modes.

To summarize, the dynamics of the slow modes in our work is given by equations (\ref{eq:eqPhiQ_full}) and (\ref{eq:phi_scaling})-(\ref{eq:Gamma_xi}). Our slow mode is the diffusion of fluctuations in entropy per baryon at constant pressure, $\delta(s/n)$; this agrees with Ref.~\cite{Stephanov:2017ghc} and, up to a normalization factor, with Ref.~\cite{Akamatsu:2018vjr} (whose authors studied $n\,\delta(s/n)$), whereas the authors of Ref.~\cite{Rajagopal:2019xwg} studied diffusion of fluctuations in the chiral condensate. Our work differs from Ref.~\cite{Stephanov:2017ghc} by using model B instead of model H for the relaxation rate $\Gamma_Q$ of specific entropy fluctuations, while Ref.~\cite{Rajagopal:2019xwg} uses model A for the relaxation rate of chiral fluctuations. Model differences exist also outside the critical region where Ref. \cite{Rajagopal:2019xwg} uses a constant relaxation rate $\Gamma_0$ while we follow Ref.~\cite{Akamatsu:2018vjr} and set it proportional to $\lambda_T/\cp$; Ref.~\cite{Stephanov:2017ghc} uses the simplified prescription $D_p=\lambda_T/\cp=T/(6\pi\eta\xi_0)$. 

A final difference arises from the fact that Ref.~\cite{Akamatsu:2018vjr} ignores advection by setting the fluid velocity to zero while we and Ref.~\cite{Rajagopal:2019xwg} include advection effects arising from the fluid's expansion in the dynamics of the slow modes. The competition between growth of critical fluctuations near the critical point and advection from hydrodynamic expansion of the fluid will be studied in Sec.~\ref{sec:results}.

%
\section{Partial-equilibrium equation of state}
\label{sec:quasieos}
%

We now discuss the influence (``back-reaction") of the critical slow modes, whose evolution was discussed in the preceding section, on the  background fluid, through the equation of state (EoS). The slow modes carry energy and entropy and thus contribute to the pressure of the system by adding to the thermal equilibrium values of these quantities in the background fluid. The resulting  ``partial-equilibrium EoS" or ``quasi-equilibrium EoS" is a central ingredient of {\sc hydro+} \cite{Stephanov:2017ghc}.

The slow modes add new, non-equilibrated degrees of freedom to the quantum states of the system. Denoting the complete-equilibrium entropy density of the d.o.f. describing the hydrodynamic background fluid by $s(e, n)$ and additional entropy density contributed by the additional non-equilibrated d.o.f. as $\D s$, the partial-equilibrium entropy density can be written as
\begin{equation}
  s_{\plus}(e, n, \phi)\equiv s(e,n)+\D s(e, n, \phi)=\log\Omega(e, n, \phi)\,,
\end{equation}
where $\Omega(e, n, \phi)$ is the number of quantum states of the system with $(e, n, \phi)$.\footnote{%
    Here $\phi$ is short for the mode spectrum $\phi_Q$.}
$\D s$ is always negative and describes how much entropy the state with non-equilibrium fluctuations or correlations is missing compared to a state in which the fluctuations are completely equilibrated \cite{Stephanov:2017ghc}. When $\phi$ relaxes to its equilibrium value $\bar\phi(e, n)$, the entropy $s_{\plus}$ should also approach its maximum value $s(e, n)$:
\begin{equation}
  \textrm{max}\; s_{\plus}(e, n, \phi) = s_{\plus}(e, n, \bar\phi) = s(e, n)\,.
\end{equation}
%

\begin{figure}[!tp]
    \hspace*{-3mm}
    \includegraphics[width=0.35\textwidth]{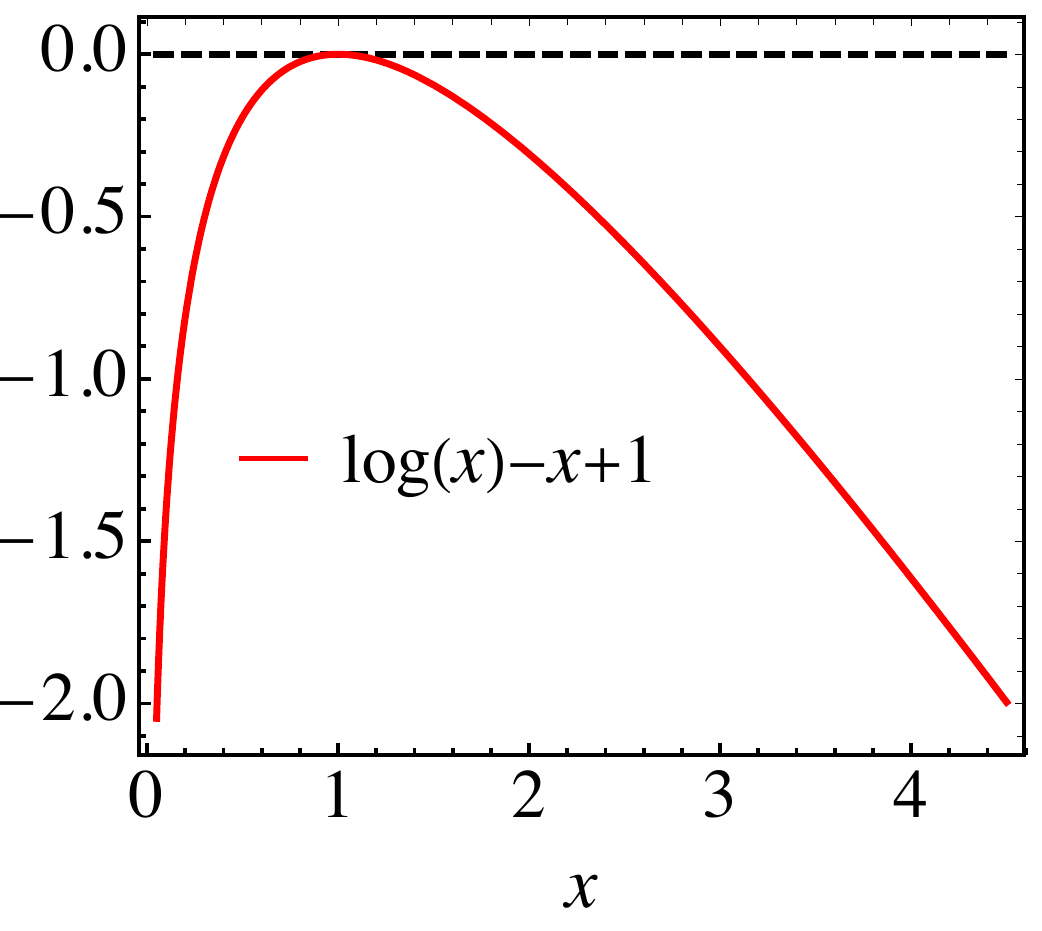}
    \caption{Illustration of the contribution $\Delta s_Q$ to the entropy density by a single slow mode with wave number $Q$ ($x{\,\equiv\,}\phi_Q/\bar\phi_Q$). $\Delta s_Q$ is negative whether $\phi_Q$ is below or above its equilibrium value ({\it cf.} Eq.~(\ref{eq:deltas}) below).}
    \label{fig:dsq}
\end{figure}

In principle the equilibrium entropy $s(e, n)$ should include the thermodynamic behavior near the critical point. In past work, however, which mostly ignored the back-reaction of the off-equilibrium fluctuations on the partial-equilibrium EoS, the critical scaling properties near the critical point 
were also ignored in the complete-equilibrium EoS.

The slow-mode contribution to the  partial-equilibrium entropy density\footnote{%
    In natural units $\hbar{\,=\,}c{\,=\,}1$, $\D s$ has units of [fm$^{-3}$] while $\D s_Q$ has units of [fm$^{-2}$].}
is given explicitly by \cite{Stephanov:2017ghc}
\begin{eqnarray}
    &&\D s(e,n,\phi) \equiv \int dQ\, \Delta s_Q 
\nonumber\\
    &&= \int dQ \frac{Q^2}{(2\pi)^2} \left[\log\frac{\phi_Q}{\bar\phi_Q(e, n)}-\frac{\phi_Q}{\bar\phi_Q(e, n)}+1\right],
\label{eq:deltas}
\end{eqnarray}
where the local equilibrium value $\bar\phi_Q$ of the slow mode $\phi_Q$ is determined by the local values of $e$ and $n$ (see Eq. (\ref{eq:eqPhiQ_full})) and we used local isotropy to simplify the integration measure (the factor $\frac{1}{2}$ arises from $\phi$ being related to the width of the fluctuations \cite{Stephanov:2017ghc})
\begin{equation}
    \frac{1}{2}\int \frac{d^3Q}{(2\pi)^3} = \frac{1}{2}\int \frac{4\pi Q^2 d Q}{(2\pi)^3} = \int \frac{Q^2 d Q}{(2\pi)^2}.
\end{equation}
Eqs.~(\ref{eq:phi_scaling}) and (\ref{eq:deltas}) show that (for purpose of calculating the back-reaction) the normalization of $\bar\phi_Q$ is irrelevant since only the ratio $\phi_Q/\bar\phi_Q$ appears. The normalization of $\bar\phi_Q$ controls the magnitude of the contribution of the critical fluctuations to any fluctuation observable, but we shall not compute such observables here. The function in square brackets, $\log(x)-x+1$, and hence the non-equilibrium entropy correction (\ref{eq:deltas}) is negative semi-definite and plotted in Fig.~\ref{fig:dsq} for illustration. 
In the derivation of Eq.~\ref{eq:deltas}, $|\Delta s|$ is assumed to be much smaller than $s$.
Therefore, $s_{(+)}$ remains positive definite within the domain of the applicability of Eq.~\eqref{eq:deltas}.
Note also
 that in deriving this expression \cite{Stephanov:2017ghc} the separation of scales $\ell^{-1}\ll Q$ was used and only the contribution of the slowest mode to $\Delta s$ was included.

The off-equilibrium contribution from the slow mode to the entropy density $s_{(+)}$ modifies the pressure to $p_{(+)}$, given by the generalized thermodynamic relation \cite{Stephanov:2017ghc}
\begin{equation}
\label{eq:splus}
    s_{\plus} = \beta_\plus p_\plus+\beta_\plus e - \alpha_\plus n\,,
\end{equation}
with modified inverse temperature and chemical potential defined by
\begin{eqnarray}
  &&\beta_{\plus} = \left(\frac{\partial s_\plus}{\partial e}\right)_{\!\!n\phi} \equiv \frac{1}{T} + \D \beta\,,
\nonumber\\
  &&\alpha_{\plus} = - \left(\frac{\partial s_\plus}{\partial n}\right)_{\!\!e\phi} \equiv \frac{\mu}{T} + \D \alpha\,,
\label{eq:alphabeta}
\end{eqnarray}
where the corrections are
\begin{equation}
  \D \beta \equiv \left(\frac{\partial \D s}{\partial e}\right)_{\!\!n\phi}\,,\quad
\D \alpha \equiv - \left(\frac{\partial \D s}{\partial n}\right)_{\!\!e\phi}\,.\label{eq:dba}
\end{equation}
Solving Eq. (\ref{eq:splus}) for $p_\plus{\,\equiv\,}p+\D p$ (where $s=\beta(e{+}p)-\alpha n$ with $\beta\equiv1/T$ and $\alpha\equiv\mu/T$) one finds
\begin{equation}
    \D p = \bigl[-(e{+}p)\D\beta + n \D\alpha + \D s\bigr]\big/\beta_{\plus},
\label{eq:dp}
\end{equation}
with  $\D \beta$ and $\D \alpha$ given by (using Eqs. (\ref{eq:deltas}) and (\ref{eq:dba}))
\begin{eqnarray}
  &&\D \beta = \int dQ \frac{Q^2}{(2\pi)^2} \frac{\phi_Q - \bar\phi_Q}{(\bar\phi_Q)^2} \left(\frac{\partial \bar\phi_Q}{\partial e}\right)_{\!\!n}\,,
\nonumber\\ 
  &&\D \alpha = - \int dQ \frac{Q^2}{(2\pi)^2} \frac{\phi_Q - \bar\phi_Q}{(\bar\phi_Q)^2} \left(\frac{\partial \bar\phi_Q}{\partial n}\right)_{\!\!e}\,.
\label{eq:dalphabeta}
\end{eqnarray}
Using $\bar\phi_Q$ from Eq.~(\ref{eq:eqPhiQ_full}) and defining the dimensionless quantities $\tilde{\bar\phi}_0 = \bar{\phi}_0 $/fm$^3$ and $\tilde\xi = \xi/$fm we find the following expressions for the two derivatives of $\bar\phi_Q$:
\begin{eqnarray}
\label{eq:deriv}
  &&\left(\frac{\partial \bar\phi_Q}{\partial e}\!\right)_{\!\!n} 
  = \left[\left(\frac{\partial \log\hatbarphi}{\partial e}\!\right)_{\!\!n} - 2\,  Q\xi\,f_2(Q\xi)\left(\frac{\partial \log\hatxi}{\partial e}\!\right)_{\!\!n} \right]\bar\phi_Q\,, 
\nonumber\\
  &&\left(\frac{\partial \bar\phi_Q}{\partial n}\!\right)_{\!\!e} 
  = \left[\left(\frac{\partial \log\hatbarphi}{\partial n}\!\right)_{\!\!e} - 2\, Q\xi\,f_2(Q\xi)\left(\frac{\partial \log\hatxi}{\partial n}\!\right)_{\!\!e}\right]\bar\phi_Q\,,
\nonumber\\
\end{eqnarray}
where, for a given EoS, the two derivatives of $\log \hatbarphi$ can be calculated from $\bar\phi_0=\cpz/n^2(\xi/\xi_0)^2$ and don't depend on $Q$, and the two derivatives of $\log\hatxi$ only depend on the parametrization of the equilibrium correlation length $\xi$. The shifts in temperature and chemical potential are obtained by plugging the expressions (\ref{eq:deriv}) into Eqs.~(\ref{eq:dalphabeta}). 

In this work, we focus on the non-equilibrium slow-mode correction $\Delta s$ (\ref{eq:deltas}) to the entropy density as a proxy for estimating the expected size of back-reaction effects on the bulk hydrodynamic evolution. More precisely, such hydrodynamical effects would be driven by the gradients of the modified pressure $p_{(+)} = \bigl(s_{(+)}{\,+\,}\alpha_{(+)}n\bigr)/\beta_{(+)}-e$ (see Eqs.~(\ref{eq:splus},\ref{eq:alphabeta})). We expect the fractional non-equilibrium slow mode corrections to the entropy density and pressure gradients to be of similar orders of magnitude.

%
\section{Setup of the framework}
\label{sec:frame}
%

We now describe our {\sc hydro+} setup as executed in this paper. Similar to \cite{Rajagopal:2019xwg} we explore a simplified expansion geometry (boost-invariant longitudinal coupled to azimuthally symmetric transverse expansion), but with an analytic solution for the hydrodynamic expansion of the hydrodynamic background (ideal Gubser flow) rather than the numerical solutions studied in \cite{Rajagopal:2019xwg}. The analytic background flow facilitates the study of the slow-mode evolution, for which we explore a different scenario than in Ref.~\cite{Rajagopal:2019xwg} by moving the critical point away from the temperature axis to a region of non-negligible net baryon density. We thus explore a situation with a different critical scaling behavior than the one studied in \cite{Rajagopal:2019xwg}, and we include finite net baryon density effects in the computation of the contribution to the entropy density caused by off-equilibrium critical fluctuations. Unlike Ref.~\cite{Rajagopal:2019xwg} we here ignore the back-reaction of the slow modes on the dynamical evolution of the background fluid. For the specific expansion geometry studied in \cite{Rajagopal:2019xwg}, the authors found very small back-reaction corrections to the background evolution. In this work we confirm that also for the (ideal) Gubser flow studied here the off-equilibrium slow-mode contribution $\D s$ to the entropy density is generically small ($\D s/s{\,\sim\,}\mathcal{O}(10^{-5}{-}10^{-4})$), so we expect similarly small modifications to the background flow caused by them. A full (3+1)-dimensional simulation that includes both dissipation and back-reaction from critical fluctuations in the evolution of the fluid medium is in progress and will be reported elsewhere. 

We start by supplying in Sec.~\ref{sec:param} the necessary parametrizations of the correlation length $\xi$, heat conductivity $\lambda_T$, and specific heat capacity $\cpz$, which control the relaxation dynamics of the critical slow modes. In Sec.~\ref{sec:gubser} we briefly review for completeness the hydrodynamic evolution equations for the background fluid undergoing ideal Gubser flow. In Sec.~\ref{sec:pheno} we show how to evaluate the off-equilibrium slow-mode correction $\D s$ to the entropy density on an isothermal hypersurface $\Sigma$ within Gubser flow; this lays the groundwork for future computations of a full set of hydrodynamic observables including non-equilibrium slow-mode corrections on the freeze-out hypersurface of a heavy-ion collision. 

%
\vspace*{-2mm}
\subsection{Parametrization}
\label{sec:param}
\vspace*{-3mm}
%

%
\begin{figure*}[!tp]
    \centering
    \includegraphics[width= 0.9\textwidth]{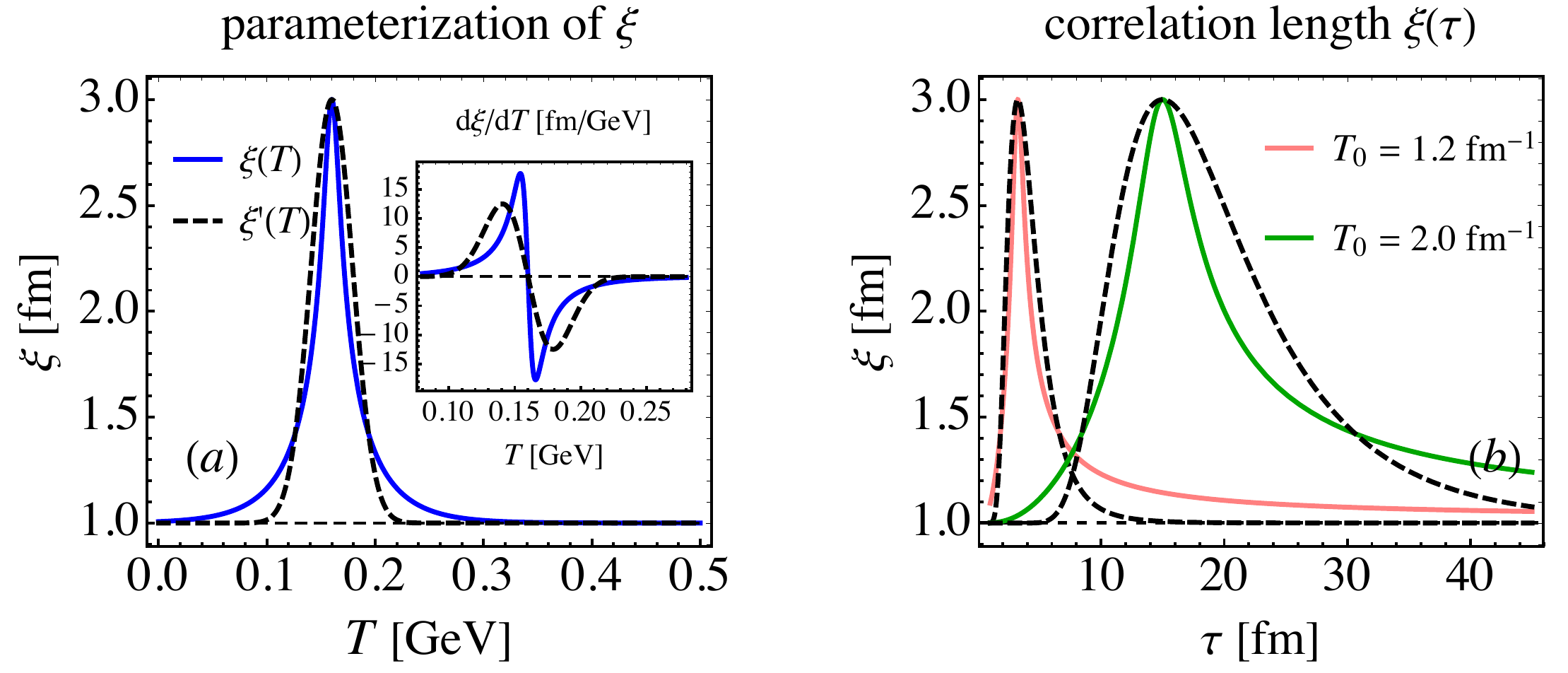}
    \vspace*{-3mm}
    \caption{ 
    (a) Parametrization of the equilibrium correlation length $\xi(T)$ according to Eqs.~(\ref{eq:xipara1}) (solid blue \cite{Rajagopal:2019xwg}) and (\ref{eq:xipara2}) (dashed black). The horizontal dashed line indicates the background correlation length $\xi_0{\,=\,}1$\,fm far away from the critical point. The inset plot shows the derivatives of $\xi(T)$ with respect to $T$. (b) Time evolution of the equilibrium correlation length $\xi\bigl(T(\tau)\bigr)$ for ideal Bjorken flow with $T(\tau) = T_0 (\tau_0/\tau)^{1/3}$ where $\tau_0=1$\,fm, for two choices of $T_0$ as indicated in the legend. Solid and dashed lines refer to the corresponding equilibrium parametrizations in panel (a).}
    \label{fig:param_xi}
    \vspace*{-4mm}
\end{figure*}
%

At finite net baryon density one should expect the equilibrium correlation length to be a function of both $e$ and $n$, $\xi(e,n)$. For our exploratory study we take it as a function of temperature only (i.e. we neglect its dependence on the baryon chemical potential $\mu$).\footnote{%
    Note that in the context of the present work this does not imply that we cannot simulate the effects of a critical {\it point}, situated at a unique location $(T_c,\mu_c)$ in the temperature-chemical potential plane: the specific background flow assumed in this work (ideal Gubser flow) will be seen to evolve at constant $\mu/T$, and we will simply assume that the critical point lies on that trajectory, i.e. $\mu/T=\mu_c/T_c$.}   
We explore two parametrizations: first, we consider the form used in \cite{Rajagopal:2019xwg},
\begin{equation}
    \xi(T) = \frac{\xi_0}{\Bigl[\tanh^2\left(\frac{T-T_c}{\Delta T}\right)\Bigl(1-\left(\frac{\xi_0}{\xi_\mathrm{max}}\right)^4\Bigr)+\left(\frac{\xi_0}{\xi_\mathrm{max}}\right)^4\Bigr]^{1/4}}\,,
\label{eq:xipara1}
\end{equation}
with parameters
\begin{equation}
    T_c = 160\,\textrm{MeV},\ 
    \Delta T = 0.4\,T_c,\ 
    \xi_0 = 1\,\textrm{fm},\ 
    \xi_\mathrm{max} = 3\,\textrm{fm}.
\end{equation}
Without the cutoff $\xi_\mathrm{max}{\,\gg\,}\xi_0$, this parametrization gives $\xi\propto|T-T_c|^{-1/2}$ for small $|T-T_c|$ as motivated by mean field theory \cite{Rajagopal:2019xwg}. We here cut off this singular growth at $\xi_\mathrm{max}=3$\,fm but in such a way that the first derivative of $\xi(T)$ still changes very rapidly near $T=T_c$. (see inset in Fig.~\ref{fig:param_xi}a). Note that our $\Delta T$ is twice that used in Ref.~\cite{Rajagopal:2019xwg}, to avoid the special treatment found necessary in \cite{Rajagopal:2019xwg} to deal with very sharp peaks in the correlation length. While general arguments say that the equilibrium value of $\xi$ should diverge at $T_c$, in an expanding system the actual correlation length will alaways remain finite due to critical slowing down \cite{Berdnikov:1999ph}. Regulating the critical divergence of the equilibrium correlation length at $T_c$ as done in Eq.~(\ref{eq:xipara1}) should therefore not make much of a difference in practice. However, Eq.~(\ref{eq:xipara1}) leads to large temporal gradients $\partial_t \bar\phi_Q/ \bar\phi_Q$, resulting from large derivatives $d\xi/dT$, near $T=T_c$.  To explore their importance for the dynamics we also study a second parametrization $\xi'(T)$ which does not share this feature,
\begin{equation}
    \xi'(T) = \xi_0+\Delta\xi\exp\left[-\frac{\left(T-T_c\right)^2}{2\sigma_T^2}\right]\,,
\label{eq:xipara2}
\end{equation}
with $\sigma_T=0.3\,\Delta T=0.12\,T_c$ and   $\Delta\xi=\xi_\mathrm{max}{\,-\,}\xi_0$, whose temperature derivative $d\xi'/dT$ now changes much more slowly near $T_c$ (see inset in Fig.~\ref{fig:param_xi}a). The two parametrizations (\ref{eq:xipara1},\ref{eq:xipara2}) are compared in  Fig.~\ref{fig:param_xi}a; one sees that for the given choice of the Gaussian width $\sigma_T$ Eqs.~(\ref{eq:xipara1}) and (\ref{eq:xipara2}) have very similar overall shapes, but Eq.~(\ref{eq:xipara2}) avoids the sharpness of the peak at $T_c$. Fig.~\ref{fig:param_xi}b shows the resulting evolution of the correlation length $\xi(\tau)$ for the two parametrizations shown in panel (a), for a fluid undergoing Bjorken expansion with two different initial temperature values $T_0$ at time $\tau_0=1$\,fm as shown in the figure. While qualitatively very similar, parametrization (\ref{eq:xipara2}) leads to smoother time dependence as the systems pass through $T_c$, but also to a faster return to the background value $\xi_0$ as the system moves away from the critical point. We will study the evolution of the slow mode with these two parametrizations in Sec.~\ref{sec:expansion}.

The non-critical heat capacity $\cpz\, [\textrm{fm}^{-3}]$ and heat conductivity $\lambda_T\, [\textrm{fm}^{-2}]$ are parametrized as follows \cite{Akamatsu:2018vjr}:
\begin{equation}
    \lambda_T = C_{\lambda} T^2\,,\quad 
    \cpz = \frac{s^2}{\alpha n}\,.
\label{eq:param}
\end{equation}
Here $C_{\lambda}$ is a unitless free parameter and $\alpha=\mu/T$. We choose $C_{\lambda}$ such that the relaxation rate $\Gamma_Q \sim {\cal O}$(1/fm) \cite{Rajagopal:2019xwg}. In principle $\cpz$ should be derived from the EoS of the medium using $\cpz=nT\bigl[\partial(s/n)/\partial T\bigr]_p$. However, in the conformally symmetric background we will be using in Sec.~\ref{sec:gubser}, any parametrization with the correct units should give the same (Gubser- or Milne-) time dependence, up to an overall normalization,\footnote{%
    This holds only as long as back-reaction is neglected because the slow-mode dynamics breaks conformal symmetry.}
and for not too large values of $\alpha$ the parametrization (\ref{eq:param}) of $\cpz$ is numerically quite accurate for the ideal massless gas EoS (\ref{EOSp},\ref{EOSn}) used in this work.

With $\cpz$ from (\ref{eq:param}) and $\bar\phi_0{\,=\,}(\cpz/n^2)(\xi/\xi_0)^2$ we can evaluate the derivatives $(\partial \log\hatbarphi/\partial e)_n$ and $(\partial \log\hatbarphi/\partial n)_e$ occurring in Eqs.~(\ref{eq:deriv}). It is then easy to improve the calculation for more realistic choices of the correlation length $\xi(\mu, T)=\xi(e,n)$, by tabulating the derivatives $(\partial \log \hatxi/\partial e)_n\,\textrm{and}\, (\partial \log \hatxi/\partial n)_e$ (which do not depend on the dynamics of the slow modes) and having the dynamical evolution code read and interpolate that table. This is what we do in \bes+; the details will be reported when presenting first results for realistic (3+1)-dimensional simulations with full backreaction in an upcoming work.

%
\subsection{Background fluid}\label{sec:gubser}
%

In this work we evolve the slow modes in an externally prescribed background fluid undergoing ideal Gubser flow \cite{Gubser:2010ze, Gubser:2010ui}, ignoring their back-reaction onto this background dynamics. In this subsection we briefly describe this background, referring the reader to Refs.~\cite{Gubser:2010ze, Gubser:2010ui, Du:2019obx} for technical details.

Although it is known how to include dissipative corrections in Gubser flow \cite{Gubser:2010ze, Gubser:2010ui, Du:2019obx} we here exploit the ideal fluid limit where the energy momentum tensor and baryon number current have the simple decompositions $T^{\mu\nu} = \ed u^{\mu}u^{\nu} - \peq\Delta^{\mu\nu}$, $N^{\mu} = \n u^{\mu}$, where $u^\mu u^\nu$ and $\Delta^{\mu\nu} \equiv g^{\mu\nu} - u^{\mu}u^{\nu}$ project on the temporal and spatial directions in the LRF, respectively. We use the Landau frame for the LRF in which the flow velocity $u^\mu$ (normalized by $u^\mu u_\mu=1$) is defined as the timelike eigenvector of $T^{\mu\nu}$, $T^{\mu\nu}u_\nu=\ed u^\mu$, with the LRF energy density $e$ as the eigenvalue. In this setup, the conservation laws for energy, momentum and baryon number, $d_{\mu}T^{\mu\nu}{\,=\,}0{\,=\,}d_{\mu}N^{\mu}$ (where $d_\mu$ denotes the covariant derivative in a generic system of coordinates), reduce to
\cite{Jeon:2015dfa} 
\begin{subequations}
\label{id-hydro}
\begin{eqnarray}
 D\n &=& -\n\theta\;,
\label{eq:vhydro-N}
\\
 D\ed &=& -(\ed{+}\peq)\theta\;,
\label{eq:vhydro-E}
\\
Du^\mu &=& \nabla^\mu\peq/(\ed{+}\peq)\;.
\label{eq:vhydro-u}
\end{eqnarray}
\end{subequations}
$D{\,=\,}u_\mu d^\mu$ stands for the time derivative in the LRF, $\theta{\,=\,}d_\mu u^\mu$ is the scalar expansion rate, $\nabla^\mu{\,=\,}\partial^{\langle\mu\rangle}$ (where generally $A^{\langle\mu\rangle}{\,\equiv\,}\Delta^{\mu\nu}A_\nu$) denotes the spatial gradient in the LRF, and $\sigma^{\mu\nu}{\,=\,}\nabla^{\langle\mu}u^{\nu\rangle}$ (where generally $B^{\langle\mu\nu\rangle} \equiv \Delta^{\mu\nu}_{\alpha\beta} B^{\alpha\beta}$, with the traceless spatial projector $\Delta^{\mu\nu}_{\alpha\beta} \equiv \frac{1}{2} (\Delta^\mu_\alpha \Delta^\nu_\beta + \Delta^\nu_\alpha \Delta^\mu_\beta) - \frac{1}{3} \Delta^{\mu\nu} \Delta_{\alpha\beta}$) is the shear flow tensor.

Gubser flow \cite{Gubser:2010ze, Gubser:2010ui} is an analytic solution to these equations that describes a static system with flow velocity components\footnote{%
    All quantities expressed in terms of Gubser coordinates are labeled with a hat. For dimensionful quantities, such as the energy density $e$, the corresponding hatted quantity is made dimensionless by multiplying it with appropriate powers of $\tau$ (e.g. $\hat{e}=\tau^4 e$).
    \label{fn12}
    } 
$\hat{u}^\mu=(1,0,0,0)$ in a particular curvilinear coordinate system (``Gubser coordinates'') $\hat x^\mu = (\rho,\vartheta,\phi, \eta_s)$, where 
\begin{align}
    \rho(\tau,r) &\equiv - \sinh^{-1}\left(\frac{1-q^2\tau^2+q^2r^2}{2q\tau}\right)\;,\label{eq-rho}\\
    \vartheta(\tau,r) &\equiv\tanh^{-1}\left(\frac{2qr}{1+q^2\tau^2-q^2r^2}\right)\;,
\label{eq-theta}
\end{align}
with an arbitrary energy scale $q$ that controls the physical size of the system, relates the Gubser coordinates to Milne coordinates $x^\mu{\,=\,}(\tau,r,\phi,\eta_s)$. Mapping the static flow $\hat{u}^\mu$ back to Milne coordinates in Minkowski space one obtains the Gubser flow profile
\begin{subequations}
\label{eq-gubser-u}
\begin{align}
  u^\tau(\tau, r) &= \cosh\kappa(\tau,r)\;, 
\label{eq-gubser-ut}
\\
  u^{x}(\tau, r) &= \frac{x}{r}\sinh\kappa(\tau,r)\;,
\label{eq-gubser-ux}
\\
  u^{y}(\tau, r) &= \frac{y}{r}\sinh\kappa(\tau,r)\;,
\label{eq-gubser-uy}
\\
  u^\phi(\tau, r) &= u^\eta(\tau, r) = 0\;,
\label{eq-gubser-uphi}
\end{align}
\end{subequations}
where $x{\,=\,}r\cos\phi$, $y{\,=\,}r\sin\phi$, and
$\kappa(\tau,r)$ is the transverse flow rapidity, corresponding to the transverse flow velocity 
\begin{equation}
    v_\perp(\tau,r)=\tanh \kappa(\tau,r) \equiv 
    \frac{2q^2\tau r}{1+q^2\tau^2+q^2r^2}\;.
\label{eq-gubser-kappa}
\end{equation}
In Cartesian coordinates this flow profile combines boost-invariant longitudinal expansion with azimuthally symmetric transverse expansion with the particular profile (\ref{eq-gubser-kappa}).

The Gubser velocity profile (\ref{eq-gubser-kappa}) is an exact solution of the ideal fluid dynamic equations (\ref{id-hydro}) as long as all macroscopic densities depend only on the time-like coordinate $\rho\in\mathbb{R}$ (``de Sitter time''). Such systems are conformally symmetric. With this symmetry the LRF time derivative $D$ reduces to $\partial_\rho$, and  the energy and baryon number conservation laws (\ref{eq:vhydro-N},\ref{eq:vhydro-E}) turn into a coupled set of ODEs in $\rho$:
\begin{subequations}
\label{eq-gubser-evolution}
\begin{eqnarray}
  \partial_\rho\hat n + 2\tanh\rho\, \hat n &=& 0\;,
\label{eq-gubser-evolution-n}
\\
  \partial_\rho\hat e + 2\tanh\rho\, \hat e &=& -2\tanh\rho\,\hat p\;.
\label{eq-gubser-evolution-e}
\end{eqnarray}
\end{subequations}
However, since critical dynamics near the critical point breaks conformal symmetry and in particular Weyl invariance \cite{Gubser:2010ze, Gubser:2010ui} by introducing the correlation length $\xi$ and the wave number $Q$, the evolution equations (\ref{eq:phi_scaling}) lose their simple form when  recast in Gubser coordinates and expressed in terms of Gubser-rescaled quantities $\hat\phi_{\hat{Q}}$, $\hat{\Gamma}_{\hat{Q}}$ (see footnote \ref{fn12}). However, since we ignore the back-reaction from the slow modes to the conformal background fluid, we can solve Eq.~(\ref{eq:phi_scaling}) in Milne coordinates (which are also the coordinates in which \bes+ is formulated), using the Gubser profile (\ref{eq-gubser-u}) for the Milne components of $u^\mu$ as externally prescribed.

For consistency with the conformal symmetry of Eqs.~(\ref{eq-gubser-evolution}) we specify for the equilibrium EoS an ideal gas of quarks and gluons with $N_c{\,=\,}3$ colors and $N_f{\,=\,}2.5$ massless quark flavors of baryon number 1/3, each carrying baryon chemical potential $\mu/3$.\footnote{%
    This agrees with EOS3 in Ref.~\cite{Du:2019obx}.}
Its pressure and baryon density are given by
\begin{eqnarray}
\label{EOSp}
    \frac{p}{T^4} &=&  
    p_0 +N_f\left[\frac{1}{2}\left(\frac{\mu}{3T}\right)^2+\frac{1}{4\pi^2}\left(\frac{\mu}{3T}\right)^4\right]\,, 
\\
\label{EOSn}
    \frac{n}{T^3}&=&{N_f}\left[\frac{1}{3}\left(\frac{\mu}{3T}\right)+\frac{1}{9\pi^2}\left(\frac{\mu}{3T}\right)^3\right]\,,
\end{eqnarray}
where $p_0=\bigl[2(N_c^2{-}1)+(7/2)N_cN_f\bigr]\frac{\pi^2}{90}$. Using $\alpha=\mu/T$ this can be rewritten as \cite{Hatta:2015era}
\begin{equation}
    e = 3p \equiv f_*(\alpha) T^4\,,\quad n = g_*(\alpha)\mu T^2 \equiv \alpha g_*(\alpha) T^3\,,\label{eq:eos1}
\end{equation}
and thus
\begin{equation}
    s = \frac{1}{T}(e+p-\mu n)=T^3\Bigl(\frac{4}{3}f_*(\alpha)-\alpha^2g_*(\alpha)\Bigr)\equiv h_*(\alpha) T^3\,,
\label{eq:eos2}
\end{equation}
with the unitless coefficients
\begin{subequations}
\label{eq:fgh}
\begin{eqnarray}
  f_*(\alpha)&=&3p_0+\frac{N_f}{6}\alpha^2+\frac{N_f}{108\pi^2}\alpha^4\,,
\label{eq:f}
\\
  g_*(\alpha)&=&\frac{N_f}{9}+\frac{N_f}{81\pi^2}\alpha^2\,,
\label{eq:g}
\\
  h_*(\alpha)&=&4p_0+\frac{N_f}{9}\alpha^2\,.
\label{eq:h}
\end{eqnarray}
\end{subequations}
One easily verifies that Eqs.~(\ref{eq-gubser-evolution}) together with the conformal EoS (\ref{eq:eos1}) are solved consistently if $\alpha=\mu/T$ is a $\rho$-independent constant.\footnote{%
    This is generally not true for dissipative Gubser flow at non-zero baryon density.}
The corresponding temperature evolution is \cite{Gubser:2010ze}
\begin{equation}
    \hat{T}(\rho) = \frac{C}{(\cosh\rho)^{2/3}}\;,
\end{equation}
or, equivalently, in Milne coordinates
\begin{equation}
    T(\tau,r) = \frac{C}{\tau}\frac{(2q\tau)^{2/3}}{\bigl[1+2q^2(\tau^2{+}r^2)+q^4(\tau^2{-}r^2)^2\bigr]^{1/3}}\,,
\label{eq-gubser_temp}
\end{equation}
where $C$ is a constant of integration. 

We see that the background fluid is defined by providing three constants, the size parameter $q$ (with larger $q$ corresponding to smaller transverse size), the chemical potential in units of the temperature $\alpha=\mu/T$ (with larger $\alpha$ corresponding to lower collision energies which are characterized by larger baryon number stopping), and the normalization constant $C$ (which should increase with collision energy). To simulate central Au+Au collisions we follow \cite{Gubser:2010ze} and set $q^{-1}{\,=\,}4.3$\,fm. For phenomenological guidance how to fix $\alpha{\,=\,}\mu/T$ we follow Ref.~\cite{Hatta:2015era} and note that, since ideal Gubser flow evolves at constant $\alpha$, we can determine the latter from experimental data at chemical freeze-out.  A thermal analysis of hadron yield ratios measured in Au+Au and Pb+Pb collisions at RHIC and LHC leads to the following collision energy dependence of the temperature and baryon chemical potential at chemical freeze-out \cite{Cleymans:2005xv}:
\begin{equation}
    T(\mu) = a - b\mu^2 - c\mu^4\,,\quad \mu(\sqrt{s}) = \frac{d}{1+e\sqrt{s}}\,,
\end{equation}
where $a{\,=\,}0.166$\,GeV, $b{\,=\,}0.139$\,GeV$^{-1}$, $c{\,=\,}0.053$\,GeV$^{-3}$, $d{\,=\,}1.308$\,GeV, $e{\,=\,}0.273$\,GeV$^{-1}$, and $\sqrt{s}$ is the collision energy per nucleon pair in GeV. This curve can be well described by \cite{Cleymans:2005xv, Hatta:2015era}
\begin{equation}
    \alpha(\sqrt{s}) = \frac{\mu}{T}(\sqrt{s}) 
    \approx d/(a\,e\sqrt{s}) \approx 29\,\mathrm{GeV}/\sqrt{s}\,;
\end{equation}
for $\sqrt{s} = 200$\,GeV this yields $\alpha \approx 0.145$.\footnote{%
    This value of $\alpha$ is not large, implying that at this collision energy the system will not pass close to the critical point if the latter is at $\mu_c\gtrsim 400$\,MeV \cite{Bazavov:2017dus,Mukherjee:2019eou}. The large-$\alpha$ regime is studied at the lower end of the range of collision energies explored in the RHIC BES program \cite{Adamczyk:2017iwn}. However, since our intent is not to do BES phenomenology but to explore the mechanisms that drive critical fluctuation dynamics, we select $\sqrt{s} = 200$\,GeV which facilitates comparison with previous work \cite{Gubser:2010ze,Hatta:2015era}. Purely for this convenience we are therefore imagining (similar to Ref.~\cite{Rajagopal:2019xwg}) a critical point at a small value of $\mu_c/T_c$; however, where Ref.~\cite{Rajagopal:2019xwg} focused on the influence of such a critical point on dynamics at $\mu=0$, we here explore its influence on fluctuation dynamics for a system passing close to the critical point on a trajectory with non-zero baryon chemical potential, $\mu\ne0$.}
Using $N_f=2.5$ in Eqs.~(\ref{eq:fgh}) this leads to
\begin{equation}
    f_* \approx 13.91\,,\quad g_* \approx 0.28\,,\quad h_* \approx 18.54\,.\label{eq:estfgh}
\end{equation}
Finally, we fix $C$ from the total entropy produced in the collision. Again, since ideal Gubser flow conserves entropy, this can be determined from the experimentally measured final charged hadron multiplicity. Following Ref.~\cite{Gubser:2010ze} we estimate the entropy per unit space-time rapidity $\eta_s$ (which is conserved in ideal fluid dynamics with longitudinal boost-invariance) from the measured pseudorapidity density of charged particles as
\begin{equation}
    \frac{dS}{d\eta_s} \approx 7.5 \frac{dN_\mathrm{ch}}{d\eta} \approx 5000\,,
\label{eq:gubser_init_s}
\end{equation}
where we used boost-invariance to identify $\eta_s{\,=\,}\eta$ (which also makes $dS/d\eta_s$ independent of $\eta_s$) and inserted $dN_\mathrm{ch}/d\eta{\,\simeq\,}660$ for the most central Au+Au collisions at $\sqrt{s}{\,=\,}200$\,GeV. $C$ is now fixed by evaluating $dS/d\eta_s$ with the entropy density (\ref{eq:eos2}) corresponding to the ideal Gubser temperature profile (\ref{eq-gubser_temp}), by integrating the entropy flux $s_\mu d^3\sigma^\mu$ through a suitably chosen hypersurface $\Sigma$ over the entire fireball:
\begin{equation}
    \frac{dS}{d\eta_s} = \int_\Sigma s_\mu(\tau,r)\, \frac{d^3\sigma^\mu(\tau,r,\eta_s)}{d\eta_s}, 
\label{eq:gubser_init_s1}
\end{equation}
where $s_\mu{\,=\,}s\,u_\mu$. Using for simplicity a $\tau{\,=\,}$const. surface with $d^3\sigma^\mu=(\tau\,2\pi r\,dr\,d\eta_s,0,0,0)$, this simplifies to \cite{Gubser:2010ze}
\begin{equation}
    \!\!\!
    \frac{dS}{d\eta_s} = 2\pi h_*
    \int_0^\infty T^3(\tau,r)\, u^\tau(\tau,r)\, \tau\,r\,dr\,
    = 4\pi h_* C^3,\ 
\label{eq:gubser_init_s2}
\end{equation}
where in the last step we inserted Eq.~(\ref{eq-gubser-ut}) in the form
$$
  u^\tau(\tau,r) = \frac{1 + q^2\tau^2 + q^2r^2}      {\bigl[1 + 2q^2(\tau^2{+}r^2) + q^4(\tau^2{-}r^2)^2\bigr]^{1/2}}
$$
as well as Eq.~(\ref{eq-gubser_temp}) and performed the integral.\footnote{%
    Due to scale invariance and entropy conservation the integral is independent of both $q$ and the value of $\tau$ that defines the hypersurface.}
Comparing this expression with Eq.~(\ref{eq:gubser_init_s}) yields $C\simeq 2.8$.

\subsection{Entropy correction on a closed hypersurface}
\label{sec:pheno}
%

After evolving the slow modes on top of this background Gubser flow we can use equations (\ref{eq:deltas}) and (\ref{eq:dp}) to calculate the off-equilibrium corrections from the slow modes to the bulk properties of the background fluid. Ref.~\cite{Rajagopal:2019xwg} studied these corrections at different proper times; in this subsection we discuss how to evaluate them on an arbitrary hypersurface  (for example an isothermal surface). 

Equation (\ref{eq:deltas}) allows to compute the non-equilibrium slow-mode correction to the entropy density, $\Delta s(x)$. The corresponding correction to the entropy current is denoted by $\delta s^\mu\equiv u^\mu\Delta s$. When integrated over a hypersurface $\Sigma$ that encloses the entire system this generates a correction to the total entropy 
\begin{equation}
    \delta S = \int_\Sigma \delta s^\mu(x)\,d^3\sigma_\mu(x) 
    = \int_\Sigma \Delta s(x)\, u(x){\,\cdot\,}d^3\sigma(x)\,.
\label{eq:fods}
\end{equation}
Parametrizing the surface $\Sigma$ in Milne coordinates
$(\tau, x, y, \eta_s)$ as
\begin{eqnarray}
    \Sigma^\mu(x)=(\tau_f(x,y,\eta_s),x,y,\eta_s)\;,
\label{eq:fosurface}
\end{eqnarray}
where $\tau_f(x,y,\eta_s)$ is the longitudinal proper time associated with the surface point located at spatial position $(x,y,\eta_s)$, the surface normal vector at point $x$ is given by
\begin{eqnarray}
    d^3\sigma_\mu(x) 
    &=& -\epsilon_{\mu\nu\kappa\lambda} \frac{\partial\Sigma^\nu}{\partial x} \frac{\partial\Sigma^\kappa}{\partial y} \frac{\partial\Sigma^\lambda}{\partial\eta_s}\, \sqrt{-g}\,dxdyd\eta_s
\nonumber\\
\label{eq:foel}
    &=& \left(1, -\frac{\partial\tau_f}{\partial x},     -\frac{\partial\tau_f}{\partial y}, -\frac{\partial\tau_f}{\partial\eta_s}\right)
    \tau_f dxdyd\eta_s\;.
\end{eqnarray}
Here $\epsilon_{\mu\nu\kappa\lambda}$ is the Levi-Civita symbol, and $\sqrt{-g}=\tau$ is the metric determinant for Milne coordinates.

The expressions (\ref{eq:fosurface}) and (\ref{eq:foel}) are completely general and allow to parametrize any hypersurface $\Sigma$ (although $\tau_f(x,y,\eta_s)$ may be multivalued). We here want to apply them to our Gubser background flow without back-reaction from non-equilibrium slow modes, which possesses longitudinal boost-invariance and azimuthal symmetry, and study the slow-mode correction to the entropy per unit space-time rapidity $d\delta S/d\eta_s$ on an isothermal hypersurface defined by $T(\tau_f(r),r)=T_f$. For this situation it is advantageous to use polar coordinates $(r,\phi)$ instead of $(x,y)$ in the transverse plane, such that Eq.~(\ref{eq:foel}) simplifies to 
\begin{equation}
    d^3\sigma_\mu(x)=
    \left(1, -\frac{\partial\tau_f}{\partial r}, 0, 0 \right)\tau_f rdrd\phi d\eta_s
\label{eq:foe2}
\end{equation}
and the entropy correction becomes
\begin{equation}
    \frac{d\delta S}{d\eta_s} = 2\pi\int_0^{r_\mathrm{max}} \Delta s
    \left(u^\tau-u^r\frac{\partial\tau_f}{\partial r} \right)\tau_f(r) r dr\;. 
\label{eq:fodsgubser}
\end{equation}
Here the finite upper limit $r_\mathrm{max}$ accounts for the finite radial extent of this isothermal surface --- for points with $r{\,>\,}r_\mathrm{max}$ the temperature never exceeds $T_f$. Using that on the surface $T\bigl(\tau_f(r),r\bigr) = T_f$ is constant and hence $dT(\tau_f, r) = (\partial T/\partial r)dr + (\partial T/\partial \tau_f)d\tau_f = 0$, the second term under the integral (\ref{eq:fodsgubser}) can be evaluated as follows \cite{Gursoy:2014aka}:
\begin{equation}
    \frac{\partial\tau_f}{\partial r} = -\left[\left(\frac{\partial T}{\partial r}\right)\left(\frac{\partial T}{\partial \tau}\right)^{-1}\right]_{T_f}\,.
\label{eq:fode}
\end{equation}
The two factors inside square brackets are easily evaluated using the ideal Gubser temperature profile (\ref{eq-gubser_temp}). 

Before moving to numerical studies let us  quickly estimate the expected slow-mode contribution to the partial-equilibrium entropy density with the setup described in this section. As argued, the typical $Q$ contributing to $\Delta s$ in Eq.~(\ref{eq:deltas}) is of the order $Q_\mathrm{neq}$. Therefore $\Delta s \simeq \frac{1}{3}\frac{1}{(2\pi)^2}\,Q^{3}_\mathrm{neq}$ or, as a fraction of the equilibrium entropy density $s$ from Eq.~(\ref{eq:eos2}), %
\begin{equation}
\label{eq:estdeltas}
    \frac{\Delta s}{s} \sim \frac{1}{3}\frac{1}{(2\pi)^2} \frac{1}{h_{*}} \left(\frac{Q_\mathrm{neq}}{T}\right)^3. 
\end{equation}
Using $h_{*}$ from Eq.~(\ref{eq:estfgh}) and  $(Q_\mathrm{neq}/T) \lesssim 1$ we arrived at $(\Delta s/s) \sim {\cal O}(10^{-4})$.

%
\section{Results and discussion}\label{sec:results}
%

We now exploit this framework to study the dynamics of the slow modes near the QCD critical point. Here, the availability of analytic expressions for the ideal Gubser flow hydrodynamic background turns out to be helpful: by taking different limits of the background flow, we can easily separate critical dynamics from flow-induced effects. For example, we can turn off transverse flow by taking the limit $q\to0$, corresponding to an infinite transverse radius of the fireball, and we can turn off critical effects by replacing the correlation length $\xi$ by a constant $\xi_0{\,=\,}1$\,fm, the assumed correlation length far away from the critical point. We focus on collective expansion effects on slow-mode evolution in Sec.~\ref{sec:expansion}, on advection effects in Sec.~\ref{sec:advection}, and on critical dynamics due to the growth of the correlation length near the critical point in Sec.~\ref{sec:corrlen}. In the last subsection \ref{sec:phenomenology} we give the reader a feeling for the expected magnitude of phenomenological effects resulting from non-equilibrium slow mode dynamics, by studying the space-time evolution of non-equilibrium corrections to the entropy of the fireball and their imprint on the freeze-out hypersurface. 

%
\begin{figure*}[!tp]
    \centering
    \includegraphics[width= \textwidth]{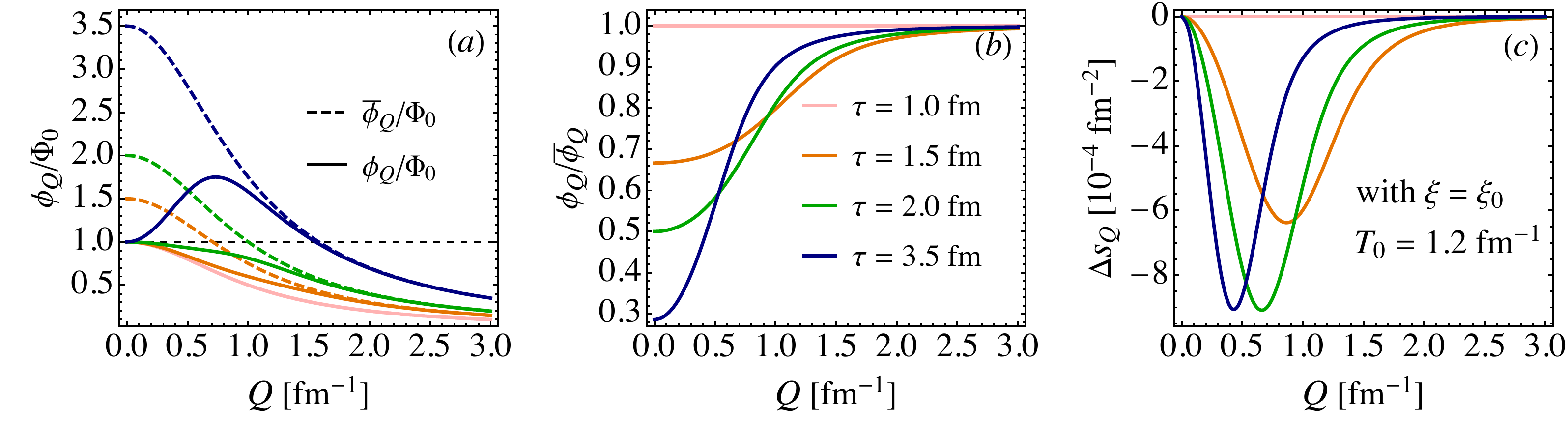}
    \caption{$Q$-dependence of the dynamics of the slow modes in a system undergoing Bjorken expansion (i.e. Gubser expansion in the limit $q\to0$) with constant $\xi=\xi_0$. Lines with different colors correspond to different times as identified in the legend. (a) Comparison of the $Q$-dependences of the scaled equilibrium ($\bar\phi_Q/\Phi_0$, dashed) and non-equilibrium ($\phi_Q/\Phi_0$, solid) values of the correlator $\phi_Q$, as functions of wave number $Q$. (b) $Q$-dependence of the non-equilibrium/equilibrium ratio $\phi_Q/\bar\phi_Q$. (c) $Q$-dependence of the non-equilibrium entropy density correction arising from the slow mode with wave number $Q$, $\Delta s_Q$. Initial conditions are $T_0{\,=\,}1.2$\,\fm, $\Gamma_0{\,=\,}1.0$\,\fm, and $\Phi_0{\,=\,}1.0$\,fm$^{-3}$ at $\tau_0{\,=\,}1$\,fm.
    \label{fig:bjorken_q}}
\end{figure*}
%

To simplify the discussion let us introduce some notation. Throughout this section we use a background medium whose hydrodynamic evolution starts at $\tau_0{\,=\,}1$\,fm, with initial velocity and temperature profiles (\ref{eq-gubser-u}) and (\ref{eq-gubser_temp}). We introduce the shorthand $\Phi{\,\equiv\,}\cpz/n^2{\,=\,}\bar{\phi}_{0,\xi_0}$ for the equilibrium value of the static ($Q{\,=\,}0$) slow mode (which depends on space-time position $x$ through the medium properties) and denote by $\Phi_0{\,\equiv\,}\Phi(\tau_0,r)$ its initial $r$-profile at $\tau_0$. We further introduce  $\Gamma{\,\equiv\,}\Gamma_{\xi_0}{\,=\,}2\Bigl(\lambda_T/(\cpz\xi_0^2)\Bigr)$ ({\it cf.} Eq.~(\ref{eq:Gamma_xi})) with $\xi_0{=}1$\,fm and denote by $\Gamma_0\equiv \Gamma_{\xi_0}(\tau_0,r)$ its initial profile at $\tau_0$. To simplify the discussion of critical effects induced by the growth of the correlation length $\xi$ near the critical point, we introduce the ``non-critical reference value" $\bar\phi_{Q,\xi_0}{\,\equiv\,}(\cpz/n^2)/(1+(Q\xi_0)^2)$, i.e. the equilibrium value for the mode with wave number $Q$ in a system with constant correlation length $\xi_0$. When focusing on effects from critical dynamics we therefore plot ratios such as 
\begin{equation}
  \frac{\bar\phi_Q}{\bar\phi_{Q,\xi_0}} = \left(\frac{\xi}{\xi_0}\right)^2 \frac{f_2(Q\xi)}{f_2(Q\xi_0)} = \left(\frac{\xi}{\xi_0}\right)^2\left(\frac{1+(Q\xi_0)^2}{1+(Q\xi)^2}\right)\,.
\end{equation}
When simultaneously looking at the $Q$-dependence we plot
\begin{equation}
  \frac{\bar\phi_Q}{\bar\phi_{0,\xi_0}} = \left(\frac{\xi}{\xi_0}\right)^2 f_2(Q\xi)  =\left(\frac{\xi}{\xi_0}\right)^2
   \left(\frac{1}{1+(Q\xi)^2}\right)\,.
\end{equation}
To facilitate comparison we follow Ref.~\cite{Rajagopal:2019xwg} and initialize the slow modes $\phi_Q$ at their equilibrium values $\bar\phi_Q$.

\subsection{Medium expansion}
\label{sec:expansion}
\subsubsection{Constant correlation length}
\label{constant_xi}
%

In this subsection we focus on effects on the fluctuation dynamics arising from the space-time dependence of the hydrodynamic fields caused by the expansion of the background fluid. We do so by tracing the non-equilibrium evolution of the slow mode $\phi_Q$, as well as its equilibrium value $\bar\phi_Q$, in a noncritical expanding medium with constant correlation length $\xi=\xi_0$. To focus on the dilution and cooling effects caused by the expansion, rather than the collective flow that accompanies it, we remove the spatial gradients in the medium (i.e. advection affects), by letting $q\to 0$. According to Eqs.~(\ref{eq-gubser-u}) this results in $u^\tau \approx 1$ and $u^r \approx 0$, i.e. 1-dimensional Bjorken expansion along the longitudinal direction with the temperature profile
(see Eq.~(\ref{eq-gubser_temp}))
\begin{equation}
    T \approx \frac{C(2q)^{2/3}}{\tau^{1/3}} \equiv T_0\left(\frac{\tau_0}{\tau}\right)^{1/3}\,,
\label{eq:bjorken_temp}
\end{equation}
where $C{\,\propto\,}q^{-2/3}{\,\to\,}\infty$ as $q{\,\to\,}0$ so that $T_0\equiv C(4q^2/\tau_0)^{1/3}$ remains finite. With our conformal EoS this implies the well-known Bjorken scaling laws
\begin{equation}
    n\propto \frac{1}{\tau}\,,\quad 
    s\propto \frac{1}{\tau}\,,\quad 
    e\propto \frac{1}{\tau^{4/3}}
\end{equation}
as $q\to 0$, and the parametrization (\ref{eq:param}) gives
\begin{equation}
  \bar\phi_Q \propto \frac{\cpz}{n^2} \propto \frac{(s^2/n)}{n^2} \propto \tau\,,\quad
  \Gamma_Q \propto \frac{\lambda_T}{\cpz} \propto \frac{T^2}{(s^2/n)} \propto \tau^{1/3}\,.
\end{equation}
One sees that in this geometry, and without critical correlations, both $\Gamma_Q$ and $\bar\phi_Q$ increase monotonically with $\tau$ as the system expands:
\begin{eqnarray}
    {\bar\phi}_Q(\tau) &=& \frac{\tau}{\tau_0}\,{\bar\phi}_Q(\tau_0) = \frac{\tau}{\tau_0}\,\Phi_0\, f_2(Q\xi_0)\,,
\label{eq:bjorken_phi_tau}
\\
    \Gamma_Q(\tau) &=& \left(\frac{\tau}{\tau_0}\right)^{1/3} \!\! \Gamma_Q(\tau_0)
    = \left(\frac{\tau}{\tau_0}\right)^{1/3}\!\!\Gamma_0\, 
    f_\Gamma(Q\xi_0)\,.\quad
\label{eq:bjorken_gamma_tau}
\end{eqnarray}
With this Bjorken flow profile the equation of motion for $\phi_Q$ turns into an ODE,
\begin{equation}
    \partial_\tau \phi_Q(\tau) = - \Gamma_Q(\tau) \Bigl(\phi_Q(\tau) - \bar\phi_Q(\tau)\Bigr)\,,
\label{eq:bjorken_eom}
\end{equation}
which we solve numerically. As the normalization of $\phi_Q$ is arbitrary, and the heat conductivity $\lambda_T$ contains a free parameter $C_\lambda$, we simply set $\Phi_0{\,=\,}1$\,fm$^3$ and $\Gamma_0{\,=\,}1$\,fm$^{-1}$ for the initial conditions in Eqs.~(\ref{eq:bjorken_phi_tau}) and (\ref{eq:bjorken_gamma_tau}) \cite{Rajagopal:2019xwg}. 

Figure~\ref{fig:bjorken_q} shows several snapshots of the $Q$-dependence of $\phi_Q$ (panels (a,b)) and of the corresponding non-equilibrium entropy correction $\Delta s_Q$ (panel (c)) that illustrate their time evolution. Panel (a) shows that the equilibrium value $\bar\phi_Q$ (dashed lines) increases with time but decreases with growing wave number $Q$, reflecting the scaling function $f_2(Q\xi_0)$ (\ref{eq:bjorken_phi_tau}). The solid lines showing the non-equilibrium value $\phi_Q$ exhibit an interesting feature: while they approach their corresponding equilibrium values at large $Q$, they stay close to their initial value $\Phi_0\, f_2(Q\xi_0)$ at small wave number. This reflects the $Q$ dependence of the relaxation rate, $\Gamma_Q\propto Q^2$ for $Q\ll\xi^{-1}$ and $\Gamma_Q\propto Q^4$ for $Q\gg\xi^{-1}$. This feature sets a scale $Q_\textrm{neq}(\tau)$ which decreases with time that separates modes that can equilibrate within time $\tau$ from those that cannot. For small $Q$, $\bar\phi_Q$ increases with time but $\phi_Q$ remains basically frozen,  causing the ratio $\phi_Q/\bar\phi_Q$ to decrease with time, as shown in Fig.~\ref{fig:bjorken_q}b. On the other hand the same panel also shows that, since the relaxation rate $\Gamma_Q$ increases and the expansion rate $\theta$ decreases with time, at sufficiently late times even the low-$Q$ modes equilibrate (i.e. $Q_\textrm{neq}$ decreases with times). As $Q$ decreases from large to smaller values, the decreasing ratio $\phi_Q/\bar\phi_Q$ implies a larger (negative) contribution $\Delta s_Q$ to the entropy density, but as $Q$ approaches zero these are cut off by the phase space factor $(Q/2\pi)^2$ (see Eq.~(\ref{eq:deltas})). The largest contribution to $|\Delta s|$ thus arises from modes with $Q{\,\sim\,}Q_\mathrm{max}{\,\sim\,}\mathcal{O}(Q_\textrm{neq})$ (which decreases with time), as shown in Fig.~\ref{fig:bjorken_q}c (see also Ref. \cite{Rajagopal:2019xwg}).  

\begin{figure}[!tp]
    \centering
    \includegraphics[width= 0.7\linewidth]{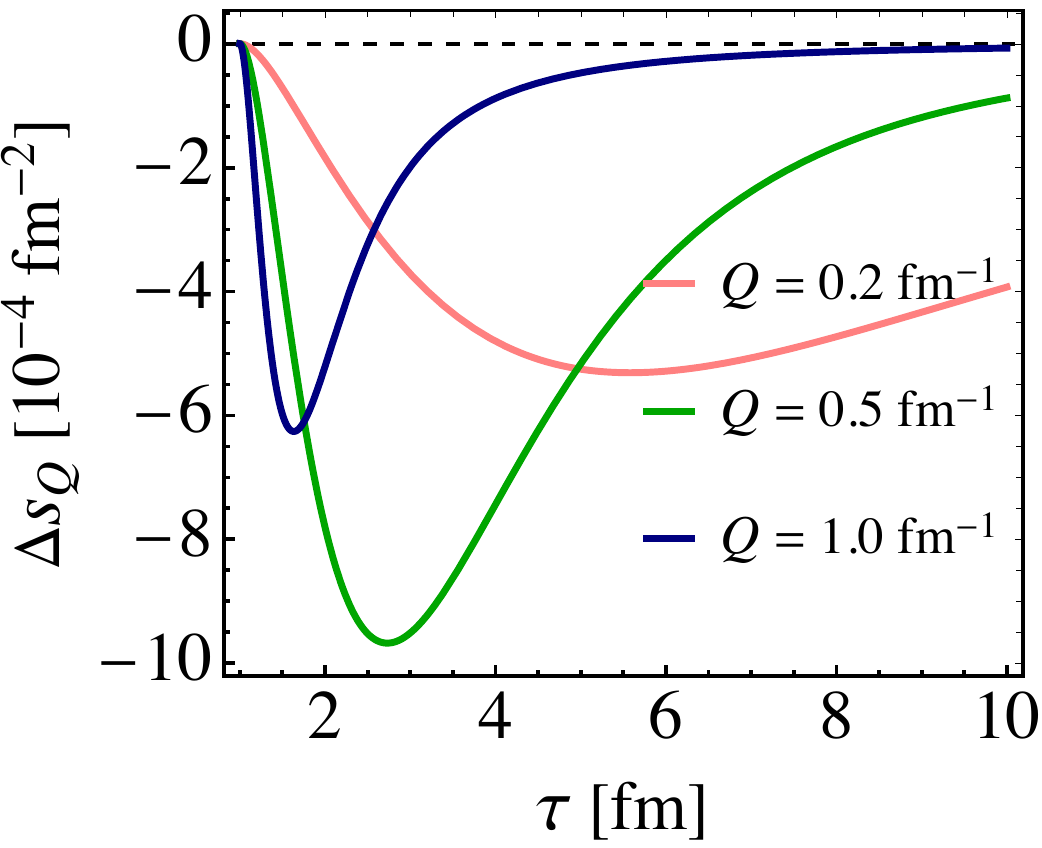}
    \caption{Time evolution of the correction $\Delta s_Q$ to the entropy density for three wave numbers $Q=0.2,\,0.5,\,1.0$\,\fm, for the same setup as in Fig.~\ref{fig:bjorken_q}.
    \label{fig:bjorken_stau}}
\end{figure}

\begin{figure*}[!tp]
    \centering
    \includegraphics[width= \textwidth]{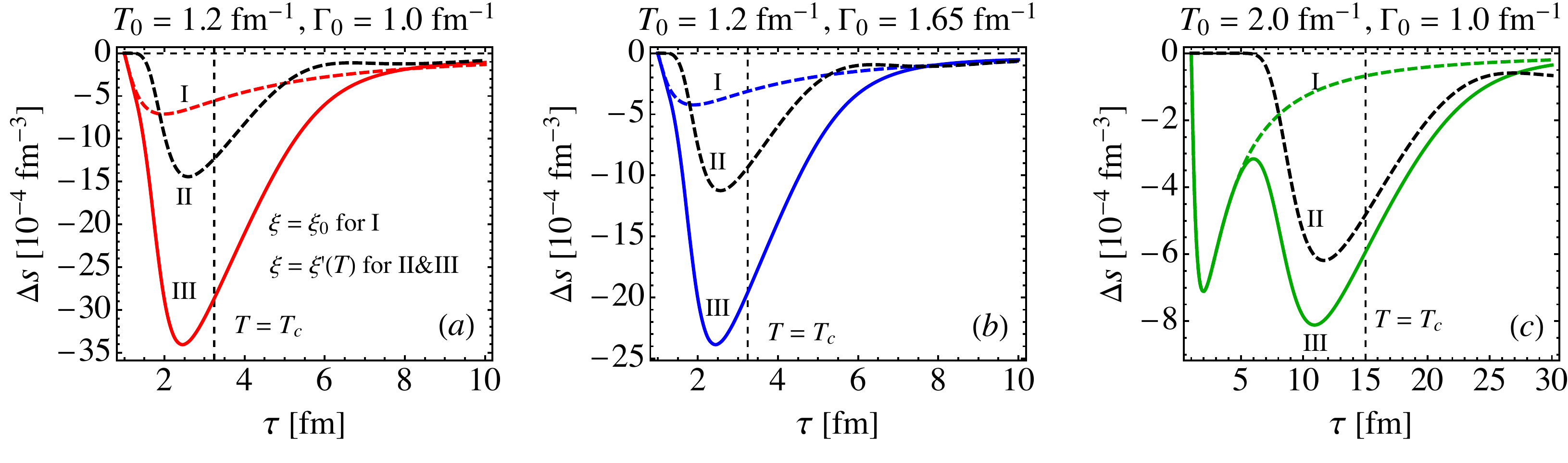}
    \caption{Comparison of the time evolution of the $Q$-integrated non-equilibrium entropy density correction $\Delta s$ for different dynamic models. I: constant $\xi = \xi_0$ (colored dashed lines); II: 
    $\xi=\xi'(T)$ (Eq.~(\ref{eq:xipara2})) with time-independent $\Phi=\Phi_0$ and $\Gamma=\Gamma_0$ (black dashed lines); III: full dynamics with evolving $\Phi$, $\Gamma$ and $\xi=\xi'(T)$ (colored solid lines). The three panels correspond to different initial conditions: (a) $T_0 = 1.2$\,\fm\ and $\Gamma_0=1.0$\,\fm; (b) $T_0 = 1.2$\,\fm\ and $\Gamma_0=1.65$\,\fm; (c) $T_0 = 2.0$\,\fm\ and $\Gamma_0=1.0$\,\fm. The vertical black dashed lines show the time $\tau{\,=\,}3.24$\,fm in (a,b), and $\tau{\,=\,}15$\,fm in (c) when the temperature passes through $T_c$ where $\xi$ peaks.
    \label{fig:bjorken_dynamics_comparison}}
\end{figure*}
%
\begin{figure*}[!tp]
    \centering
    \includegraphics[width= \textwidth]{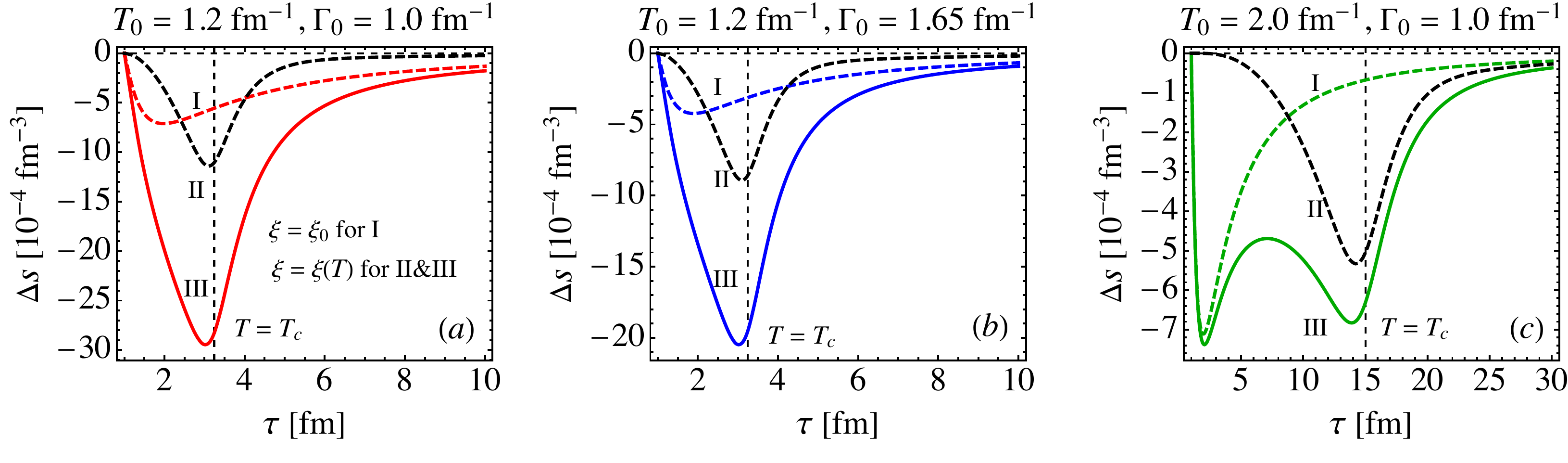}
    \caption{Same as Fig.~\ref{fig:bjorken_dynamics_comparison}, but with the correlation length $\xi(T)$ parametrized by Eq.~(\ref{eq:xipara1}).
    \label{fig:bjorken_dynamics_comparison2}}
\end{figure*}

Another way to illustrate this is shown in Fig.~\ref{fig:bjorken_stau} where we plot the time evolution of $\Delta s_Q$ for three typical $Q$ values: $Q{\,=\,}0.5$\,\fm (which is close to $Q_\textrm{max}$) 
as well as $Q{\,=\,}0.2$ and 1.0\,\fm (which are below and above $Q_\textrm{max}$). As time evolves, the modes first drop out of equilibrium (and thus begin to contribute to $\Delta s$) but later re-equilibrate (with their contribution to $\Delta s$ peaking and later decreasing). We note that the entropy contributed by the non-equilibrium slow modes, both for a fixed wave number $Q$ and integrated over $Q$, initially decreases as the slow modes are driven out of equilibrium by the large initial longitudinal expansion rate. As already explained, the negative sign of $\Delta s$ simply reflects the fact that, as long as the slow modes are out of equilibrium, the system has not yet reached a state of maximum entropy. For large $Q$, $|\Delta s_Q|$ peaks early and ceases rather quickly, and for small $Q$ the entropy contribution peaks and decays later. At first sight it looks as if the time integral of $\Delta s_Q$ might diverge as $Q\to0$, but we checked numerically that infrared convergence is ensured by the phase-space factor $(Q/2\pi)^2$.  

\vspace*{-3mm}
\subsubsection{Critical correlations in an expanding medium}
\label{critical_xi}
\vspace*{-2mm}

In this subsection we add critical effects, by generalizing the results from the preceding subsection for Bjorken expansion to include a temperature-dependent correlation length which peaks at a critical temperature $T_c$. We start with the Gaussian parametrization (\ref{eq:xipara2}), showing the corresponding $Q$-integrated non-equilibrium entropy density corrections $\Delta s(\tau)$ in Fig.~\ref{fig:bjorken_dynamics_comparison}, and then compare with the parametrization (\ref{eq:xipara1}) in Fig.~\ref{fig:bjorken_dynamics_comparison2}, to get a feeling for how strongly different parametrizations of $\xi$ might affect the evolution of the critical fluctuations. In each panel we compare three dynamical scenarios for the slow modes (the hydrodynamic background remains always the same): a constant correlation length $\xi{\,=\,}\xi_0$ as in the preceding subsection (I); a correlation length $\xi(T)$ which changes with the time evolving temperature $T$ while the temperature- and resulting time-dependence of $\cpz/n^2$ and $\lambda_T/\cpz$ is ignored, i.e. $\Phi{\,=\,}\Phi_0$ and $\Gamma{\,=\,}\Gamma_0$ are frozen at their initial values \cite{Rajagopal:2019xwg} (II); and (III) a fully dynamical scenario where not only $\xi$, but also the ratios $\cpz/n^2$ and $\lambda_T/\cpz$ (i.e. $\Phi$ and $\Gamma$) change with the evolving temperature. The three panels in each figure correspond to three different initial conditions: a lower initial temperature $T_0{\,=\,}1.2$\,\fm\ in panels (a,b) and a 60\% higher initial $T_0{\,=\,}2$\,\fm\ in panel (c), combined with a somewhat slower relaxation controlled by $\Gamma_0=1.0$\,\fm\ in panels (a,c) and a faster relaxation $\Gamma_0=1.65$\,\fm\ in panel (b). 

Comparison of the colored dashed and solid lines for scenarios I and III shows that at times corresponding to temperatures far above or below $T_c$ (where the time corresponding to $T=T_c$ is identified by a thin vertical black dashed line (see Fig. \ref{fig:param_xi}b)) the non-equilibrium entropy density corrections agree --- they differ only around $T_c$ where the correlation length $\xi$ is critically enhanced, leading to larger entropy corrections $|\Delta s|$. Panel (c) with the higher initial temperature $T_0{\,=\,}2.5\,T_c$ is interesting: without critical slowing down (scenario I) the non-equilibrium entropy density correction peaks early and has largely decayed (i.e. the slow modes have largely equilibrated) by the time the system passes through $T_c$; in scenario III, on the other hand, critical slowing down near $T_c$ allows the slow modes to fall out of equilibrium for a second time (starting when $T\lesssim T_c{\,+\,}\Delta T$), leading to a secondary peak of $|\Delta s|$ near $T_c$. In scenario II both the relaxation rates $\Gamma_Q$ and equilibrium values $\phi_Q$ change only because $\xi$ evolves with temperature, and therefore $|\Delta s|$ closely tracks the evolution of $\xi$, with a single peak near $T_c$ \cite{Rajagopal:2019xwg}. The shift of the $\xi$-driven peaks in $|\Delta s|$ towards temperatures $T>T_c$ is the result of a competition between relaxation towards equilibrium of the slow modes and the rate of expansion of the hydrodynamic medium which drives the slow modes away from equilibrium. Since the expansion rate falls like $1/\tau$, the balance is shifted away from equilibrium at $T>T_c$ (earlier times) and towards equilibrium at $T<T_c$ (later times), giving rise to the observed asymmetry and shift of the peak in $|\Delta s|$. 

Comparison of Figs.~\ref{fig:bjorken_dynamics_comparison} and \ref{fig:bjorken_dynamics_comparison2} shows that this asymmetry and shift is less pronounced for the parametrization (\ref{eq:xipara1}) for $\xi(T)$ which peaks more sharply at $T_c$, making the peak in $|\Delta s|$ less sensitive to the changing hydrodynamic expansion rate between $T\gtrsim T_c$ and $T\lesssim T_c$.\footnote{%
    The sharper peak of $\xi(T)$ causes $d\xi/dT$ to get large near $T_c$, causing disequilibrating expansion effects on the slow mode $\sim \partial_\tau \bar\phi_Q/\bar\phi_Q$ to dominate over equilibrating relaxation dynamics especially close to $T_c$, thereby causing the peak in 
    $|\Delta s|$ near $T_c$.}$^,$\footnote{%
    Note that the colored dashed lines describing scenario I (with a constant $\xi{\,=\,}\xi_0$ are, of course, identical in Figs.~\ref{fig:bjorken_dynamics_comparison}
    and \ref{fig:bjorken_dynamics_comparison2}.}
At a sufficiently detailed level, the choice of parametrization for $\xi$ is thus seen to have a noticeable effect on the evolution of the off-equilibrium modes, especially close to the critical point. Since Eq.~(\ref{eq:xipara1}) is the more realistic parametrization we will from now on use it as our default.

%
\subsection{Transverse flow effects on slow-mode dynamics}
\label{sec:advection}
%

We now turn our attention to transverse flow effects on slow-mode dynamics. As already mentioned in the introduction, advection by transverse flow can affect the evolution of the non-equilibrium fluctuations by carrying them outward from the middle to the edge of the fireball \cite{Rajagopal:2019xwg}. The analytically known structure of the ideal Gubser solution makes it possible to study this effect semi-analytically, without having to solve a (3+1)-dimensional set of coupled differential equations. Things become particularly simple and clear in the early time regime $\tau \ll 1/q$ \cite{Hatta:2014upa, Hatta:2014jva, Hatta:2015era}. In this limit, the flow velocity can be approximated by
\begin{equation}
\label{eq:flow_approx}
    u^\tau \approx 1+\mathcal{O}(\tau^2)\,,\quad u^r\approx\frac{2q^2\tau r}{1+q^2r^2}\,.
\end{equation}
Since we will ignore the $\mathcal{O}(\tau^2)$ corrections this approximation breaks down when $\tau \sim 1/q$ \cite{Hatta:2014upa, Hatta:2014jva, Hatta:2015era}. The expansion rate is approximated by $\theta \approx \partial_r u^r + 1/\tau + u^r/r$, and the equations of motion become $(\partial_\tau + u^r\partial_r)\phi_Q = - \Gamma_Q (\phi_Q{-}\bar\phi_Q)$, or equivalently,
\begin{equation}
    \partial_\tau \phi_Q = - \Gamma_Q (\phi_Q - \bar\phi_Q) - u^r\partial_r\phi_Q \,.
\label{eq:gubser_smallt_eom}
\end{equation}
The last term, driven by the transverse radial flow $u^r$, modifies the Bjorken dynamics studied in the previous subsection, by contributing with a negative sign to the time derivative of the slow mode $\phi_Q$ if its gradient $\partial_r\phi_Q$ points along the flow direction. The approximate temperature profile is
\begin{equation}
    \!\!\!
    T(\tau,r)\approx\frac{C}{\tau^{1/3}}\frac{(2q)^{2/3}}{(1+q^2r^2)^{2/3}}\equiv T_0\left(\frac{\tau_0}{\tau}\right)^{1/3}
    \!\!\mathcal{F}^{2/3}(r)\,,\ 
\label{eq:temp_smallt}
\end{equation}
with $\mathcal{F}(r)\equiv1/(1+q^2r^2)$ and $T_0\equiv C(4q^2/\tau_0)^{1/3}$. The function $\mathcal{F}(r)$ encodes the $r$-dependence of the temperature profile. From the temperature and the condition $\alpha{\,=\,}\mu/T{\,=\,}$const. the other thermodynamic quantities can be derived using the EoS. Comparing Eqs. (\ref{eq:temp_smallt}) and (\ref{eq:bjorken_temp}) we see that at the early times the $\tau$-dependence of the profile agrees with the one for Bjorken flow while the radial profile is modified by the factor $\mathcal{F}(r)$. This is expected since for $\tau \ll 1/q$ the expansion is dominantly along the longitudinal direction \cite{Gubser:2010ze, Gubser:2010ui}. In this limit, the background fluid can be considered as a superposition of fluid cells undergoing Bjorken expansion with different initial conditions. Still, the evolution equation (\ref{eq:gubser_smallt_eom}) for the critical fluctuations now has the additional source term $- u^r\partial_r\phi_Q$ on its right hand side, and  comparing the following results to those presented in Sec. \ref{sec:expansion} therefore allows us to isolate advection effects caused by transverse expansion flow through this term. To gain intuition about transverse flow effects, we will in this subsection use this picture to late times ($\approx 15$ fm) as a simplified background, even though the approximation breaks down for $\tau > 1/q$, i.e. it no longer represents Gubser flow at such late times.

%
\begin{figure*}[!tp]
    \centering
    \includegraphics[width=\textwidth]{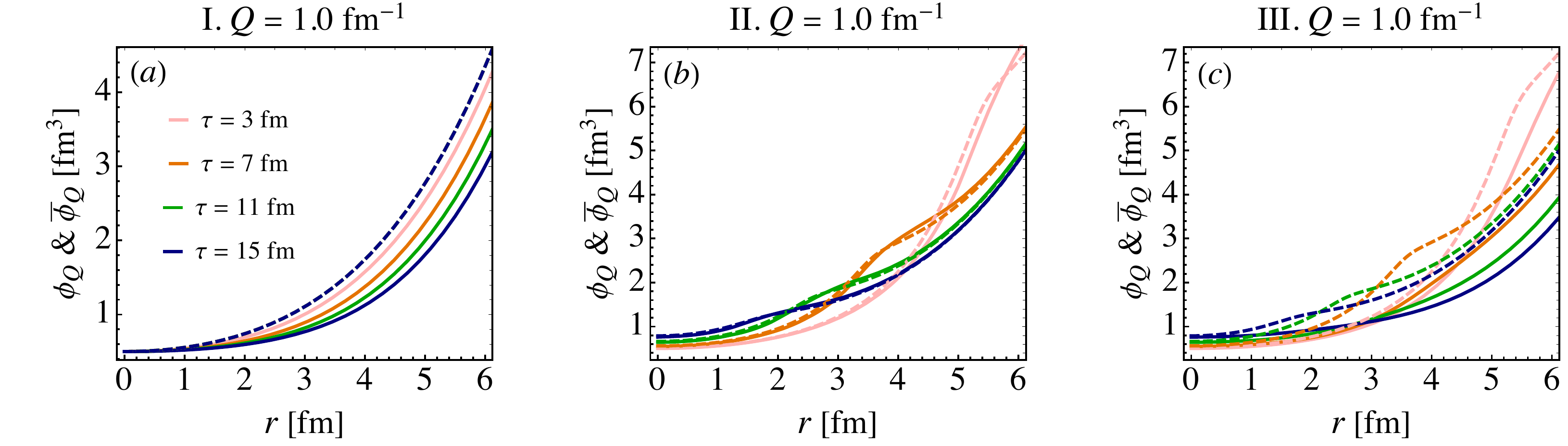}
    \caption{Evolution of the mode with $Q = 1.0$\,fm$^{-1}$ in three dynamic models: (I) with transverse flow $u^r$ and constant $\xi=\xi_0$; (II) without transverse flow (i.e. setting $u^r = 0$) but $\xi=\xi(T)$;  (III) with transverse flow $u^r$ and $\xi=\xi(T)$. Dashed lines show the equilibrium values $\bar\phi_Q$, solid lines the non-equilibrium values $\phi_Q$. Different colors indicate different times, $\tau = 3,\, 7,\, 11,\, 15$\,fm. The initial temperature and relaxation rate at $r=0$ are $T_0=2.2$\,\fm\ and $\Gamma_0=0.9$\,\fm, respectively.
    \label{fig:gubser_phievo_r_0}}
\end{figure*}%
%

We simplify the dynamics further by setting $\Phi{\,=\,}\Phi_0$ and $\Gamma{\,=\,}\Gamma_0$, i.e. by ignoring the time dependence of $\cpz/n^2$ and $\lambda_T/\cpz$ \cite{Rajagopal:2019xwg}, in order to focus on flow and suppress effects simply caused by cooling through expansion. The time dependence of $\bar\phi_Q$ and $\Gamma_Q$ will then arise solely from the temperature- and resulting time-dependence of $\xi$:
\begin{eqnarray}
    \bar\phi_Q &=& \Phi_0 \mathcal{F}^{-2}(r)\left(\frac{\xi}{\xi_0}\right)^2f_2(Q\xi)\,,
\label{eq:adv_initial1}\\
    \Gamma_Q &=& \Gamma_0\,\mathcal{F}^{-2/3}(r)\left(\frac{\xi_0}{\xi}\right)^4 f_\Gamma(Q\xi)\,.
\label{eq:adv_initial2}
\end{eqnarray}
Even if $\xi$ were a constant, $\bar\phi_Q$ and $\Gamma_Q$ still depend on $r$ because they depend on the temperature of the medium whose spatial variation is described by the profile $\mathcal{F}(r)$. In the general case these quantities acquire additional $r$-dependence through the $T(\tau,r)$-dependence of $\xi(T)$. Initial conditions for $\bar\phi_Q$ and $\Gamma_Q$ are computed with $\Phi_0{\,=\,}1.0$\,fm$^{-3}$ and $\Gamma_0{\,=\,}0.9$\,\fm, using an initial temperature $T_0{\,=\,}2.2$\,\fm\ at $\tau_0{\,=\,}1$\,fm. The corresponding initial temperatures for cells at transverse positions $r{\,=\,}1.69$ and 5.64\,fm are $T(r{=}1.69\,\mathrm{fm},\tau_0){\,=\,}2.0$\,\fm\ and $T(r{=}5.64\,\mathrm{fm},\tau_0)=1.2$\,\fm, with initial relaxation rates $\Gamma(r{=}1.69\,\mathrm{fm}, \tau_0){\,=\,}1.0$\,\fm\ and $\Gamma(r{=}5.64\,\mathrm{fm}, \tau_0)=1.65$\,\fm. The latter agree with the initial conditions studied in Fig.~\ref{fig:bjorken_dynamics_comparison2}b,c in the preceding subsection, to facilitate comparison.

%
\begin{figure*}[!tp]
    \centering
    \includegraphics[width=\textwidth]{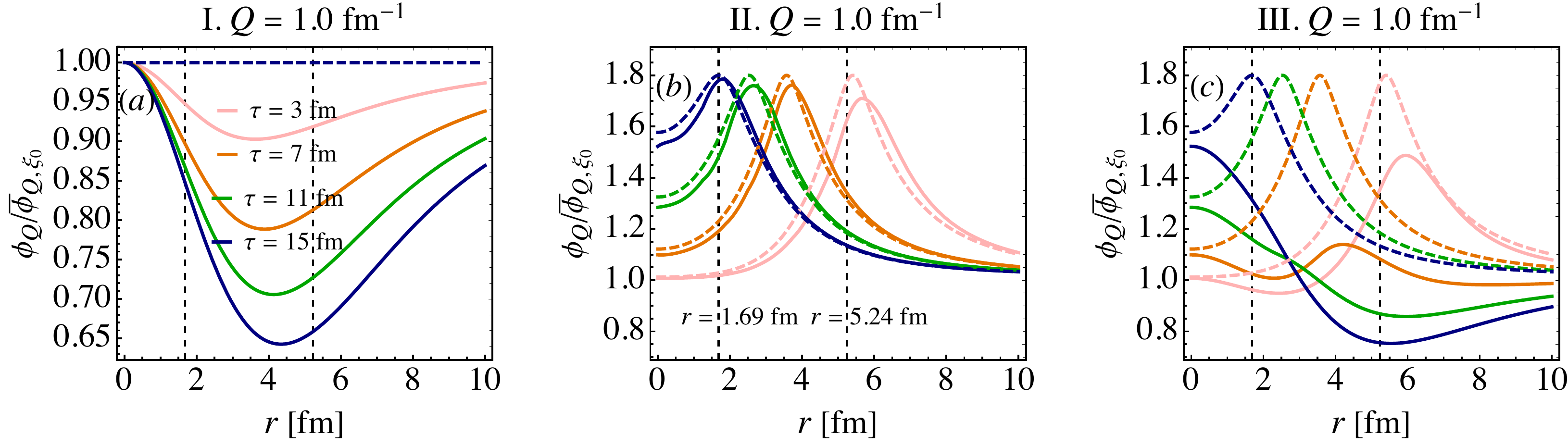}
    \caption{Same as Fig.~\ref{fig:gubser_phievo_r_0}, but using unitless ratios for the vertical axes and an extended range up to $r{\,=\,}10$\,fm for the horizontal axes. Dashed lines show the rescaled equilibrium values $\bar\phi_Q/\bar\phi_{Q, \xi_0}$, solid lines the corresponding non-equilibrium values $\phi_Q/\bar\phi_{Q, \xi_0}$. Here $\bar\phi_{Q, \xi_0}$ denotes $\bar\phi_Q$ for $\xi=\xi_0$ for each respective $r$ at $\tau_0$ (i.e. the dashed line in Fig.~\ref{fig:gubser_phievo_r_0}a). The two vertical dashed lines show the two radial distances $r{\,=\,}5.64$ and 1.69\,fm where the initial temperatures are 1.2 and 2.0 fm$^{-1}$ and the initial relaxation rates are 1.65 and 1.0\,\fm, respectively, corresponding to the cases studied in Fig.~\ref{fig:bjorken_dynamics_comparison2}b,c. 
    \label{fig:gubser_phievo_r}}
\end{figure*}%
%

We shall see that the additional $r$-dependence introduced by the profile factor $\mathcal{F}(r)$ in Eqs.~(\ref{eq:adv_initial1},\ref{eq:adv_initial2}) results in qualitatively different evolution of the slow modes in this work compared to Ref.~\cite{Rajagopal:2019xwg}. In Fig.~\ref{fig:gubser_phievo_r_0} we compare, for a single slow mode with wave number $Q{\,=\,}1$\,\fm, three dynamic models: I (panel a), constant correlation length $\xi{\,=\,}\xi_0$ with non-zero radial flow $u^r$; II (panel b), temperature dependent correlation length $\xi{\,=\,}\xi(T)$ without radial flow, $u^r{\,=\,}0$; and III (panel c), temperature dependent $\xi{\,=\,}\xi(T)$ combined with non-zero radial flow $u^r$.\footnote{%
    Note that we do not change the flow velocity of the background fluid --- we only turn on or off the $u^r$ term in the evolution equation for the slow modes.}
Note that by setting $u^r{\,=\,}0$ in (II), fluid cells at different transverse positions $r$ do not affect each other and, with the profile (\ref{eq:temp_smallt}), evolve independently from each other following Bjorken dynamics; in other words, transverse flow is turned off in scenario (II). For each of these three scenarios, the plots show snapshots at 4 different times of the radial profiles of the equilibrium (dashed lines) and non-equilibrium (solid lines) values of the slow mode, $\bar\phi_{Q}$ and $\phi_Q$, respectively.

In Fig.~\ref{fig:gubser_phievo_r_0} we first draw the reader's attention to the generic upward-sweeping behavior at large $r$ of both equilibrium and non-equilibrium values of the slow mode. This dependence arises directly from the profile factor $\mathcal{F}(r)$ in Eq.~(\ref{eq:adv_initial1}) and reflects the fact that in the expression $\bar\phi_0=\cpz/n^2$ the square of the baryon density $n^2$ decreases faster with decreasing temperature and hence increasing $r$ than the heat capacity $\cpz$. From Eq.~(\ref{eq:adv_initial2}) it is clear that $\Gamma_Q$ shares with $\bar\phi_Q$ this monotonic rise with $r$, at a somewhat slower rate. Since $\cpz \propto T^3 \propto n$, meaning that $\bar\phi_0\propto 1/n$, the upward rise of this measure of fluctuations at large $r$ corresponds to the increase in fluctuations in a region where there are few particles.  In any future phenomenological analysis, it will contribute little to observables. In our model study, however, we shall see that this $r$-dependence is useful as a device that will enable us to visualize important physical effects.

In scenario I (Fig.~\ref{fig:gubser_phievo_r_0}a) the equilibrium value $\bar\phi_Q$ (dashed line) remains frozen at its initial value because $\xi{\,=\,}\xi_0$ is constant and independent of temperature. However, the radial gradient of $\phi_Q$ couples to the non-zero radial flow $u^r$ and causes $\phi_Q$ to evolve differently at different radial positions $r$. Since the gradient $\partial_r\phi_Q$ points along $u^r$, the time derivative of the slow mode $\phi_Q$ gets a negative contribution in Eq.~(\ref{eq:gubser_smallt_eom}) from $-u^r\partial_r\phi_Q$, and thus $\phi_Q$ is pushed downward (or rather outward) further and further as time progresses. This outward transport of $\phi_Q$ by radial flow is known as ``advection'' \cite{Rajagopal:2019xwg} although (due to the different profile functions $\mathcal{F}$) it manifests itself differently here than in Ref.~\cite{Rajagopal:2019xwg}. As expected, at $r=0$ the transverse flow vanishes, $u^r(0){\,=\,}0$, and the slow mode does not evolve. 

In scenario II without transverse flow (Fig.~\ref{fig:gubser_phievo_r_0}b), cells at different $r$ evolve independently. In this scenario the temperature dependence is included for $\xi(T)$. Since the temperature (\ref{eq:temp_smallt}) is highest at $r{\,=\,}0$, the location of $T_c$ moves inward (due to cooling by longitudinal expansion) as time proceeds. In Figs.~\ref{fig:gubser_phievo_r_0}b,c this is reflected by the leftward movement of the 
``bumps" on the dashed curves which correspond to the location where the temperature of the medium is in the range where $\xi(T)$ has its peak. When plotted in this way, the upward sweep of the curves is more apparent to the eye than the bumps. Recalling, though, that this upward sweep occurs by definition in regions with small $n$ that would contribute little to observable consequences, we plot in the following Fig.~\ref{fig:gubser_phievo_r} a unitless ratio which serves to eliminate this $r$-dependence, making the effects of the critical fluctuations --- which arise in regions with larger $n$ --- more apparent. Figs.~\ref{fig:gubser_phievo_r}b,c are in this sense the better way to visualize the consequences of our analysis, but the dynamics in the equations that govern $\phi$ are more easily understood from Figs.~\ref{fig:gubser_phievo_r_0}b,c so we shall inspect these first. We see that the dashed lines move around with time, as the temperature of the plasma changes in space and time and as the region where $\xi$ peaks moves inward. We then see that in Fig.~\ref{fig:gubser_phievo_r_0}b the solid lines follow the dashed lines and become closer and closer to the dashed lines as time proceeds. This clearly demonstrates relaxation towards equilibrium in the absence of any effects of transverse flow.

%
\begin{figure*}[!tp]
    \centering
    \includegraphics[width=0.7\textwidth]{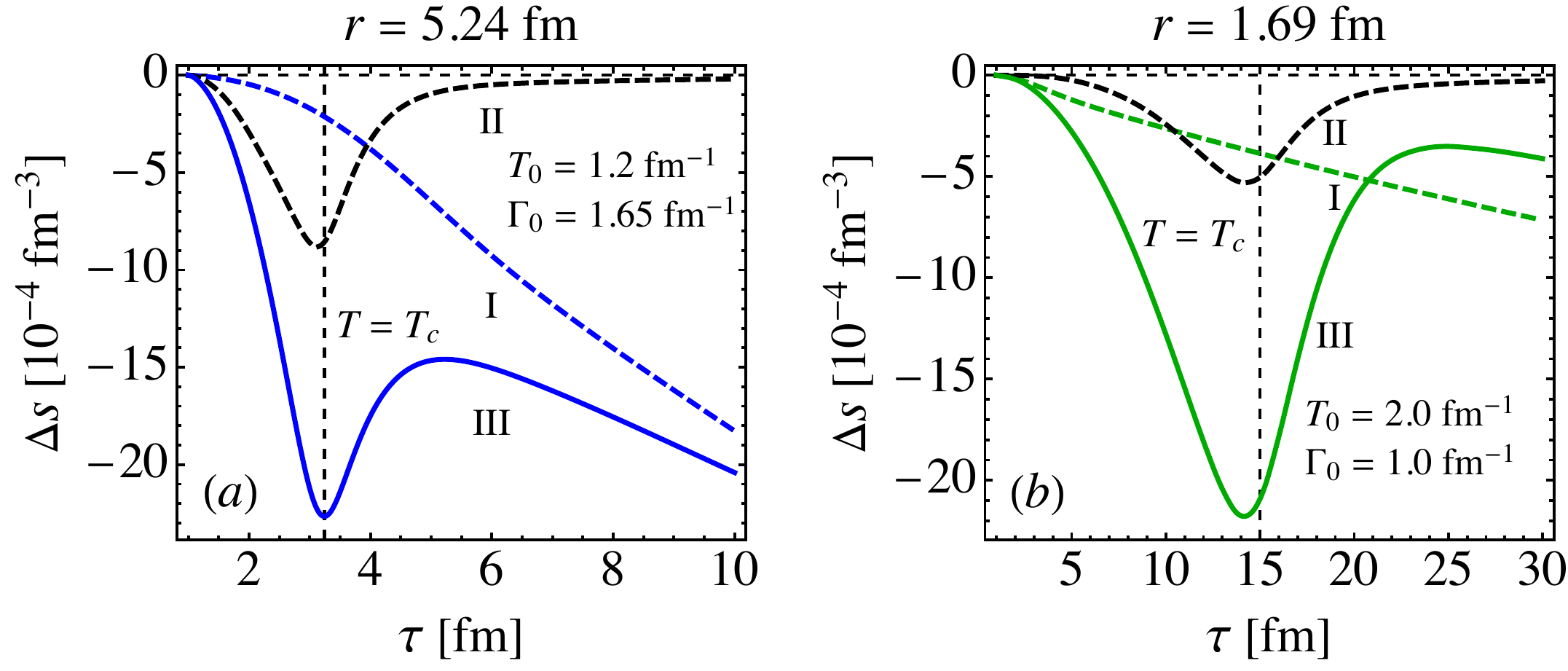}
    \caption{Evolution of the $Q$-integrated non-equilibrium entropy density correction $\Delta s$ at two radial distances, $r = 5.64$ fm (left) and $r = 1.69$ fm (right), for the scenarios I, II, III shown in Fig.~\ref{fig:gubser_phievo_r}a,b,c, respectively. The dashed vertical lines indicate the times when the fluid passes through $T = T_c$. Lines labeled by II correspond to the same evolution as those with the same label shown in Fig.~\ref{fig:bjorken_dynamics_comparison2}b,c.
    \label{fig:gubser_sevo_twors}}
\end{figure*}
%

Scenario III, shown in Fig.~\ref{fig:gubser_phievo_r_0}c, combines the dynamical effects included in scenarios I and II. Transverse flow effects can be uniquely identified by comparing panels b and c: the radial profiles showing snapshots of the non-equilibrium evolution of $\phi_Q$ (solid lines) are pushed outward by advection and thereby away from the corresponding equilibrium profiles $\bar\phi_Q$ (dashed lines) --- dis-equilibration caused by radial flow gradients wins over equilibration by relaxation. 
As we noted above, the growth of $\phi_Q$ towards the dilute periphery of the fireball can be intuitively understood by remembering the increasing relative importance of density fluctuations when the average density gets small. A unitless fluctuation measure that correctly absorbs this trivial Poisson-statistical effect would be the product $n\,\phi_Q$. With our parametrization of $\cpz=s^2/(\alpha n)$, removing the profile function ${\cal F}(r)$ by dividing $\phi_Q$ by the factor $\bar\phi_{Q, \xi_0}$ denoting the initial value at $\tau_0$ at each position $r$ of $\bar\phi_Q$ for fixed $\xi=\xi_0$ (shown as the dashed line in Fig.~\ref{fig:gubser_phievo_r_0}a) achieves the same end. This unitless ratio is shown, for both equilibrium (dashed) and non-equilibrium (solid) slow modes, in Fig.~\ref{fig:gubser_phievo_r}. This way of plotting $\phi_Q$ de-emphasizes the peripheral, low-density regions and brings out more clearly those features that will be phenomenologically relevant in future computations of experimental fluctuation signals.

With this in mind we now discuss Fig.~\ref{fig:gubser_phievo_r}. In scenario I (Fig.~\ref{fig:gubser_phievo_r}a) the normalized equilibrium value $\bar\phi_Q/\bar\phi_{Q,\xi_0}$ (dashed horizontal line) remains frozen at 1. Furthermore, as expected from Eq.~(\ref{eq:gubser_smallt_eom}), the minimum value of the scaled $\phi_Q$ matches the maximum of the transverse flow $u^r$, resulting from a maximum (negative) contribution from the advection term $-u^r\partial_r\phi_Q$ in (\ref{eq:gubser_smallt_eom}).

In scenarios II and III, shown in Figs.~\ref{fig:gubser_phievo_r}b,c, the $r$-dependence from ${\cal F}(r)$ drops out from the normalized equilibrium ratio $\bar\phi_Q/\bar\phi_{Q, \xi_0}$. In consequence, the peaks of the dashed lines in Figs.~\ref{fig:gubser_phievo_r}b,c unambiguously reflect the peak at $T_c$ of the correlation length $\xi(T)$ which enters in the numerator of that ratio. Following the location of the critical point at $T_c$, these peaks in Fig.~\ref{fig:gubser_phievo_r}b move inward as time proceeds and the system cools by expansion.

The difference between panels b and c of Fig.~\ref{fig:gubser_phievo_r} is in the dynamical evolution of the (normalized) non-equilibrium slow mode $\phi_Q/\bar\phi_{Q,\xi_0}$ (solid lines), {\em with} (panel c) and {\em without} (panel b) advection by transverse radial flow. Panel b isolates and nicely illustrates the effects of critical slowing-down: for example, at $r{\,=\,}1.69$\,fm, where $\xi(T)$ increases monotonically from the red (bottom left) to the blue (top left) lines, we see that $\phi_Q$ trails behind the evolution of $\bar\phi_Q$ and always below the equilibrium value; at $r{\,=\,}5.64$\,fm, on the other hand, which for $\tau>3$\,fm sits on the falling side of the $\xi(T)$ curve, we see that $\phi_Q$ first trails below $\bar\phi_Q$ when $\xi$ is still growing but moves above $\bar\phi_Q$ (again trailing behind the equilibrium value) once $T$ drops below $T_c$ and $\xi(T)$ begins to decrease again at this radial position. Generically, the transverse expansion rate is smaller and the slow-mode relaxation rate is larger at large $r$, so that at large $r$ the slow modes relax faster to their equilibrium value than near the center; this, too, is clearly visible in this panel.

In scenario III (Fig.~\ref{fig:gubser_phievo_r}c), we see that when transverse flow is included, the non-equilibrium value $\phi_Q$ falls behind $\bar\phi_Q$ further and further as time proceeds, except at $r=0$ where $u^r=0$ and thus the evolution is the same as in Fig.~\ref{fig:gubser_phievo_r}b. At larger $r\gtrsim 8$ fm, on the other hand, the advection effects seen and discussed in Fig.~\ref{fig:gubser_phievo_r_0} cause the $r$-dependence of the normalized ratio $\phi_Q/\bar\phi_{Q,\xi_0}$ in panel c to develop similarities with what is seen in panel a, especially at late times when the system has passed the critical region and $\xi$ goes back to $\xi_0$. In the intermediate region, $r\approx4$\,fm where the flow is the largest, the $r$-dependence of the solid lines is complicated by the fact relaxation effects are gradually being overshadowed by the transverse flow and advection effects which grow with time.

Note that, unlike Ref.~\cite{Rajagopal:2019xwg}, we do not see clear signs of a second peak in Fig.~\ref{fig:gubser_phievo_r}c, i.e. the critical fluctuation peak is not transported outward to larger $r$ by the transverse flow. The main model feature responsible for this difference is the profile function $\mathcal{F}(r)$ in our expression (\ref{eq:adv_initial2}) for the relaxation rate $\Gamma_Q$ of the slow mode which accelerates relaxation of $\phi_Q$ at large $r$. We confirmed that when the $r$ dependence is removed in Eqs.~(\ref{eq:adv_initial1},\ref{eq:adv_initial2}) we observe a second outgoing peak as shown in Fig.~7 of Ref.~\cite{Rajagopal:2019xwg}. We note that, even if such an outward-moving (advected) second peak in $\phi_Q$ were to show up in our model, it would be dwarfed at large $r$ by the upward-sweeping noncritical fluctuations and would become essentially invisible after normalization with $\bar\phi_{Q,\xi_0}$ as done in Fig.~\ref{fig:gubser_phievo_r_0}. Furthermore, in our model the profile function ${\cal F}(r)$ increases the relaxation rate in the dilute periphery.

%
\begin{figure*}[!tp]
    \centering
    \includegraphics[width= \textwidth]{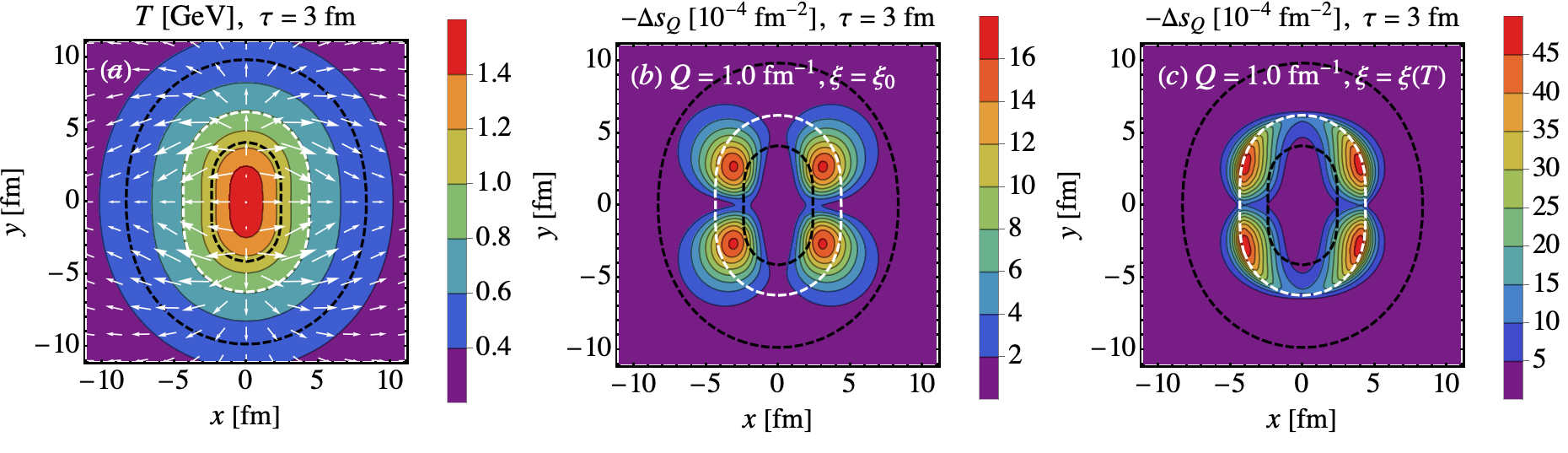}
    \caption{Transverse flow effects in an anisotropic profile at $\tau=3$\,fm. (a) Temperature contours in the transverse plane, elongated along the $y$ direction. Arrows indicate the anisotropic transverse flow. (b) and (c) show the correction $\Delta s_Q$ to the entropy density arising from the mode with wave number $Q=1.0$\,\fm; (b) assumes constant $\xi=\xi_0$ (scenario I) while (c) assumes temperature dependent $\xi(T)$ (scenario III). In all three panels the white dashed contour indicates $T{\,=\,}T_c$ while the two black dashed contours show $T_H=T_c+\Delta T=224$\,MeV and $T_L=T_c-\Delta T=96$\,MeV (where $\Delta T{\,=\,}64$\,MeV), respectively.
    \label{fig:gubser_advection_aniso}}
\end{figure*}
%
 
The back-reaction of this non-equilibrium slow-mode dynamics on the entropy density of the medium (i.e. the non-equilibrium entropy correction $\Delta s$) is shown as a function of time in Fig.~\ref{fig:gubser_sevo_twors}. The two panels show how this plays out at two different transverse distances from the fireball center, a larger one at $r{\,=\,}5.24$\,fm (Fig.~\ref{fig:gubser_sevo_twors}a) which passes through $T_c$ first at $\tau\approx 3.24$\,fm, and a smaller one at $r{\,=\,}1.69$\,fm (Fig.~\ref{fig:gubser_sevo_twors}b) which passes through $T_c$ later at $\tau\approx 15$\,fm (vertical dashed lines). The black dashed lines describe scenario II without transverse flow and reproduce the identically labeled lines from Figs.~\ref{fig:bjorken_dynamics_comparison2}b,c. The colored dashed lines for scenario I show that transverse flow can induce large $|\Delta s|$ even for a constant correlation length $\xi = \xi_0$, i.e., without critical slowing-down. When critical behavior of $\xi(T)$ is added in scenario III (solid colored lines), the magnitude of $\Delta s$ increases strongly in the critical region around $T_c$. The crossing of the dashed and solid lines at $\tau \approx 20$\,fm in Fig.~\ref{fig:gubser_sevo_twors}b must be attributed to critical slowing-down in scenario III which keeps $\phi_Q$ from reacting to the time-increasing transverse flow effects as quickly as it can when $\xi{\,=\,}\xi_0$ is a (small) constant. 

A qualitative feature of the entropy density evolution shown in Fig.~\ref{fig:gubser_sevo_twors} is that (for the background flow pattern assumed in this subsection) transverse flow appears to cause a non-equilibrium entropy correction from slow-mode dynamics that increases approximately linearly with time at late times.\footnote{%
    Physically this is, of course, an artefact because the flow pattern studied in this subsection results from an approximation that should not be used at large times
    $\tau > 1/q$.}
This can be understood from Fig.~\ref{fig:gubser_phievo_r}, panels a and c, by following the ratio $\phi_Q/\bar\phi_Q$ (i.e. the ratio between the solid and dashed colored lines) in time along the two vertical dashed lines indicating the $r$ positions studied in Fig.~\ref{fig:gubser_sevo_twors}: One sees that, for both constant (a) and temperature dependent (c) correlation length $\xi$, $\phi_Q/\bar\phi_Q$ decreases monotonically with time, explaining the growing magnitude of the entropy correction $\Delta s$ at late times seen in Fig.~\ref{fig:gubser_sevo_twors}.

We close this subsection with a discussion of transverse flow effects on anisotropic perturbations of the transverse profile. In the early time limit, $\tau\ll 1/q$, the perturbed solution is available analytically \cite{Gubser:2010ui,Hatta:2014jva, Hatta:2015era}:\footnote{%
    Here and below the undeformed (``isotropic") profiles (e.g. those in Eqs.~(\ref{eq:flow_approx},\ref{eq:temp_smallt}))
    are labeled by a subscript ``iso".}
\begin{eqnarray}
\label{eq:gubser_aniso}
    &&T=T_\mathrm{iso} (1 - \epsilon_n{\cal A}_n\delta)\,,
\nonumber\\
    &&u^r = u^r_\mathrm{iso} - \epsilon_n\nu_s (\delta u^r)\,,
\\\nonumber
    &&u^\phi = -\epsilon_n\nu_s (\delta u^\phi)\,.
\end{eqnarray}
Here $\epsilon_n$ is related to the eccentricity, $\delta$ and $\nu_s$ are parameters controlling the fluctuations of temperature and flow velocity, and the deformation profile is
\begin{equation}
\label{eq:A}
    {\cal A}_n(r,\phi) \equiv \left(\frac{2qr}{1+(qr)^2}\right)^{\!\!n}\cos(n\phi)\,,
\end{equation}
which is $\propto Y_{n,n}(\vartheta,\phi)+Y_{n,-n}(\vartheta,\phi)$ (here $(\vartheta,\phi)$ are polar coordinates in de Sitter space \cite{Gubser:2010ze}). The flow profile deformations are related to ${\cal A}_n$ by \cite{Hatta:2014jva, Hatta:2015era}
\begin{eqnarray}
    \delta u^r &=& \frac{2q\tau}{1+(qr)^2}\partial_\vartheta{\cal A}_n 
\\
    &=& n \,\frac{(2qr)^{n-1}(2q\tau)}{\bigl(1{+}(qr)^2\bigr)^n}
        \left(\frac{1{-}(qr)^2}{1{+}(qr)^2}\right)\cos(n\phi)\,,
\nonumber\\
    \delta u^\phi &=& \tau\partial_\phi{\cal A}_n = -n\tau\left(\frac{2qr}{1{+}(qr)^2}\right)^{\!\!n}\sin(n\phi)\,.
\end{eqnarray}

We assume $\epsilon_n\ll1$ so that we can linearize in $\epsilon_n$, e.g. $e=e_\mathrm{iso}(1 - \epsilon_n{\cal A}_n\delta)^4\approx e_\mathrm{iso}(1-4\epsilon_n{\cal A}_n\delta)$, and similarly for the slow modes (remembering $\Phi_0 \propto T^{-3}$ and $\Gamma_0 \propto T^{-1}$):
\begin{equation}
    \Phi_0 \approx \Phi_{0,\mathrm{iso}}(1+3\epsilon_n{\cal A}_n\delta)\,,\quad \Gamma_0 \approx \Gamma_{0,\mathrm{iso}}(1+\epsilon_n{\cal A}_n\delta)\,.
\label{eq:gubser_aniso_phi}
\end{equation}
Using these linearized expressions in Eqs.~(\ref{eq:adv_initial1},\ref{eq:adv_initial2}) we obtain  deformed profiles for $\bar\phi_Q$ and $\Gamma_Q$. Here we only consider elliptic deformations ($n=2$) and follow \cite{Hatta:2014jva, Hatta:2015era} by setting $\delta{\,=\,}1$, $\nu_s{\,=\,}-3/2$ and $\epsilon_2{\,=\,}0.15$. Using Eq.~(\ref{eq:gubser_aniso_phi}) together with the temperature and flow profiles in Eqs.~(\ref{eq:gubser_aniso}), solving the equations of motion for the slow modes as before, one can explore the anisotropic evolution.

The results are shown in Fig.~\ref{fig:gubser_advection_aniso}. Panel (a) shows the temperature distribution in the transverse plane (which is clearly elongated in $y$ direction) and the anisotropic transverse flow. In panels (b,c) we compare, for the same dynamical scenarios I and III studied above, the entropy density modification $\Delta s_Q$ arising from the non-equilibrium evolution of the slow mode with wave number $Q{\,=\,}1$\,\fm. In scenario I with constant correlation length $\xi{\,=\,}\xi_0$, shown in panel (b), $\Delta s_Q$ does not feel the critical temperature and is sensitive only to effects arising from the anisotropic transverse flow; the location of the maximum of $|\Delta s_Q|$ basically coincides with that of the maximum flow velocity which can be identified in panel (a) by scanning for the longest flow arrows. In scenario III with a temperature dependent correlation length $\xi(T)$, shown in panel (c), the $|\Delta s_Q|$ maxima are clearly shifted closer to the $T=T_c$ critical contour. As time procedes, this contour moves inward due to cooling by longitudinal expansion. This means that the peak of $\bar\phi_Q$ associated with the critical peak of $\xi(T)$ moves inward, too, and that $\phi_Q$ tries to catch up with it. At $\tau = 3$\,fm, the flow is not yet very strong, so $\phi_Q$ does not lag too far behind its equilibrium value, and the  non-equilibrium effects causing $|\Delta s_Q|$ are tightly constrained to the critical contour. However, since the relaxation rate $\Gamma_Q$ increases with $r$ through ${\cal F}(r)$ in Eq.~(\ref{eq:adv_initial2}), the deviations from equilibrium tend to be smaller outside than inside the $T=T_c$ contour, explaining the slight inward shift of the maxima of $|\Delta s_Q|$ from the critical contour.\footnote{%
    This reasoning is supported by studying the $\tau{\,=\,}3$\,fm radial profiles of $\phi_Q$ and $\bar\phi_Q$ in Fig.~\ref{fig:gubser_phievo_r}c.} 
We see that, similar to panel (b), the azimuthal variation of the flow velocity causes the appearance of four ``hot spots''\footnote{%
    More accurately these should be called ``cold spots'' because $\Delta s_Q$ is negative and thus reduces the effective temperature.
    \label{fn20}}
of $|\Delta s_Q|$ at angles corresponding to flow maxima, but that the radial position of these maxima is strongly biased towards $T_c$ by the critical peaking of the correlation length.

The anisotropic entropy density correction $\Delta s$ from the slow modes ``reacts back'' on the expanding medium and affects its geometric eccentricity. We can define a slow-mode induced change of ellipticity, $\Delta \epsilon_2$, by using the definition of $\epsilon_2$ in terms of the expectation value of $\cos(2\phi)$, with a conformally weighted entropy density \cite{Hatta:2014jva,Hatta:2015era} as weight function:
\begin{equation}
    \epsilon_2 +\Delta\epsilon_2 = 
    -\frac{\int rdr d\phi\, (s{+}\Delta s)\,         
           \left[(qr)^2/\bigl(1{+}(qr)^2\bigr)\right]
           \cos(2\phi)}
          {\int rdr d\phi\, (s{+}\Delta s)\, \left[(qr)^2/\bigl(1{+}(qr)^2\bigr)\right]}\,.
\end{equation}
The uncorrected background gives $\epsilon_2{\,=\,}0.182284$. Including only the contribution to the correction $\Delta s$ from the slow mode with $Q=1.0$\,\fm\ and weighting it by a $Q$-bin width $\Delta Q{\,=\,}1$\,\fm, we find that at $\tau{\,=\,}3$\,fm the entropy correction changes the ellipticity to $\epsilon_2 + \Delta\epsilon_2 = 0.182293$ for scenario I in panel (b) and to $\epsilon_2 + \Delta\epsilon_2 = 0.182302$ for scenario III in panel (c). Transverse flow thus leads to a slight increase of the ellipticity, resulting from the slightly reduced particle emission (lowered entropy density) from the ``hot spots''$^{\ref{fn20}}$ indicated Fig.~\ref{fig:gubser_advection_aniso}b,c; the ellipticity correction is larger for scenario III which includes critical behavior of the correlation length $\xi$. [Note that altogether the ellipticity correction is tiny, of relative order $\lesssim 10^{-4}$ (consistent with the estimate (\ref{eq:estdeltas})), reflecting the smallness of the non-equilibrium entropy correction on the scale of the overall entropy density of the background fluid.]

%
\subsection{Combined non-equilibrium dynamical effects in full Gubser flow}
\label{sec:corrlen}
%

%
\begin{figure*}[!tp]
    \centering
    \includegraphics[width= \textwidth]{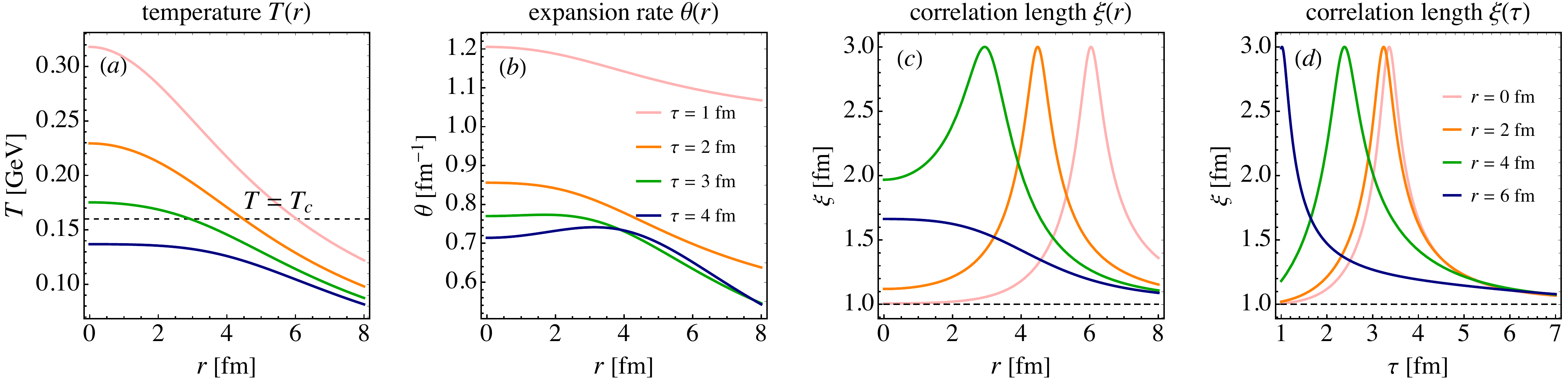}
    \caption{Radial profiles of the temperature $T(r)$ (a), scalar expansion rate $\theta(r)$ (b), and correlation $\xi(r)$ (c) at four different times $\tau{\,=\,}1,\,2,\,3,\,4$\,fm, as well as the time evolution $\xi(\tau)$ of the correlation length (d) at four different transverse distances $r{\,=\,}0,\,2,\,4,\,6$\,fm, for ideal Gubser flow.
    The black dashed lines in (c,d) indicate the constant value $\xi_0{\,=\,}1$\,fm.
    \label{fig:gubser-setup}}
\end{figure*}
%

\begin{figure*}[!tp]
    \centering
    \includegraphics[width= \textwidth]{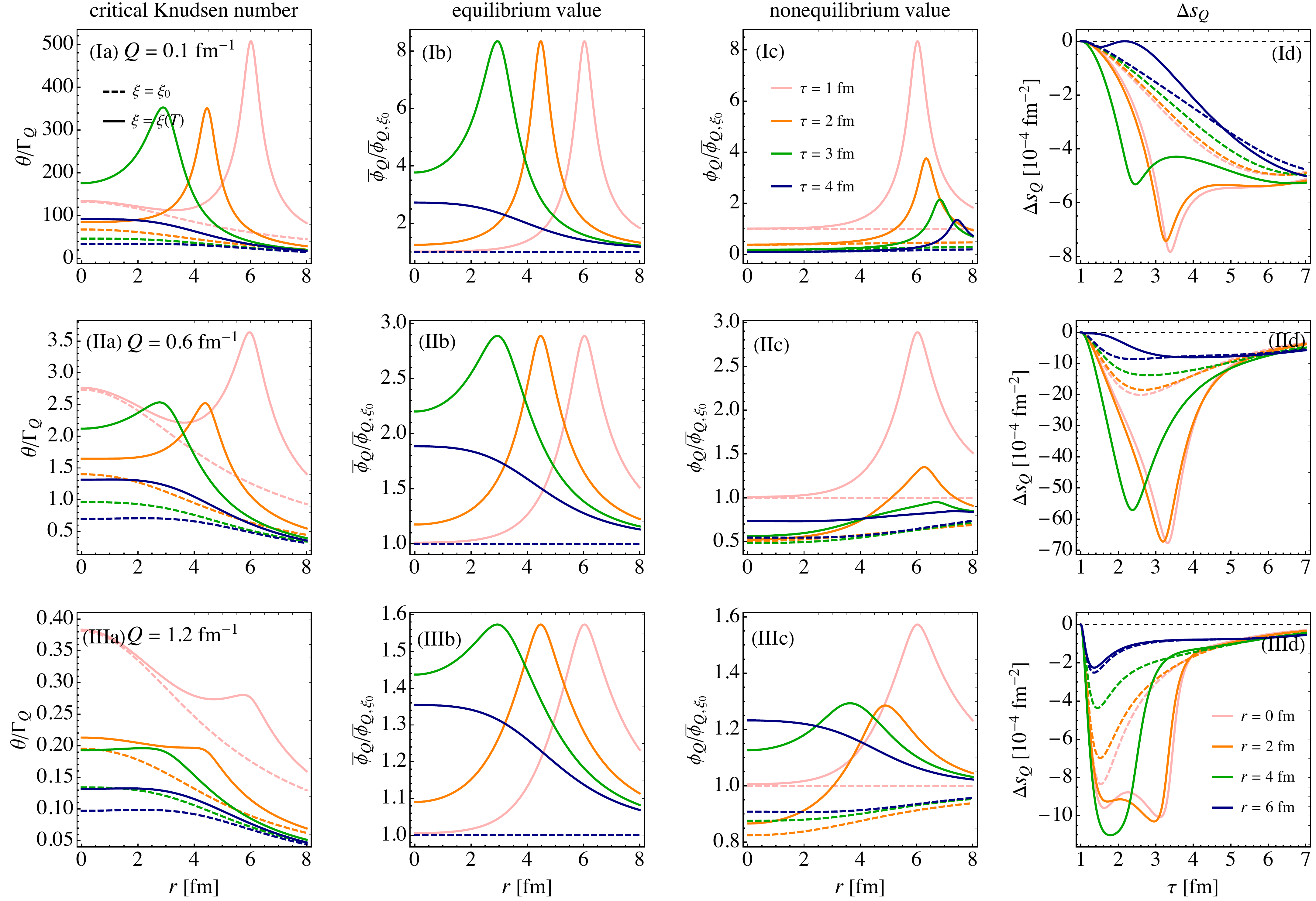}
    \caption{Comparison for the dynamics of three slow modes with widely different wave numbers: $Q = 0.1$\,\fm\ (I, top row), $Q = 0.6$\,\fm\ (II, middle row) and $Q = 1.2$\,\fm\ (III, bottom row). Dashed colored lines indicate evolution with constant correlation length $\xi = \xi_0$, solid colored lines use a temperature-dependent correlation length $\xi(T)$ that peaks at $T_c$. Similar to Fig.~\ref{fig:gubser-setup}, the three left columns (a-c) show radial profiles at different times, for the critical Knudsen number (a) and the (normalized) slow-mode equilibrium (b) and actual non-equilibrium values (c). The right column (panels (d)) shows the time evolution at four different transverse distances of the contribution $\Delta s_Q$ to the non-equilibrium entropy density correction arising from these three slow modes. Characteristic differences for small and large wave numbers are seen when comparing the top and bottom rows while the middle row shows a mixture of these characteristics at an intermediate wave number (see text for details). 
    \label{fig:gubser-dynamics}}
\end{figure*}

In the last two subsections we focused on the effects of background medium expansion and advection on slow-mode dynamics, and these studies were facilitated by taking certain limits of the background flow ((0+1)-dimensional Bjorken flow in Sec.~\ref{sec:expansion}, early-time limit for ideal Gubser flow in Sec.~\ref{sec:advection}). While some discussion of the specific effects caused by critical growths of the correlation length $\xi$ near the critical point was already included in these subsections, we will now extend the discussion of correlation length effects to the full (unapproximated) ideal Gubser flow.  

The exact temperature profile for ideal Gubser flow was given in Eq.~(\ref{eq-gubser_temp}). For the discussion in this subsection we take $C=2.78$ from Sec. \ref{sec:gubser} for the normalization. For the equilibrium values and the damping rates of the slow modes we use
\begin{eqnarray}
    \bar\phi_Q &=& \frac{\Phi_0}{\mathcal{F}^3(\tau,r)} \left(\frac{\xi}{\xi_0}\right)^{\!\!2} f_2(Q\xi)\,,\\
    \Gamma_Q &=& \frac{\Gamma_0}{\mathcal{F}(\tau,r)} \left(\frac{\xi_0}{\xi}\right)^{\!\!4} f_\Gamma(Q\xi)\,,
\label{eq:general_initial}
\end{eqnarray}
where $\mathcal{F}(\tau,r){\,\equiv\,}T(\tau,r)/T(\tau_0,0)$, $\Phi_0{\,=\,}1.0$\,fm$^{-3}$, and $\Gamma_0{\,=\,}0.9$\,\fm.

Figure~\ref{fig:gubser-setup} shows snapshots of the key temperature (a), expansion rate (b) and correlation length (c) profiles for different times, as well as the time evolution of correlation length in panel (d) for different transverse positions. Panel (a) shows that fireball remains hottest at $r{\,=\,}0$ until all of it has cooled below $T_c$, and that (due to cooling by a combination of longitudinal and transverse expansion) the critical surface $T(\tau,r){\,=\,}T_c$ moves inward with time. Panel (b) shows that, generically, the expansion rate {\it decreases} with time, driven by the slowing rate of longitudinal expansion, $\theta_\parallel\sim1/\tau$; this facilitates equilibration of the slow modes at later times. However, the bump at $r\approx3.5$\,fm of the expansion rate at $\tau{\,=\,}4$\,fm in panel (b) also demonstrates the increasing contribution $\theta_\perp$ from transverse flow as time increases, causing the expansion rate to {\it increase} with time for $\tau>3$\,fm in the periphery $r>4$\,fm. For our initial conditions transverse expansion does not, however, dominate the expansion rate until the entire fireball has cooled below $T_c$. Panel (c) illustrates that the peak of the correlation length moves inward together with $T_c$ as time proceeds, and panel (d) shows that fluid cells pass through the critical point earlier at large $r$ than at smaller $r$, with cells at $r>6$\,fm starting out and remaining subcritical.

In Fig.~\ref{fig:gubser-dynamics} we study the dynamical evolution of the actual and equilibrium values (columns (b,c)) of slow modes with  three different wave numbers, $Q{\,=\,}0.1\,\mathrm{fm}^{-1}<\xi_\mathrm{max}^{-1}$ (top row, I), $\xi_\mathrm{max}^{-1}<Q{\,=\,}0.6\,\mathrm{fm}^{-1}<\xi_0^{-1}$ (middle row, II) and $Q{\,=\,}1.2\,\mathrm{fm}^{-1}\gtrsim\xi_0^{-1}$ (bottom row, III), as well as of two related $Q$-dependent quantities, the critical Knudsen number $\theta/\Gamma_Q$ (column (a)) reflecting the competition between disequilibrating collective expansion and equilibrating mode relaxation, and the non-equilibrium slow-mode correction $\Delta s_Q$ to the entropy density in column (d). Throughout, dashed lines reflect non-critical dynamics with constant correlation length $\xi{\,=\,}\xi_0$ while solid lines show the results for critical dynamics with a correlation length $\xi(T)$ that peaks at $T_c$. 

For noncritical dynamics ($\xi{\,=\,}\xi_0$, dashed lines) the critical Knudsen number $\theta/\Gamma_Q$ in column (a) is seen to decrease with time throughout the fireball, basically following the scalar expansion rate $\theta$ shown in Fig.~\ref{fig:gubser-setup}b. Due to the factor $Q^2$ in $\Gamma_Q$, the critical Knudsen numbers are roughly 100 times larger in the top row than in the bottom row, severely hindering relaxation of the slow mode towards equilibrium. In comparison with the dashed lines, the solid lines show the additional effect of critical slowing down caused by the critical enhancement of the correlation length $\xi(T)$ near $T_c$. This effect can be up to a factor ${\sim\,}10$ (mostly due to the factor $\xi_0^2/\xi^2$ in $\Gamma_\xi$) in the top panel (small $Q$) but is seen to be significantly smaller for more typical $Q$ values like the one shown in the bottom panel (due to the compensating factor $1{+}Q^2\xi^2$ in $f_\Gamma$).  

\begin{figure*}[!tp]
    \centering
    \hspace*{-8mm}
    \includegraphics[width= 1.03\textwidth]{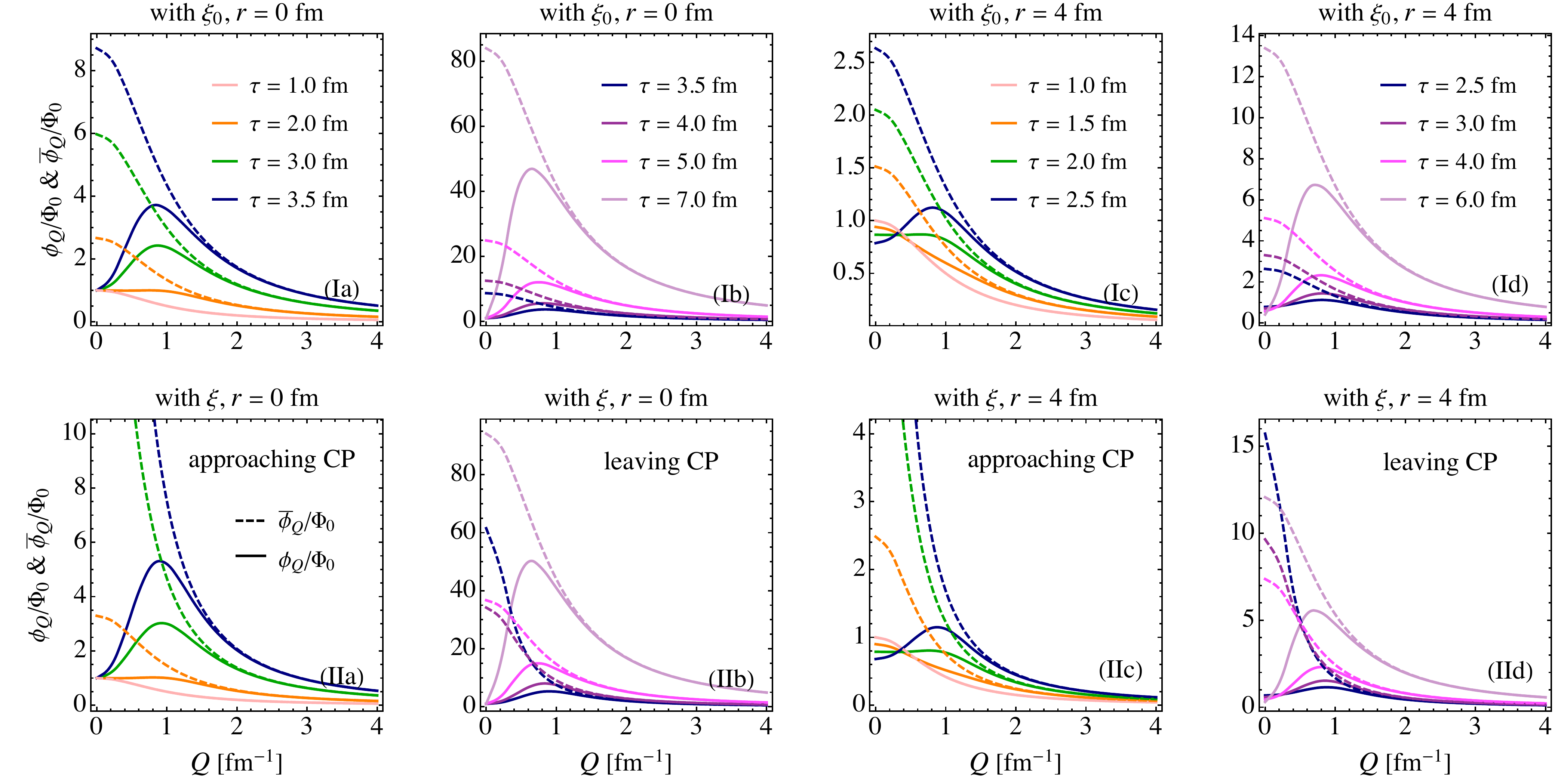}
    \caption{$Q$-dependence of the dynamics of the slow modes with $\xi = \xi_0$ (top row, I) and $\xi = \xi(T)$ (bottom row, II), at $r{\,=\,}0$\,fm (left two columns (a,b)) and $r{\,=\,}4$\,fm (right two columns (c,d)). At $r=0$\,fm (4 fm) $\xi$ approaches its maximum at $\tau{\,\approx\,}3.5$\,fm (2.5\,fm) (see Fig.~\ref{fig:gubser-setup}d). The dynamics at $r{\,=\,}0$\,fm (4\,fm) before $\tau{\,=\,}3.5$\,fm (2.5\,fm) (``approaching CP'') is shown in panels (a) (panels (c)); for later times (``leaving CP'') it is shown in panels (b) and (d), respectively. [Note that the non-critical dynamics shown in the top row does not feel the critical point CP.] In all eight panels dashed (solid) lines show the equilibrium (nonequilibrium) values of $\phi_Q$, both scaled by $\Phi_0(r)$. See text for discussion.
    \label{fig:gubser-nearcp}
    \vspace*{-3mm}}
\end{figure*}

Due to the small $Q$ value in the top row, the critical Knudsen numbers shown in panel (Ia) of Fig.~\ref{fig:gubser-dynamics} are huge (ranging from the tens to the hundreds). This implies that relaxation towards equilibrium plays practically no role in the dynamical evolution of this particular slow mode, and that the evolution patterns seen in panels (Ib) and (Ic) are entirely due to advection. For non-critical dynamics (dashed lines) the normalized equilibrium value shown in panel (Ib) remains frozen at 1 by definition while the normalized non-equilibrium value in panel (Ic) decreases with time, similar to Fig.~\ref{fig:bjorken_q}b at small $Q$ and Fig.~\ref{fig:gubser_phievo_r}a. For critical dynamics (solid lines) comparison of panels (Ib) and (Ic) shows how advection pushes the critical peak of $\phi_Q$ outward while the peak of its equilibrium value would follow the critical temperature inward as time proceeds. For the modes with the larger wave numbers shown in the two lower rows, the equilibrium value of the slow mode in panels (IIb) and (IIIb)  follows qualitatively the same pattern as for the small-$Q$ mode shown in (Ib), although the critical effects are significantly reduced at the higher $Q$ values. Due to the much smaller critical Knudsen numbers for the mode with $Q{\,=\,}1.2$/fm in the bottom row, panels (IIIb) and (IIIc) show quite different evolution patterns, reflecting the competition between relaxation (thermalization) and advection: the critical peak of $\phi_Q$ is now no longer pushed outward by advection, but moves inward via relaxation towards equilibrium. For the mode with an intermediate wave number $Q{\,=\,}0.6\,$\fm (row II), on the other hand, the dynamics shows a mixture of the characteristics seen in rows I and III --- one can still recognize in panel IIc at $r\gtrsim6$\,fm the initial peak being advected outward while being damped by relaxation (this is the dominant feature in panel Ic) while at the same time the relaxation dynamics that dominates in panel IIIc causes the solid lines in panel IIc at intermediate $r$ values to rise above the dashed lines as time proceeds.

Taken together, the three rows of Fig.~\ref{fig:gubser-dynamics} illustrate that there are two effects at play: (i) the {\it initial} peak in the fluctuations is carried outwards by advection while it is at the same time damped by relaxation; and (ii) as the location where critical fluctuations would occur in equilibrium (indicated by the peaks in the curves shown in column b) moves inward toward smaller $r$ values, the actual out-of-equilibrium fluctuations at these smaller values of $r$ increase, with the solid curves in column c relaxing upward toward the same-colored curves in column b, but more slowly due to critical slowing down. In principle both effects are present in all three rows, but the first effect is invisible in row III because at larger values of $Q$ the initial peak dissipates more rapidly and in addition it is rapidly dwarfed by the increase in the noncritical fluctuations $\phi_Q \propto 1/n$ at larger $r$, and the second effect is invisible in row I because relaxation is very slow at such a small value of Q.

The right column (d) of Fig.~\ref{fig:gubser-dynamics} (which should be compared with Figs.~\ref{fig:bjorken_dynamics_comparison2} and \ref{fig:gubser_sevo_twors}) shows the time evolution of the non-equilibrium entropy density correction $\Delta s_Q$ arising from the three modes studied in the three rows, at four different radial positions. Again, dashed (solid) lines reflect noncritical (critical) dynamics. The differences in (Id) are qualitatively similar to those observed between scenarios I and III in Fig.~\ref{fig:gubser_sevo_twors}, while the differences between dashed and solid lines in panel (IIId) are more similar to those observed between scenarios I and III in Fig.~\ref{fig:bjorken_dynamics_comparison2}. This is because low-$Q$ modes are more affected by advection (which was included in Fig.~\ref{fig:gubser_sevo_twors}) than high-$Q$ modes which can successfully fight advection effects (which Fig.~\ref{fig:bjorken_dynamics_comparison2} did not include). Depending on when the system enters the critical regime, the evolution of $|\Delta s_Q(\tau)|$ can feature two peaks, one due to expansion before reaching $T_c$ and another arising from critical slowing-down when entering the critical region. While both peaks are seen in panel (IIId) at $r{\,=\,}0$ and 2\,fm ({\it cf.} Fig.~\ref{fig:bjorken_dynamics_comparison2}c), only a single peak is observed for $r{\,\geq\,}4$\,fm ({\it cf.} Fig.~\ref{fig:bjorken_dynamics_comparison2}a,b). Consistent with the dynamics plotted in panel IIc we see in panel IId that the time evolution of $|\Delta s_Q|$ for the intermediate-$Q$ mode interpolates smoothly between panels Id and IIId. A notable feature, however, is the much larger magnitude of $|\Delta s_Q|$ in (IId) compared to both (Id) and (IIId): It is explained by referring to Fig.~\ref{fig:bjorken_stau} where we noted that, while off-equilibrium dynamical effects are stronger at small $Q$, their contribution to $|\Delta s(\tau)|$ peaks at an intermediate wave number $Q_\mathrm{max}{\,\sim\,}\mathcal{O}(Q_\textrm{neq})$, due to phase-space suppression by the factor $(Q/2\pi)^2$ at small $Q$.

We close this subsection by showing in Fig.~\ref{fig:gubser-nearcp} for two radial positions ($r{\,=\,}0$ (a,b) and 4\,fm (c,d)) seven different time snapshots (as detailed in the legend) of the entire $Q$-spectrum of the slow modes, for both noncritical (top row) and critical dynamics (bottom row). Solid lines show the dynamically evolving slow modes spectra, dashed lines their corresponding equilibrium spectra.\footnote{%
    Note that at each $r$ we normalize $\bar\phi_Q$ and $\phi_Q$ by $\Phi_0(r)$, i.e. by the initial equilibrium value for $\xi=\xi_0$ at the same position.} 
For clarity, the dynamics is shown separately for the system approaching $T_c$ (columns a,c) and receding from $T_c$ (columns b,d); for $r=0$ (columns a,b) $T_c$ is reached at $\tau\approx3.5$\,fm, at $r{\,=\,}4$\,fm (columns c,d) this happens somewhat earlier at $\tau\approx 2.5$\,fm. In the bottom row, the equilibrium expectation for the magnitude of the slow modes (dashed curves) rises with time while approaching the critical point, and then begins to drop as the critical point is passed.  Note, however, that at very late times it rises again since, as the density keeps decreasing, the equilibrium $\bar\phi_Q$ grows like $1/n$, as we have discussed in Section~\ref{sec:advection}. In all plots, high-$Q$ modes are seen to closely follow their equilibrium values, hardly affected by advection. At $r{\,=\,}0$ (columns (a),(b)) low-$Q$ modes are basically frozen at their initial values while at $r{\,=\,}4$\,fm (columns (c), (d)) they are visibly affected by advection: even though $\Gamma_Q{\,=\,}0$ at $Q{\,=\,}0$, $\phi_0$ is seen to decrease with time instead of being frozen because advection moves the smaller value of $\phi_0$ at smaller $r$ to $r{\,=\,}4$\,fm.

The short summary of this subsection is that low-$Q$ slow modes are most strongly affected by the phenomenon of critical slowing down near a critical point. 

\begin{figure*}[!tp]
    \hspace*{-3mm}
    \includegraphics[width= 1.02\textwidth]{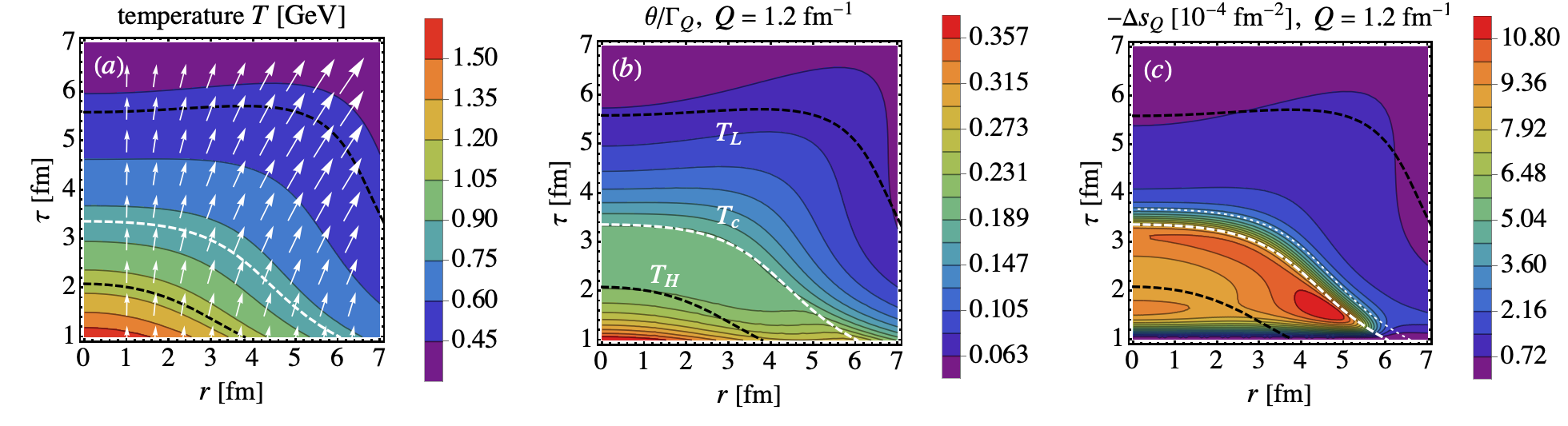}
    \caption{Full ideal Gubser space-time evolution of (a) temperature $T$ (contour lines) and flow velocity $(u^\tau, u^r)$ (white arrows), (b) critical Knudsen number $\theta/\Gamma_Q$ for $Q{\,=\,}1.2$\,\fm\ (contour lines), and (c) entropy modification $-\Delta s_Q$ contributed by this mode (contour lines), for a temperature dependent correlation length $\xi(T)$. The white dashed line in these panels denotes the $T(\tau,r){\,=\,}T_c{\,=\,}160$\,MeV isotherm; the black dashed lines show isotherms for  $T(\tau,r){\,=\,}T_H{\,=\,}224$\,MeV and $T(\tau,r){\,=\,}T_L{\,=\,}96$\,MeV which enclose the critical region. The dotted white line in panel (c) shows the ``freeze-out contour'' with temperature $T(\tau,r){\,=\,}T_f{\,=\,}148$\,MeV.
    \label{fig:gubser-pheno1}}
\end{figure*}

%
\subsection{Space-time evolution of non-equilibrium slow mode effects and modified particle emission}
\label{sec:phenomenology}
%

In this subsection we will study, for the same setup as in the preceding subsection, the space-time structure of the non-equilibrium slow mode contribution to the entropy density, $\Delta s(\tau,r)$. Since this observable integrates over all slow-mode wave numbers $Q$, each of which evolves differently, it provides us with a global view of the interplay between off-equilibrium effects caused by expansion and advection before and after reaching the critical region, as well as their additional enhancement by critical slowing-down in the critical region.

Figure~\ref{fig:gubser-pheno1} shows the space-time evolution of the background fluid (a) and of the critical Knudsen number (b) and non-equilibrium entropy density modification $-\Delta s_Q$ for a single representative slow mode with wave number $Q{\,=\,}1.2\,\mathrm{fm}^{-1}{\,\simeq\,}\xi^{-1}$ (c). The left panel (a) shows the evolution of the temperature contours and of the hydrodynamic flow, indicated by vectors. The middle panel (b)  demonstrates that critical Knudsen number is initially very large, due to the $1/\tau$ divergence of the longitudinal expansion rate at early times, and afterwards decays monotonically in time and also almost monotonically in radial direction. At later times ($\tau \gtrsim 4$ fm) growing radial flow causes the critical Knudsen number surface in $(\tau,r)$ to develop a weakly pronounced ridge along the direction pointing to the upper right corner. The right panel (c) shows that the slow-mode entropy correction $\Delta s_Q$ vanishes on the initial condition surface at $\tau_0{\,=\,}1$\,fm: this is a reflection of our (model-dependent) equilibrium initial conditions. Quickly thereafter, however, the large longitudinal expansion rate causes the slow mode to go out of equilibrium and generate a sizable amount of $|\Delta s_Q|$, which then does not, however, decrease with the critical Knudsen number as naively expected but, owing to the effects of critical slowing-down, remains high until the system has cooled below $T_c$ (denoted by the thick dashed white line), at which point it starts decreasing precipitously. On the $T_f{\,=\,}148$\,MeV ``freeze-out surface'' is already very small and will hardly affect the particle emission rate; the dominant phenomenological effects will likely be of second order, arising from the integrated effects of $\Delta s_Q(\tau,r)$ on the evolution history once the back-reaction onto the medium is taken into account (see Ref. \cite{Rajagopal:2019xwg}).    

\begin{figure}[!tp]
    \hspace*{-5mm}
    \includegraphics[width=1.08\linewidth]{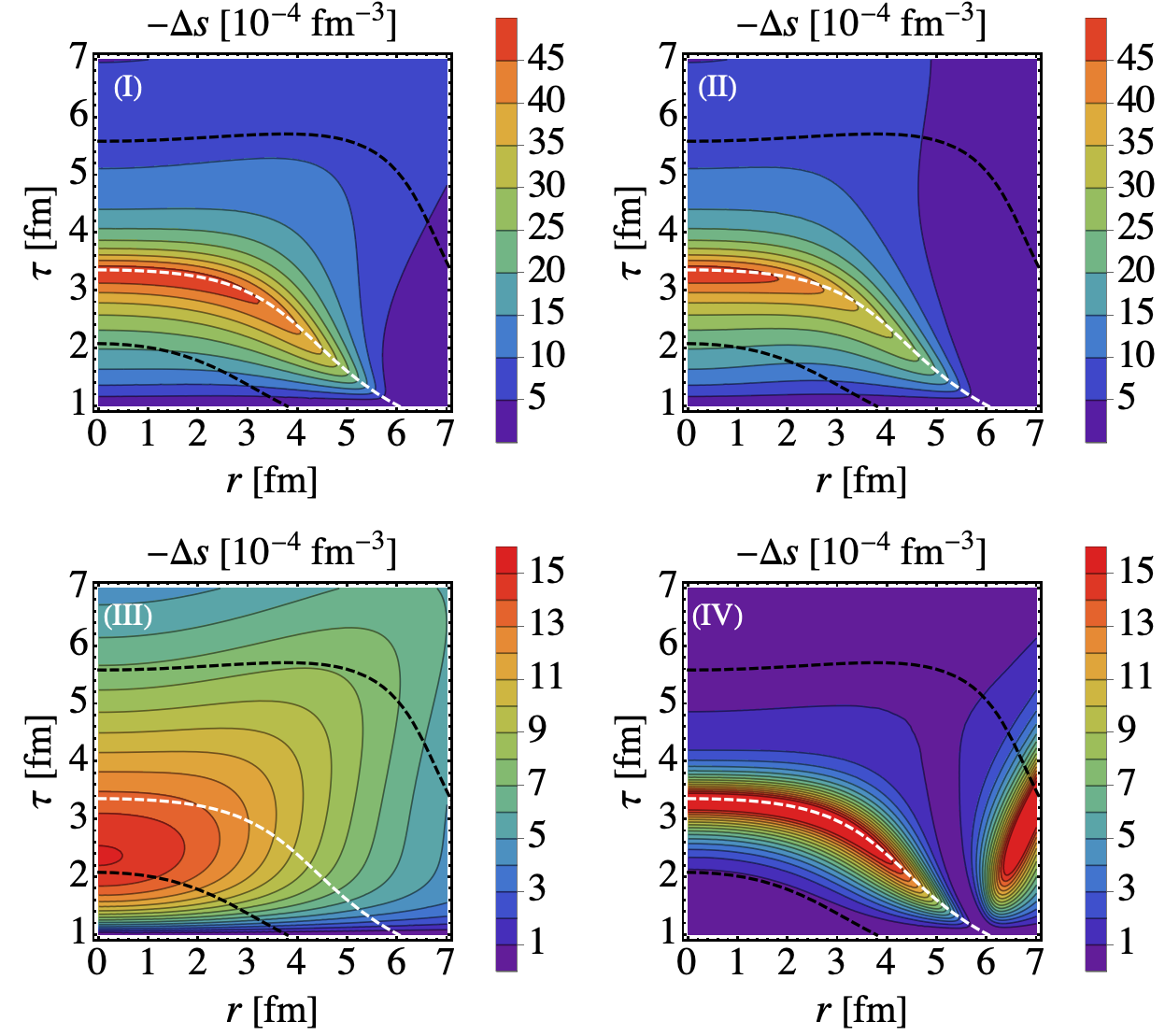}
    \caption{Full evolution of the $Q$-integrated entropy density modification $-\Delta s(\tau,r)$ for four different dynamic models: (I) full dynamics of the slow modes, (II) dynamics with the advection effects turned off by setting $u^r = 0$ in the slow-mode evolution equations, (III) dynamics with  constant correlation length $\xi = \xi_0$, and (IV) dynamics with $\Phi{\,=\,}\Phi_0$ and $\Gamma{\,=\,}\Gamma_0$, i.e. without accounting for the evolution of $\cpz/n^2$ and $\lambda_T/\cpz$. The black and white dashed lines are the same as in Fig.~\ref{fig:gubser-pheno1}.
    \label{fig:gubser-pheno2}}
\end{figure}

The $Q$-integrated entropy density modification effects from non-equilibrium slow-mode dynamics are illustrated in Fig.~\ref{fig:gubser-pheno2}. The four panels show four different dynamical models, as described in the caption. For the full dynamics shown in panel (I), the non-equilibrium entropy modification is strongly peaked near the critical isotherm. Turning off advection in panel (II) (which emulates the dynamics studied in \cite{Akamatsu:2018vjr}) pushes the contours of constant $\Delta s$ closer to the center, showing that conversely advection moves some of the non-equilibrium entropy effects to larger radii $r$. Removing the critical behavior of the correlation length $\xi(T)$ in panel (III), by setting $\xi{\,=\,}\xi_0{\,=\,}1$\,fm, strongly reduces and largely washes out the entropy modification effects; removing instead the temporal evolution of  $\cpz/n^2$ and $\lambda_T/\cpz$ from the calculation of the slow-mode damping rate and their equilibrium values in panel IV (emulating the dynamics studied in \cite{Rajagopal:2019xwg}) leads to both a reduction (by about a factor 3) and a tightening around $T_c$ of the non-equilibrium entropy modification effects. Interestingly, panel (IV) features a second branch of large entropy modification at $r{\,\gtrsim\,}6$\,fm, moving outward with the expanding fluid. This is because when the peak of $\bar\phi_Q/\bar\phi_{Q, \xi_0}$ moves inward with $T_c$, the peak of $\phi_Q/\bar\phi_{Q, \xi_0}$ from the initial condition moves outwards by advection (see Fig. \ref{fig:gubser-dynamics}, panel (Ic)). Since near the peak $\phi_Q$ is much larger than the local $\bar\phi_Q$, this yields a peak of $-\Delta s$. Something similar was also observed in Ref.~\cite{Rajagopal:2019xwg}.  

Finally, to obtain a quantitative idea about how much, in the absence of back-reaction onto the medium, non-equilibrium slow mode dynamics might be able to affect particle emission from the freeze-out hypersurface, we use Eq.~(\ref{eq:fodsgubser}) to compute the total change in entropy per unit space-time rapidity, $d\delta S/d\eta_s$, integrated over the freeze-out surface. Since $\Delta s$ drops rather precipitously below $T_c$, we can ballpark the uncertainty of this prediction by working it out on three isotherms with $T=T_c=160,\,155$, and 148\,MeV. Table~\ref{tab:modifiedS} shows the entropy modification per unit rapidity on the three isotherms $T{\,=\,}160$, 155 and 148\,MeV. We see that in all cases the absolute modifications are tiny, of order $10^{-4}$ of the unmodified value. Using the variation among the results obtained on the three different isotherms we estimate the uncertainty in our calculation of the magnitude of this tiny effect to be at the few tens of percent level. We also note that the smallness of this effect is similar in magnitude to the changes in ellipticity caused by  non-equilibrium slow-mode effects studied in Sec.~\ref{sec:advection} and consistent with the rough estimate (\ref{eq:estdeltas}).

We note that although the slow-mode contribution to the entropy density is very small, its space-time evolution still provides an interesting reflection of the off-equilibrium slow mode dynamics: Since the magnitude of $\Delta s$ traces the magnitude of the expected critical point signatures (such as cumulants of fluctuations of produced particle yields \cite{Stephanov:1998dy, Stephanov:1999zu, Hatta:2003wn, Stephanov:2008qz, Stephanov:2011pb, Luo:2017faz}), Fig.~\ref{fig:gubser-pheno2} indicates which space-time regions of the fireball might contribute most prominently to such signals.

\begin{table}[!tp]
    \centering
    \begin{tabular}{|c|c|c|c|c|c|c|c|c|}
    \hline
    \hline
     & \multicolumn{2}{|c|}{Case I} & \multicolumn{2}{|c|}{Case II} & \multicolumn{2}{|c|}{Case III} & \multicolumn{2}{|c|}{Case IV}\\
    \hline
    $|d\delta S/d\eta_s|$ & abs & $10^{-4}$ &  abs & $10^{-4}$ & abs & $10^{-4}$ & abs & $10^{-4}$ \\
    \hline
    \hline
     $T=160$\,MeV    & $1.230$  & $2.46$ & $1.019$ & $2.04$  & $0.319$ & $0.64$ & $0.446$ & $0.89$\\
     \hline
     $T=155$\,MeV  & $1.202$  & $2.40$ & $0.969$ & $1.94$  & $0.354$ & $0.71$ & $0.340$ & $0.78$\\
     \hline
     $T=148$\,MeV  & $1.056$  & $2.11$ & $0.804$ & $1.61$  & $0.409$ & $0.82$ & $0.254$ & $0.51$\\
    \hline
    \hline
    \end{tabular}
    \caption{The entropy modification $-d\delta S/d\eta_s$ from Eq.~(\ref{eq:fodsgubser}) at mid-rapidity on three isotherms with $T=160,\,155$, and 148\,MeV, for the four dynamical scenarios I-IV shown in Fig.~\ref{fig:gubser-pheno2}. ``abs'' stands for the absolute modification while the number next to it gives the relative modification (in units of $10^{-4}$), obtained by dividing by the initial total entropy content $dS/d\eta_s (\tau_0)\approx5000$ at mid-rapidity.
    \label{tab:modifiedS}}
\end{table}

%
\subsection{Limits of the {\sc hydro+} framework}
%

In this final subsection we return to the groundwork of this study laid in Sec. \ref{sec:fluct}. The {\sc hydro+} framework is based on the assumption of a separation of scales, namely $\xi\ll\ell$ where $\xi$ is the correlation length and $\ell$ the hydrodynamic homogeneity length. For a quasi-1-dimensional expansion geometry such as Gubser flow there is only one macroscopic length scale parameter describing the (in-)homogeneity of the system, related to the scalar expansion rate: $\ell \sim 1/|\theta|$. The necessary scale separation (see Sec.~\ref{sec:hydrofluct}) thus requires $\xi\theta \ll \mathcal{O}(1)$. 

\begin{figure}[!htp]
    \hspace*{-5mm}
    \includegraphics[width=1.05\linewidth]{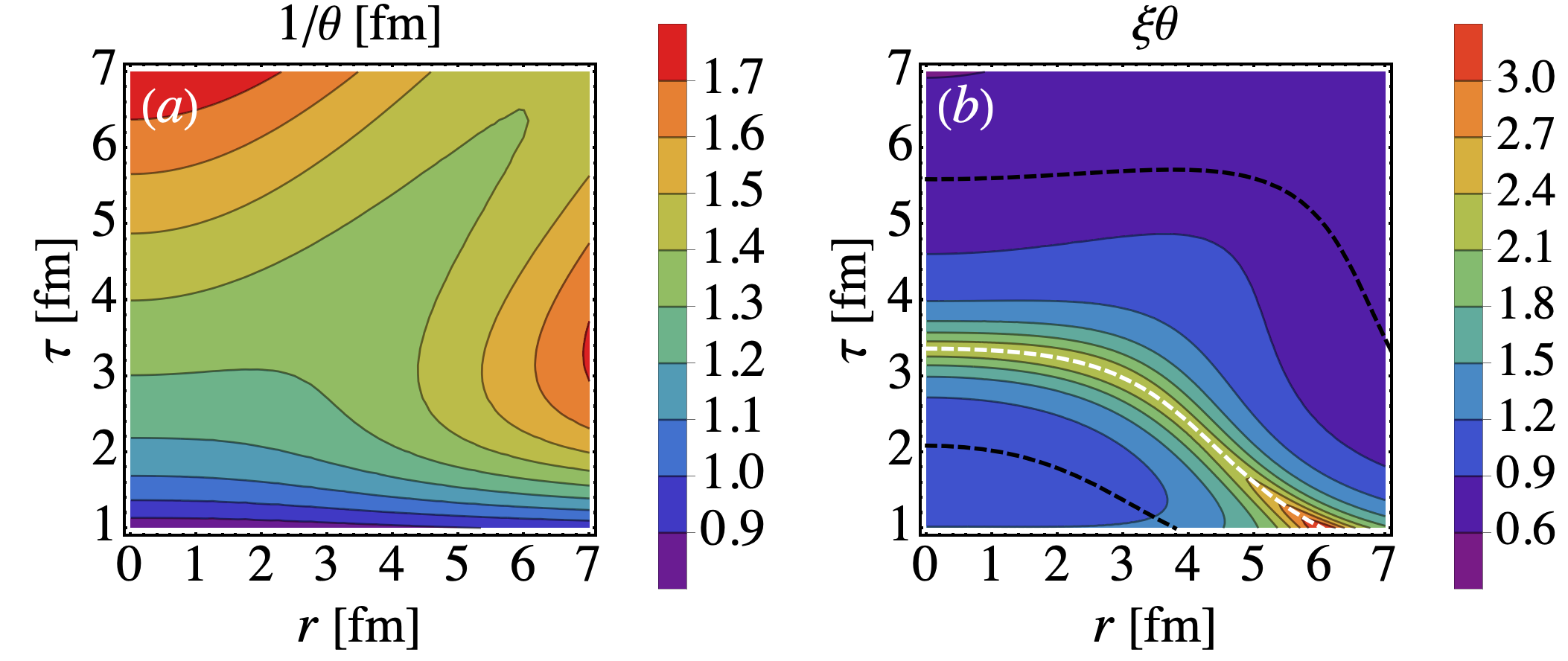}
    \caption{(a) Estimation of the hydrodynamic homogeneity length $\ell\sim 1/\theta$ for the ideal Gubser flow used in Secs. \ref{sec:corrlen} and \ref{sec:phenomenology}. (b) Estimation of the ratio $\xi/\ell \sim \xi\theta$. The black and white dashed lines are the same as in Fig.~\ref{fig:gubser-pheno1}.
    \label{fig:gubser-xitheta}}
\end{figure}

For ideal Gubser flow the scalar expansion rate (\ref{eq:expansion}) can be computed from the flow profile (\ref{eq-gubser-u}). Its inverse, proportional to the hydrodynamic homogeneity length $\ell\sim 1/\theta$, is plotted in Fig.~\ref{fig:gubser-xitheta}a. The right panel, Fig.~\ref{fig:gubser-xitheta}b, shows the ratio  $\xi/\ell \sim \xi\theta$ which needs to be sufficiently small ($<\mathcal{O}(1)$) for the {\sc hydro+} framework to be valid. One sees that the framework gets stressed mostly in a narrow region around $T_c$ but elsewhere it works well, even at very early times where the homogeneity length $\ell$ is short. If we had used a more realistic parametrization of the correlation length $\xi$ as a function of both $T$ and $\mu$, which exhibits critical growth only near a critical point $(T_c,\mu_c)$ instead of on a critical isotherm $T_c$ as here, the $(\tau,r)$ region where {\sc hydro+} might break down would shrink correspondingly.

Real heavy-ion collisions exhibit an additional feature that is not shared by Gubser flow and therefore not reflected in Fig~\ref{fig:gubser-xitheta}: event-by-event quantum fluctuations in the initial spatial energy density profile (a.k.a. ``bumpiness''). This bumpiness arises from small ratios between the nucleon and nuclear radii and between the color correlation length inside a nucleon and the nucleon's radius (see Refs.~\cite{Shen:2017bsr, Du:2018mpf} for relevant recent studies at BES energies). Initial-state density fluctuations on subnucleonic length scales are of particular phenomenological importance in collisions involving small nuclei, such as proton-proton and proton-nucleus collisions \cite{Welsh:2016siu}. They have the potential of reducing the range of validity of the {\sc hydro+} framework in small collision systems below what is shown in Fig.~\ref{fig:gubser-xitheta}b, by locally shrinking the homogeneity length $\ell$ below the inverse expansion rate plotted in Fig.~\ref{fig:gubser-xitheta}a and thereby generating local bumps for the ratio $\xi/\ell$.

%
\section{Summary and conclusions}
\label{sec:conclusion}
%

We presented a systematic study of critical slow mode evolution in an expanding quark-gluon plasma (QGP) that passes close to a critical point in the QCD phase diagram. To achieve conceptual clarity of the mechanisms controlling the critical slow mode dynamics we used an analytical model, ideal Gubser flow, for the expansion of the QGP background fluid which includes key features of the dynamics of the hot and dense medium created in relativistic heavy-ion collisions, in particular simultaneous and mutually coupled longitudinal and transverse flow. While the use of such a simplified expansion model robs us of the opportunity to make direct comparison with experimental data (this will be left for future work employing the (3+1)-dimensional numerical \bes+\ code developed in the context of the present work and briefly described in the Appendix), it provides us with the opportunity to selectively zoom in onto key mechanisms driving the critical slow mode evolution, by tuning the background flow analytically.

Just like other dissipative phenomena in a relativistic fluid, critical slow-mode dynamics is controlled by the competition between the rate of macroscopic hydrodynamic expansion (which drives the critical slow modes away from thermal equilibrium) and relaxation processes on length scales of order the correlation length $\xi$ and shorter, encoded in a wave number dependent relaxation rate $\Gamma_Q$, that help the slow modes to thermalize. Slow mode relaxation, as well as any other dissipative effects to which slow mode relaxation contributes, is affected by critical slowing down, i.e. by a dramatic reduction of the relaxation rate $\Gamma_Q$ close to the critical point where the correlation length $\xi$ for order parameter fluctuations becomes large.\footnote{%
    In fact, the critical slowing down of slow mode non-equilibrium dynamics is known to contribute, through its correction $\Delta p$ to the equilibrium pressure, to the critical slowing down of the relaxation of the bulk viscous pressure \cite{Stephanov:2017ghc, Rajagopal:2019xwg}.}
This comes in addition to a leading quadratic ($\sim Q^2$) wave number dependence of $\Gamma_Q$ which slows down the thermalization of long wavelength fluctuations already in the absence of a critical point. The competition between macroscopic expansion and microscopic relaxation is captured by the ($Q$-dependent) critical Knudsen number Kn$=\theta/\Gamma_Q$ which was shown in this work to be a good predictor for the (in-)ability of the critical slow modes to follow the dynamical evolution (via expansion of the background fluid) of their space-time dependent equilibrium value.

An important aspect of critical slow-mode dynamics in an expanding background is the phenomenon of advection, i.e. the outward transport of the slow mode with the expanding fluid by collective transverse flow which was ignored in some earlier work (e.g. \cite{Akamatsu:2018vjr}): As the system cools by longitudinal expansion, the critical surface $T(\tau,r)=T_c$ moves inward but, especially for small $Q$ where relaxation is anyhow suppressed, the critical maximum of the slow mode doesn't follow closely that inward motion of the critical surface but may instead even move outward, driven by outward radial flow transverse to the beam direction.

The present work is, to the best of our knowledge, the first one that studies {\it critical slow mode dynamics} in its full complexity\footnote{%
    Albeit not its back-reaction on the expanding fluid itself.} 
in a more or less realistic setting for relativistic heavy-ion collisions. We presented the space-time evolution of the full spectrum of wave numbers $Q$, as temporal profiles at fixed locations and as snapshots of spatial profiles at varying times, and we also computed their contribution to the overall entropy balance, in both space and time, in order to gauge the importance of feedback effects of critical slow mode dynamics on the hydrodynamical bulk evolution. While the critical slow modes $\phi_Q$ are expected to make a substantial contribution to fluctuation observables, we found that their corrections to the bulk entropy density and pressure, as well as to other macroscopic characteristics of the expanding fireball such as its elliptic geometric deformation $\epsilon_2$ in non-central heavy-ion collisions, are exceedingly small, of relative order $10^{-4}$. This confirms similar estimates presented in Ref.~\cite{Rajagopal:2019xwg} for a simpler dynamical setting for the slow modes. We therefore expect this feature to survive in upcoming fully realistic (3+1)-dimensional numerical simulations of the coupled macroscopic near-equilibrium expansion and the microscopic non-equilibrium kinetic slow mode dynamics that include all back-reaction effects; the tools for performing such a comprehensive study were developed and presented in this work. 

Confirming the smallness of back-reaction effects from non-equilibrium critical fluctuation dynamics onto the bulk properties of the fluid will be important for two reasons: it will firmly direct our attention away from bulk hydrodynamic features and towards direct fluctuation measurements when searching for critical point signatures, and it will simplify the description of the hydrodynamic fireball evolution by allowing us to ignore back-reaction effects without noticeable loss of accuracy.


\section*{Acknowledgements} 
The authors thank Xin An, Chandrodoy Chattopadhyay, Mike McNelis, Greg Ridgway and Ryan Weller for fruitful discussions. 
This work was supported by the U.S. Department of Energy, Office of Science, Office for Nuclear Physics under Awards No. \rm{DE-SC0004286} (L.D. and U.H.) and \rm{DE-SC0011090} (K.R. and Y.Y.) as well as within the framework of the Beam Energy Scan Theory (BEST) Collaboration. L.D. and U.H. acknowledge additional partial support by the National Science Foundation (NSF) within the framework of the JETSCAPE Collaboration under Award No. \rm{ACI-1550223}. 
Y.Y. acknowledges additional support from the Strategic Priority Research Program of Chinese Academy of Sciences, Grant No. XDB34000000.
L.D. appreciates the kind hospitality and support of the MIT Center for Theoretical Physics during the initial stages of this work. U.H. acknowledges the kind hospitality of the Institut f\"{u}r Theoretische Physik of the J.\,W.\,Goethe-Universit\"{a}t during parts of this project, supported by a Research Prize from the Alexander von Humboldt Foundation.

\appendix
%
\section{Validation of \bes+}\label{sec:validation}
%

Although in this work we studied the evolution of the slow modes on top of a fixed ideal Gubser flow, with conformal EoS and without back-reaction, the {\sc hydro+} framework is embedded in the \bes~code \cite{Du:2019obx}, which can simulate the dissipative hydrodynamics at non-zero baryon density with realistic EoS for any expansion geometry; it also properly couples the evolution of the background fluid and the slow modes by including the back-reaction. In this Appendix we illustrate some numerical methods and the validation of \bes+.

At non-zero baryon density, one needs solve the ``root finding" problem \cite{Du:2019obx}, where one computes $(e, n)$ in LRF and $u^\mu$ from the energy-momentum tensor and net baryon current in the global computational frame
\begin{eqnarray}
    T^{\mu\nu} &=& e\, u^{\mu}u^{\nu}-p_\plus\Delta^{\mu\nu}+\pi^{\mu\nu}\;,\\
    N^{\mu} &=& n\, u^{\mu} +\V^\mu \;, 
\end{eqnarray}
where $p_\plus$ is given by,
\begin{equation}
    p_\plus(e, n) = p(e, n) + \D p(e, n, \phi(e, n))\,,
\end{equation}
which includes the back-reaction $\Delta p$ from Eq.~(\ref{eq:dp}). Here $p(e,n)$ implicitly includes the bulk viscous pressure, i.e. $p(e, n)=p_\mathrm{eq}(e, n)+\Pi$, and the equilibrium pressure $p_\mathrm{eq}(e, n)$ is given by the EoS.

The modified root finder in Ref.~\cite{Du:2019obx} can be extended to include contributions from the slow modes to solve the root finding problem in \bes+. We introduce
\begin{eqnarray}
  M^{\tau} &=& T^{\tau\tau}-\pi^{\tau\tau} = (e+p_\plus) \bigl(u^{\tau}\bigr)^{2} - p_\plus\,,
\\
  M^{i} &=& T^{\tau i}-\pi^{\tau i} = (e+p_\plus) u^{\tau} u^{i} \quad (i = x, y, \eta_s)\,,\quad
\\
  J^\tau &=& N^{\tau}-n^\tau =\n u^{\tau}\,,
\end{eqnarray}
and the flow speed, $v\equiv\sqrt{1{\,-\,}1/(u^\tau)^2}$. Then we solve iteratively
\begin{equation}
    v \equiv \frac{M}{M^\tau + p_\plus} = \frac{M}{M^\tau + p(e, n) + \D p(e, n, \phi(e, n))}\;,\label{root-v}
\end{equation}
where $M \equiv \sqrt{(M^x)^2+(M^y)^2+\tau^2(M^\eta)^2}$ and $(e, n)$ are both functions of $v$:
\begin{equation}
  \ed(v) = M^\tau - vM\,,\quad
  \n(v) = J^\tau\sqrt{1{\,-\,}v^2}\,.
\label{root-n}
\end{equation}
When $\Delta p$ is not added to Eq.~(\ref{root-v}) (as done in this work) the back-reaction is off and has no effects on the evolution of the background fluid.

%
\begin{figure}[b]
    \hspace*{-5mm}
    \includegraphics[width= 0.5\textwidth]{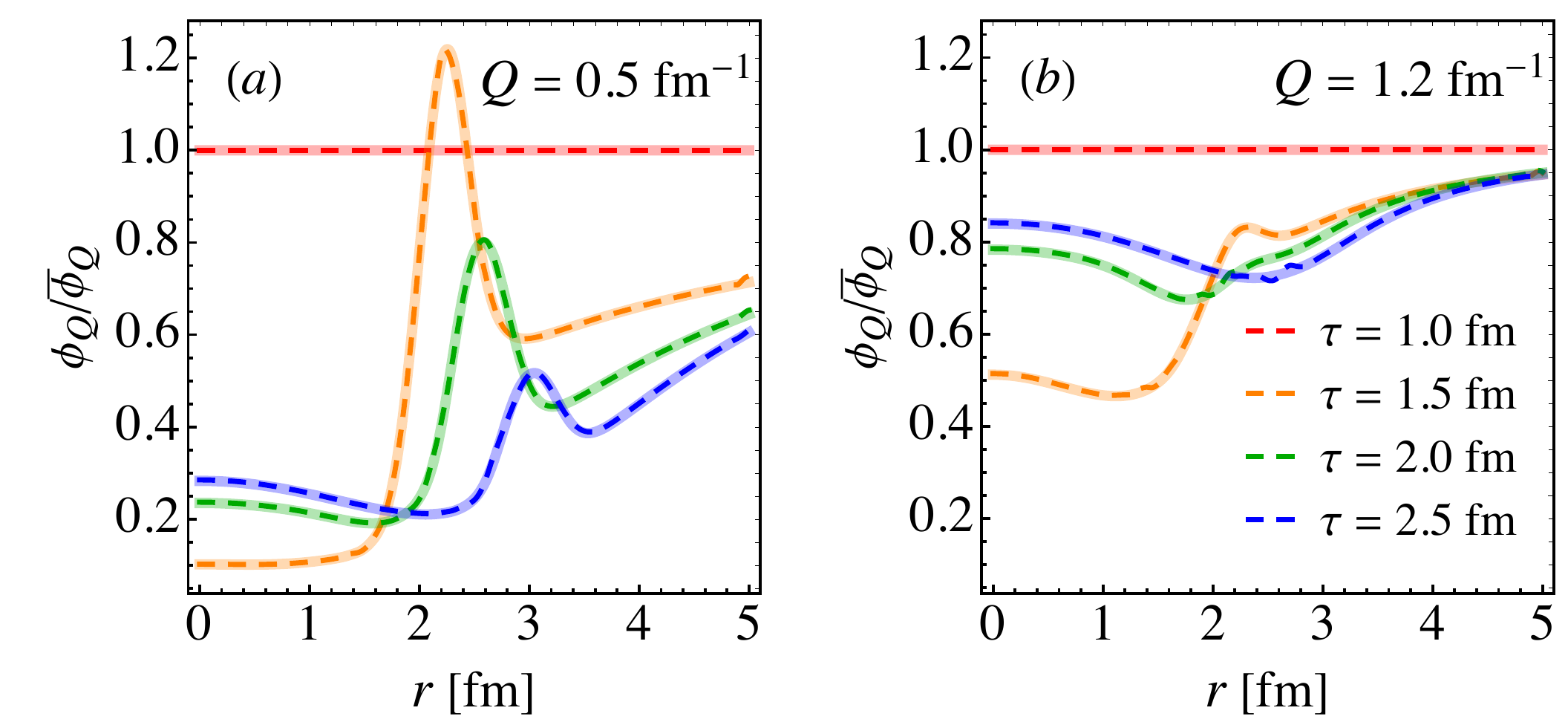}
    \caption{Comparison of $\phi_Q/\bar\phi_Q$ from semi-analytical solutions (solid lines) and numerical results from \bes+ (dashed lines) for two slow modes with (a) $Q{\,=\,}0.5$\,\fm\ and (b) $Q{\,=\,}1.2$\,\fm. For the comparison, the ideal Gubser flow provided by \bes \cite{Du:2019obx} was used with $\tau_0{\,=\,}1$\,fm, $q{\,=\,}1$\,\fm\ and $C{\,=\,}1.2$; for critical dynamics, the parametrization of the correlation length in Eq.~(\ref{eq:xipara2}) was used; otherwise the setup is the same as in Sec.~\ref{sec:corrlen}. The overall agreement is excellent, though some small wiggles can be observed in the numerical solution which originate from the numerical methods used to simulate the background fluid (see also Ref.~\cite{Du:2019obx}).
    \label{fig-validation}}
\end{figure}
%

Another numerical issue involves solving the equations of motion (\ref{eq:phi_scaling}) for the slow modes which share similarities with the evolution equations for the dissipative flows, including $\V^\mu$, $\pi^{\mu\nu}$ and especially $\Pi$. For the dissipative flows those equations are numerically solved by the Kurganov-Tadmor (KT) algorithm \cite{KURGANOV2000241}, with a second-order explicit Runge-Kutta (RK) ordinary differential equation solver \cite{leveque_2002} for the time integration step in \bes~\cite{Du:2019obx}, after being written in first-order flux-conserving form. Similarly, Eq.~(\ref{eq:phi_scaling}) can be rewritten in the same form as
\begin{eqnarray}
\!\!\!\!
    \partial_\tau \phi_Q + \partial_x (v^x \phi_Q) + \partial_y (v^y \phi_Q) + \partial_\eta (v^\eta \phi_Q) = S_Q,\quad 
\label{eq:eomre}
\end{eqnarray}
where $v^i \equiv u^i/u^\tau$ ($i = x, y, \eta_s$) is the 3-velocity of the fluid and $S_Q$ is the source term
\begin{equation}
    S_Q=-\frac{1}{u^\tau}\Gamma_Q\left(\phi_Q-\bar\phi_Q\right)+\phi_Q\partial_i v^i\;.
\end{equation}
Here we used $\partial_{i}v^{i} \equiv \partial_{x}v^{x} + \partial_{y}v^{y} + \partial_{\eta}v^{\eta}$. The slow mode equations can then be solved using the same KT-RK algorithm by straightforwardly extending \bes.

\bes~has been tested by comparing to semi-analytical solutions \cite{Du:2019obx}, and in the same spirit, we validate the numerical methods of the extened root finder (\ref{root-v}) and equations of motion (\ref{eq:eomre}) involving the slow modes, using the setup described in the caption of Fig. \ref{fig-validation}. As one can see from the figure, the agreement is excellent. Once such a numerically precise evolution of the slow modes has been achieved it is easy to derive the remaining off-equilibrium corrections, e.g. those to the pressure and entropy. The code is open source and can be freely downloaded from \url{https://github.com/LipeiDu/BEShydro}. Interested readers are encouraged to repeat the test with the setup described in the \texttt{HydroPlus} branch, especially after they make improvements to the code.

\bibliography{hydroplus}

\end{document}